\documentclass[prd, 10pt,aps,nofootinbib, floatfix, notitlepage,twocolumn,longbibliography]{revtex4-1}

\usepackage{placeins}
\usepackage{amsmath,amsfonts,amssymb}
\usepackage{mathrsfs}
\usepackage{graphicx}
\usepackage[english]{babel} 
\usepackage{ulem}
\usepackage{url} 
\usepackage{color}
\usepackage{bbold}
\usepackage{adjustbox}
 \usepackage[utf8]{inputenc} 
\usepackage[colorlinks=true,citecolor=blue,urlcolor=blue]{hyperref}

\newcommand{\Faxdm}{r_a}
\newcommand{\zax}{z_a}

\newcommand{\zpt}{z_{\text{dde}}}
\newcommand{\fpt}{F_{\text{dde}}}
\newcommand{\mt}{m_{\text{t}}}
\newcommand{\Tds}{T_{\text{ds}}}
\newcommand{\mpl}{M_{\text{\tiny{Pl}}}}
\newcommand{\ax}{a}

\definecolor{green2}{RGB}{34,139,34}

\begin{document}

\title{Dark Sector to Restore Cosmological Concordance}

\author{Itamar~J.~Allali}
\email{itamar.allali@tufts.edu}
\author{Mark P.~Hertzberg}
\email{mark.hertzberg@tufts.edu}
\author{Fabrizio Rompineve}
\email{fabrizio.rompineve@tufts.edu}
\affiliation{Institute of Cosmology, Department of Physics and Astronomy, Tufts University, Medford, MA 02155, USA
\looseness=-1}

\date{\today}

\begin{abstract}

\noindent We develop a new phenomenological model that addresses current tensions between observations of the early and late Universe. 
Our scenario features: (i) a decaying dark energy fluid (DDE), which undergoes a transition at $z \sim 5,000$, to raise today's value of the Hubble parameter -- addressing the ``$H_0$ tension,” and (ii) an ultra-light axion (ULA), which starts oscillating at $z\gtrsim 10^4$, to suppress the matter power spectrum -- addressing the ``$S_8$ tension." Our Markov Chain Monte Carlo analyses show that such a Dark Sector model fits a combination of Cosmic Microwave Background (CMB), Baryon Acoustic Oscillations, and Large Scale Structure (LSS) data slightly better than the $\Lambda$CDM model, while importantly reducing both the $H_0$ and $S_8$ tensions with late universe probes ($\lesssim 3\sigma$).
Combined with measurements from cosmic shear surveys, we find that the discrepancy on $S_8$ is reduced to the $1.4\sigma$ level, and the value of $H_0$ is further raised. Adding local supernovae measurements, we find that the $H_0$ and $S_8$ tensions are reduced to the $1.4\sigma$ and $1.2\sigma$ level respectively, with a significant improvement $\Delta\chi^2\simeq -18$ compared to the $\Lambda$CDM model. With this complete dataset, the DDE and ULA are detected at $\simeq4\sigma$ and $\simeq2\sigma$, respectively. We discuss a possible particle physics realization of this model, with a dark confining gauge sector and its associated axion, although embedding the full details within microphysics remains an urgent open question. Our scenario will be decisively probed with future CMB and LSS surveys.

\end{abstract}

\maketitle

{\bf Introduction}---Observations by the Planck satellite of the Cosmic Microwave Background (CMB) indicate a Universe expanding today at a (Hubble) rate of $H_0=67.27\pm 0.60$ km/s/Mpc~\cite{Aghanim:2019ame}, assuming the ``$\Lambda$CDM model". This is in strong ($4.4\sigma$) tension with local measurements based on supernovae from the SH$_0$ES collaboration~\cite{Riess:2019cxk}, which report a faster rate $H_0=74.03\pm 1.42$ km/s/Mpc (see also~\cite{Riess:2020fzl}). The discrepancy between early and late Universe determinations of $H_0$ appears to be supported by several other probes (e.g., lensing time delays~\cite{Wong:2019kwg}, and Baryon Acoustic Oscillations (BAO) and BOSS galaxy clustering data analyzed with the Effective Field Theory of Large Scale Structure \cite{Baumann:2010tm, Carrasco:2012cv, Hertzberg:2012qn, Colas:2019ret, DAmico:2019fhj, Ivanov:2019pdj} (EFTofLSS)).

Further disagreement arises in the determination of the amplitude of the matter power spectrum at late times, which is often parameterized by means of the combination $S_8\equiv \sigma_8\sqrt{\Omega_m/0.3}$ ($\Omega_m$ and $\sigma_8$ being respectively the total relic abundance of non-relativistic matter and the variance of matter fluctuations in a sphere of radius $8~\text{Mpc}/h$ today). In particular, a recent combination of data from cosmic shear surveys finds $S_8=0.755^{+0.019}_{-0.021}$~\cite{Asgari:2019fkq, Joudaki:2019pmv}, in $3.2\sigma$ tension with the value inferred by the Planck collaboration (see also~\cite{Heymans:2013fya, Hildebrandt:2018yau, Abbott:2017wau, Hikage:2018qbn, Heymans:2020gsg} for other measurements).

A resolution of these tensions based on systematic errors is currently lacking. It is possible that the above discrepancies may be resolved instead by modifying the cosmological ($\Lambda$CDM) model used to infer values of parameters from early Universe probes. A notable attempt in this direction is the addition of an Early Dark Energy (EDE) component~\cite{Poulin:2018cxd} (see also \cite{Agrawal:2019lmo, Lin:2019qug}) that is very rapidly diluted after the epoch of matter radiation equality. This scenario significantly alleviates the Hubble tension when fitted to a combination of Planck, BAO, and supernovae data, but exacerbates the $S_8$ tension. Therefore, when cosmic shear as well as EFTofLSS data are included, the resolving power of EDE is reduced~\cite{Hill:2020osr, Ivanov:2020ril, DAmico:2020ods, Smith:2020rxx, Murgia:2020ryi}. On the fundamental physics side, EDE relies on an ultra light scalar field  with a highly tuned potential~\cite{Gonzalez:2020fdy}. A somewhat more particle-physics-oriented scenario is that of a strong first order phase transition in a weakly coupled scalar field at the eV scale, as proposed in the New Early Dark Energy (NEDE) scenario~\cite{Niedermann:2019olb, Niedermann:2020dwg, Niedermann:2020qbw}. However, this similarly increases the $S_8$ tension, while other well-motivated scenarios, such as decay through resonance~\cite{Gonzalez:2020fdy} (see also~\cite{Kaloper:2019lpl}) or modified gravity models (see e.g.~\cite{Lin:2018nxe, Sola:2019jek, Sakstein:2019fmf, Zumalacarregui:2020cjh, Ballesteros:2020sik, Braglia:2020iik, Braglia:2020auw}) struggle to provide convincing solutions.

In this Letter, we propose a new phenomenological Dark Sector (DS) model, which is instead able to more fully restore cosmological concordance by predicting both a larger expansion rate and a suppressed matter power spectrum at late times. Our model features both a decaying dark energy (DDE) component, which addresses the Hubble tension similarly to the EDE and NEDE scenarios, as well as an ultra-light axion (ULA) field with a standard potential and generic initial conditions. By virtue of the misalignment mechanism, this axion contributes a fraction of the relic abundance of DM today. However, in contrast to CDM, it causes a suppression of power on small scales, due to the scale-dependent sound speed of its perturbations~\cite{Khlopov:1985jw, Hu:1998kj, Hwang:2009js}, and thus addresses the $S_8$ tension (see also~\cite{Fung:2021wbz} for a different ULA model with similar goals). 
We test our DS model against a wide array of cosmological datasets, by performing dedicated Markov Chain Monte Carlo (MCMC) analyses.

This phenomenological model is useful to demonstrate the ingredients that appear to be required for all observations to fit together.
As an example, our DS model may arise microscopically within a dark gauge theory sector, which confines via a first order phase transition (PT) slightly above the eV scale. The associated axion then naturally receives a small mass from gauge instantons, very much like the well-known case of the QCD axion. Assessing whether the required rapid dilution of the DS after the PT can be realized in a fundamental physics scenario remains a very interesting and urgent task.

{\bf Phenomenological Model}---Our DS model is composed of two ingredients:
\begin{itemize}
\item[$\mathbf{1}$]{A DDE fluid which undergoes a sharp transition at some redshift $\zpt$, after which its equation of state parameter changes from $w=-1$ to $w=w_f>1/3$.}
\item[$\mathbf{2}$]{An ULA field $\ax$ with a potential of the standard form $V = m_a^2 f_a^2~(1-\cos\ax/f_a)$, where $m_a$ and $f_a$ are the axion mass and decay constant, respectively.}
\end{itemize}

For the DDE fluid, we adopt the effective fluid modeling put forth in the NEDE scenario of~\cite{Niedermann:2019olb, Niedermann:2020dwg} to perform a concrete numerical analysis. This model is general enough to capture the effective behavior of several possible microscopic scenarios. Its crucial features are: At the background level, the transition at $\zpt$ is assumed to occur in much less than a Hubble time, and thus modeled as instantaneous. The redshift $\zpt$ is set by a subdominant ``trigger'' scalar field of mass $\mt$ once the rolling condition $H\simeq\mt$ is satisfied. Cosmological perturbations of the DDE fluid are initially set to vanish and then re-initialized around $\zpt$ by using as initial conditions the perturbations of the trigger field. Subsequently, they are treated as those of an ideal cosmic fluid with adiabatic sound speed $c_s^2=w_f$. Overall, the NEDE/DDE fluid introduces four extra parameters to the $\Lambda$CDM model: (i) the fraction $\fpt$ of the energy density in the DDE fluid at $z\geq \zpt$, (ii) the mass of the trigger field, or equivalently, the redshift $\zpt$ of the transition\footnote{Strictly speaking, $\zpt$ also depends mildly on $H(z)$, which is affected by the presence of the DDE, but since $\fpt\ll 1$, $\zpt$ is nearly fixed once a value of $\mt$ is specified. A similar remark applies to the ULA component.}, (iii) the equation of state parameter $w_f$, and (iv) the precise value of the ratio $H/\mt$ at which the trigger field starts rolling. However, the latter two parameters would be fixed once a particle physics model is specified. Thus, we fix $H/\mt=0.2$, justified by the dynamics of a generic scalar field. To set ideas, we also fix $w_f=2/3$, as in \cite{Niedermann:2020dwg}, although our conclusions are not strongly affected by the precise value of $w_f$, as long as $w_f\gtrsim 0.5$, while for smaller values the model falls short of alleviating the $H_0$ tension, see also~\cite{Lin:2019qug, Niedermann:2020dwg}. 
Overall, this leaves just two free parameters from the DDE component. In our numerical analysis, this is treated exactly like the NEDE fluid of~\cite{Niedermann:2019olb, Niedermann:2020dwg}, while its microphysical origin may be different; see the discussion, where we also present some tentative ideas to achieve $w_f>1/3$.

\begin{table*}
\begin{tabular} {| l | c| c| c| c| c| c| c| c|}
\hline\hline
 \multicolumn{1}{|c|}{ Parameter} & \multicolumn{2}{|c|}{~~~~~P18+BAO~~~~~} & \multicolumn{2}{|c|}{~~~~P18+BAO+EFT~~~~} & \multicolumn{2}{|c|}{~~P18+BAO+EFT+S$_8$~~} & \multicolumn{2}{|c|}{~P18+BAO+EFT+S$_8$+SN+H$_0$~}\\
\hline\hline
$100 \omega_b              $ & \multicolumn{2}{|c|}{$2.267~(2.277)^{+0.022}_{-0.026}   $} & \multicolumn{2}{|c|}{$2.265~(2.289)^{+0.020}_{-0.027}   $} & \multicolumn{2}{|c|}{$2.274~(2.28)^{+0.020}_{-0.026}   $} & \multicolumn{2}{|c|}{$2.303~(2.295)^{+0.023}_{-0.025}   $}\\
$\omega_{cdm }             $ & \multicolumn{2}{|c|}{$0.1241~(0.1261)^{+0.0031}_{-0.0044}$} & \multicolumn{2}{|c|}{$0.1227~(0.127)^{+0.0027}_{-0.0040}$} & \multicolumn{2}{|c|}{$0.1191~(0.12)^{+0.0025}_{-0.0035}$} & \multicolumn{2}{|c|}{$0.1235~(0.1238)^{+0.0030}_{-0.0029}$}\\
$\ln 10^{10}A_s            $ & \multicolumn{2}{|c|}{$3.057~(3.051)^{+0.015}_{-0.015}   $} & \multicolumn{2}{|c|}{$3.054~(3.058)^{+0.015}_{-0.015}   $} & \multicolumn{2}{|c|}{$3.050~(3.047)^{+0.015}_{-0.015}   $} & \multicolumn{2}{|c|}{$3.062~(3.057)^{+0.015}_{-0.015}   $}\\
$n_{s }                    $ & \multicolumn{2}{|c|}{$0.9761~(0.9784)^{+0.0074}_{-0.0089}$} & \multicolumn{2}{|c|}{$0.9743~(0.9864)^{+0.0067}_{-0.0087}$} & \multicolumn{2}{|c|}{$0.9738~(0.9748)^{+0.0065}_{-0.0083}$} & \multicolumn{2}{|c|}{$0.9860~(0.9828)^{+0.0065}_{-0.0066}$}\\
$\tau_{reio }              $ & \multicolumn{2}{|c|}{$0.0565~(0.0518)^{+0.0068}_{-0.0075}$} & \multicolumn{2}{|c|}{$0.0561~(0.0551)^{+0.0068}_{-0.0075}$} & \multicolumn{2}{|c|}{$0.0557~(0.0545)^{+0.0071}_{-0.0071}$} & \multicolumn{2}{|c|}{$0.0574~(0.0562)^{+0.0069}_{-0.0077}$}\\
$H_0\,[\text{km/s/Mpc}]                       $ & \multicolumn{2}{|c|}{$69.3~(69.3)^{+1.0}_{-1.4}        $} & \multicolumn{2}{|c|}{$69.09~(70.58)^{+0.86}_{-1.4}      $} & \multicolumn{2}{|c|}{$69.37~(70.02)^{+0.85}_{-1.4}      $} & \multicolumn{2}{|c|}{$71.56~(70.99)^{+0.98}_{-0.98}     $}\\
\hline
$F_{dde}                   $ & \multicolumn{2}{|c|}{$< 0.137~[95\%]~(0.077)$} & \multicolumn{2}{|c|}{$< 0.124~[95\%]~(0.11)$} & \multicolumn{2}{|c|}{$< 0.127~[95\%]~(0.073)$} & \multicolumn{2}{|c|}{$0.124~(0.123)^{+0.034}_{-0.029}   $}\\
$z_{dde}                   $ & \multicolumn{2}{|c|}{$5168~(5452)^{+1100}_{-1300}  $} & \multicolumn{2}{|c|}{$5193~(5352)^{+1300}_{-1600}  $} & \multicolumn{2}{|c|}{$5055~(4440)^{+1300}_{-1600}  $} & \multicolumn{2}{|c|}{$4749~(4894)^{+640}_{-820}  $}\\
$r_a\equiv\Omega_a/\Omega_{\text{dm}}                       $ & \multicolumn{2}{|c|}{$< 0.032~[95\%]~(0.005)$} & \multicolumn{2}{|c|}{$< 0.039~[95\%]~(0.014)$} & \multicolumn{2}{|c|}{$< 0.069~[95\%]~(0.037)$} & \multicolumn{2}{|c|}{$0.048~(0.052)^{+0.017}_{-0.017}   $}\\
\hline
$\log_{10}z_a$ &\multicolumn{2}{|c|}{fixed to: 4.2} & \multicolumn{2}{|c|}{fixed to: 4.2} & \multicolumn{2}{|c|}{fixed to: 4.2} & \multicolumn{2}{|c|}{fixed to: 4.2}\\
$m_a\,[10^{-26}\,\mbox{eV}]                       $ & \multicolumn{2}{|c|}{$(1.15)$} & \multicolumn{2}{|c|}{$(1.15)$} & \multicolumn{2}{|c|}{$(1.14)$} & \multicolumn{2}{|c|}{$(1.15)$}\\
$f_a\,[10^{16}\,\mbox{GeV}]                       $ & \multicolumn{2}{|c|}{$< 9.565~[95\%]~(3.816)$} & \multicolumn{2}{|c|}{$< 10.438~[95\%]~(6.114)$} & \multicolumn{2}{|c|}{$< 14.34~[95\%]~(9.908)$} & \multicolumn{2}{|c|}{$11.2~(12.0)^{+2.4}_{-1.9}        $}\\
$S_8                       $ & \multicolumn{2}{|c|}{$0.827~(0.838)^{+0.016}_{-0.013}   $} & \multicolumn{2}{|c|}{$0.820~(0.826)^{+0.017}_{-0.014}   $} & \multicolumn{2}{|c|}{$0.788~(0.783)^{+0.016}_{-0.015}   $} & \multicolumn{2}{|c|}{$0.784~(0.789)^{+0.014}_{-0.014}   $}\\
\hline
 &~~$\Lambda$CDM~~ & ~~DS~~ & ~~$\Lambda$CDM~~ & ~~DS~~ & ~~$\Lambda$CDM~~ & ~~DS~~ & ~~$\Lambda$CDM~~ & ~~DS~~\\
Tension with SH$_0$ES & $4.4\sigma $ & $2.7\,\sigma$ & $4.3\sigma $ & $3.0\,\sigma$ & $4.0\sigma $ & $2.8\,\sigma$ & $3.7\sigma $ & $1.4\,\sigma$\\
Tension with S$_8$ & $3.3\sigma $ & $3.1\,\sigma$ & $3.2\sigma $ & $2.8\,\sigma$ & $2.6\sigma $ & $1.4\,\sigma$ & $2.2\sigma $ & $1.2\,\sigma$\\
\hline
$\chi^2_{DS}-\chi^2_{\Lambda CDM}$ & \multicolumn{2}{|c|}{$-4.0$} & \multicolumn{2}{|c|}{$-1.6$} & \multicolumn{2}{|c|}{$-7.7$} & \multicolumn{2}{|c|}{$-17.9$}\\
\hline
\end{tabular}
  \caption{The mean (best-fit) $\pm1\sigma$ error of the cosmological parameters obtained by fitting our three-parameter DS model to the four cosmological datasets described in the text. Upper bounds are presented at 95\% CL. The discrepancy of the inferred values of $H_0$ and $S_8$ (both for $\Lambda$CDM and for the DS model) with respect to SH$_0$ES and the combined analysis of~\cite{Asgari:2019fkq} respectively is shown, as well as the improvement in $\chi^2$ with respect to $\Lambda$CDM (using the same datasets, see Appendix~\ref{sec:numerics}).}
  \label{table:results}
\end{table*}

Let us now comment on the ULA component. Similar to the DDE fluid above, at early times such an axion field behaves as dark energy with $w=-1$. Once $H\lesssim m_a$, the axion starts oscillating and eventually behaves as a dark matter component at late times, according to the misalignment mechanism. However, its effects on the growth of structures can deviate crucially from those of cold dark matter (DM). Indeed, the effective sound speed of ULA perturbations is scale-dependent~\cite{Khlopov:1985jw, Hu:1998kj, Hwang:2009js}. Therefore, in a Universe where the DM is made of ULAs, sub-horizon matter perturbations with wavenumbers above the axion Jeans wavenumber $k_J/\ax = 6^{1/4}\sqrt{H~m_a}$ do not grow during matter domination, but rather oscillate (see also~\cite{Kobayashi:2017jcf} for a recent discussion). In a Universe where an ULA makes up a fraction $\Faxdm\equiv \Omega_a/\Omega_{\text{dm}}$ of the DM, it can be shown that the suppression of the matter power spectrum is roughly $\left(\mathcal{P}^{k}_{a+cdm}/\mathcal{P}^k_{cdm}\right)_{k>k_{J,0}}\sim \left(k_{J, \text{eq}}/k\right)^{8(1-\gamma)}$~\cite{Kobayashi:2017jcf}, where $\gamma=(-1+\sqrt{25-24 \Faxdm})/4$ and $k_{J, \text{eq}}/a_0 \simeq 0.09~\text{Mpc}^{-1}(m_{a}/10^{-26}~\text{eV})$ is the Jeans wavenumber at equality. This estimate suggests that suppression of $\sim 7\%$ of the matter power spectrum at the scales probed by the $S_8$ parameter can be obtained if the Universe contains an axion with $m_a\lesssim 10^{-26}$~eV and $\Faxdm\sim 0.05$.

At the particle physics level, an ULA is fully described by the additional three parameters: (v) its mass $m_a$, (vi) its decay constant $f_a$, and (vii) its initial field value $\theta_i=a_i/f_a$. However, this last parameter is most reasonably $\mathcal{O}(1)$, unless further tuning or model building is invoked. We choose a typical value $\theta_i = 2$, although the precise choice does not alter our conclusions. Once this parameter is fixed, $m_a$ and $f_a$ can be traded for the redshift at which axion oscillations begin, $\zax$, and the fraction of the total energy density in the axion field at $\zax$. The latter can alternatively be replaced by $r_a$.
This leaves us with two parameters from this component also.

The two ingredients that make up our DS should both feature a transition a little before the epoch of matter-radiation equality, if they are to address cosmological tensions. This opens up the possibility of a common fundamental origin, which we will discuss later.


{\bf Datasets and Results}---We have numerically implemented the DS model presented in the previous section, by merging two publicly available extensions of the Boltzmann code {\tt CLASS}~\cite{Blas:2011rf}: {\tt TriggerCLASS}\footnote{\href{https://github.com/flo1984/TriggerCLASS}{\tt https://github.com/flo1984/TriggerCLASS}}, developed in~\cite{Niedermann:2019olb, Niedermann:2020dwg} to study the NEDE scenario, and {\tt AxiCLASS}\footnote{\href{https://github.com/PoulinV/AxiCLASS}{\tt https://github.com/PoulinV/AxiCLASS}}, developed in~\cite{Poulin:2018dzj, Smith:2019ihp}. This latter code uses the state-of-the-art effective fluid model of~\cite{Poulin:2018dzj} to compute the cosmological implications of ULAs.
We have then performed an MCMC analysis of our DS model, using the {\tt MontePython} sampler~\cite{Audren:2012wb, Brinckmann:2018cvx}\footnote{\href{https://github.com/brinckmann/montepython_public}{\tt https://github.com/brinckmann/montepython\_public}} also to find the $\chi^2$, while we analyzed and plotted posterior distributions using {\tt GetDist}~\cite{Lewis:2019xzd}\footnote{\href{https://getdist.readthedocs.io}{\tt https://getdist.readthedocs.io}}. 

\begin{figure}[t]
\includegraphics[width=\columnwidth]{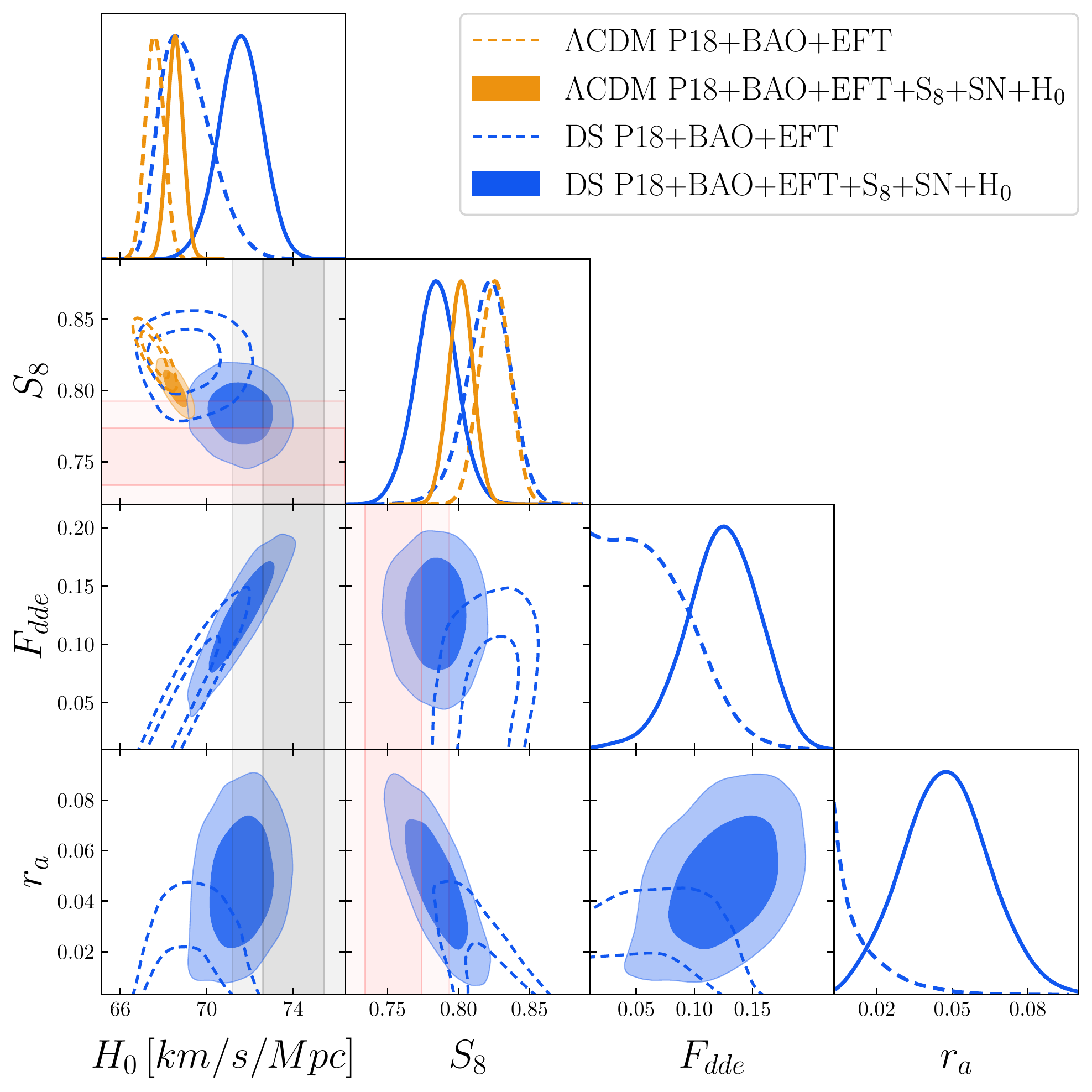}
\caption{Marginalized 1D and 2D posteriors for $H_0$ and $S_8$ in the $\Lambda$CDM and the DS models, for two representative datasets. For the latter, $\fpt$ and $\Faxdm$ posteriors are also shown. 
In grey are shown the 1-$\sigma$ (darker) and 2-$\sigma$ (lighter) ranges for $H_0$ from SH$_{0}$ES, and similarly the $S_8$ value from the joint analysis of~\cite{Asgari:2019fkq} is shown in pink.}
\label{fig:results}
\end{figure}

After the choices described above, our DS model features four free parameters in addition to the six parameters of the $\Lambda$CDM model: $\fpt,\,\zpt,\,\Faxdm$ and $\zax$. In order to obtain reliable results, we find it necessary to fix the parameter $\zax$ in the MCMC analysis (see also~\cite{Hlozek:2014lca,Poulin:2018dzj, Lague:2021frh} for a similar strategy). We then choose $\zax\simeq 10^{4.2}$, which corresponds to $m_a\simeq 10^{-26}~\text{eV}$, since this alleviates cosmological tensions most significantly. We keep the remaining three parameters free to vary, and comment on how our results are affected by a different choice of $\zax$ or by also fixing $\zpt$ in Appendix~\ref{sec:numerics}. In addition, we model neutrinos as two massless plus one massive species with $m_{\nu}= 0.06~\text{eV}$, following the Planck collaboration.

We consider four different combinations of cosmological datasets in this work:
\begin{itemize}
\item{\textbf{P18+BAO}: Planck 2018 high-$\ell$ and low-$\ell$ TT, TE, EE, and lensing data~\cite{Aghanim:2019ame}; BAO measurements from 6dFGS at $z = 0.106$~\cite{Beutler:2011hx}, SDSS MGS at $z = 0.15$~\cite{Ross:2014qpa} (BAO smallz), and CMASS and LOWZ galaxy samples of BOSS DR12 at $z = 0.38$, $0.51$, and $0.61$~\cite{Alam:2016hwk}. For the latter, we use the ``consensus'' BAO+FS likelihood which also includes measurement of the growth function $f\sigma_8(z)$ (FS) from the same samples.}
\item{\textbf{P18+BAO+EFT}: the datasets above with the addition of information from the full shape of the power spectrum of galaxies in the BOSS/SDDS sample, extracted by means of the EFTofLSS~\cite{DAmico:2019fhj, Ivanov:2019pdj, Colas:2019ret}. This is implemented with the publicly available {\tt PyBird} code~\cite{DAmico:2020kxu}\footnote{\href{https://github.com/pierrexyz/pybird}{\tt https://github.com/pierrexyz/pybird}} as a combined likelihood with BAO data from the same sample.}
\item{\textbf{P18+BAO+EFT+S$_8$}: the datasets above with the addition of a split-normal prior on $S_8$, chosen according to the recent analysis of DES data in combination with KiDS/Viking~\cite{Asgari:2019fkq}, i.e. $S_8=0.755^{+0.019}_{-0.021}$.}\footnote{The joint analysis of KIDS1000+BOSS+2dfLenS~\cite{Heymans:2020gsg} finds $S_8=0.766^{+0.02}_{-0.014}$. However, this is obtained using BOSS data and is thus not independent from the EFT and BAO likelihoods used in this work.}
\item{\textbf{P18+BAO+EFT+S$_8$+SN+H$_0$}: the datasets above with the addition of the Pantheon Supernovae data sample~\cite{Scolnic:2017caz}(SN) and the SH$_{0}$ES measurement of the Hubble parameter $H_0 = 74.03\pm 1.42$ km/s/Mpc~\cite{Riess:2019cxk}.\footnote{Using instead the recent measurement $H_0 = 73.2\pm 1.3$ km/s/Mpc~\cite{Riess:2020fzl} would not significantly alter our conclusions.}}
\end{itemize}

Before presenting our numerical results, an important caveat on the $S_8$ prior is in order. The use of such a prior as an approximation for the full weak-lensing likelihoods has been shown to be justified in the $\Lambda$CDM and EDE models~\cite{Hill:2020osr}. For ULAs, assessing the impact of the full likelihoods requires a dedicated treatment of nonlinearities. Lacking such tools (see e.g.~\cite{Hlozek:2016lzm} for a discussion), we restrict our analysis to the linear power spectrum, except for nonlinearities computed in the {\tt PyBird} likelihood, and assume that the use of a prior on $S_8$ correctly captures the constraints from the full DES and KiDS/Viking likelihoods on our DS model.

Our results for cosmological parameters are reported in Table~\ref{table:results}, while posterior distributions are plotted in Fig.~\ref{fig:results}. They have been obtained with at least eight chains per dataset, and $R-1 < 0.03$ to satisfy the Gelman-Rubin criterion~\cite{10.1214/ss/1177011136}. Detailed model comparisons and results for the matter and temperature anisotropy power spectra are reported in Appendix~\ref{sec:numerics}. We assess tensions by computing $\lvert A-B\rvert/(\sqrt{\sigma_{A}^2+\sigma_{B}^2})$, where $A$ and $B$ are the mean values of the parameter of interest ($H_0$ or $S_8$) inferred from the MCMC analysis and from the measurements  respectively, while $\sigma_{A,B}$ are the $1\sigma$ errors in these values (for asymmetric errors, we use the upper and lower errors for $H_0$ and $S_8$ inferred from the MCMC analysis respectively, and the upper error from the $S_8$ prior), see Appendix~\ref{sec:numerics}.

Let us first comment on results obtained with the Planck+BAO dataset only: The abundances of both DS components are consistent with zero at $2\sigma$, yet this dataset allows for non-negligible fraction of the DM to be in the form of an ULA, up to $\sim 3\%$ at $2\sigma$ (see~\cite{Hlozek:2014lca,Poulin:2018dzj} for previous similar bounds). The same is true for the DDE component, whose fraction of the total energy density at the redshift $\zpt\gtrsim 5,000$ is allowed to be as large as $\sim 14\%$ at $2\sigma$, with a mild preference for $\fpt\sim 7\%$. These features lead to a significant alleviation of the $H_0$ tension as compared to $\Lambda$CDM: the value of $H_0$ inferred in the DS model is only in $2.7\sigma$ tension with SH$_0$ES, in contrast to $4.4\sigma$ for $\Lambda$CDM. At the same time, the $S_8$ tension is also ameliorated, albeit less dramatically. Overall, the DS model improves the fit to this dataset as compared to $\Lambda$CDM, although only very mildly, having $\Delta\chi^2\simeq -4$ with three free extra parameters. The crucial point, however, is that in the DS model, both the $H_0$ and the $S_8$ tensions can be interpreted as moderate statistical fluctuations, weaker than in both the $\Lambda$CDM and  EDE/NEDE models; see also Appendix~\ref{sec:numerics}. This conclusion is only minimally altered by the addition of the EFT likelihood, with both tensions falling to the $3\sigma$ or below level in the combined dataset, $\Delta\chi^2\simeq -2$, and the upper bound on $r_a$ relaxed to $4\%$,\footnote{Previous constraints on ULAs were updated by~\cite{Lague:2021frh}, using a combination of BOSS and Planck 15 data. Our results are obtained with more recent CMB data. Models with significantly lighter ULAs than considered in our work, such as~\cite{Fung:2021wbz}, are strongly constrained by the datasets above, see Appendix~\ref{sec:numerics}.} with a best-fit value of $r_a\sim 1\%$, which corresponds to $f_a\sim 6\times10^{16}$\,GeV. These are the first important results of this work.

It therefore seems justified to combine the Planck+BAO+EFT dataset with a prior on $S_8$. Very interestingly, while the DDE component is almost unaffected by this addition, we notice that the best fit value of $r_a$ is raised to $\mathcal{O}(4\%)$, while fractions up to $\sim 7\%$ are allowed at $95\%$ CL, see also Appendix~\ref{sec:numerics}. As a consequence, we obtain the second important result of this work: the tension with cosmic shear measurements is very significantly reduced to $1.4\sigma$ level, as compared to $2.6\sigma$ under $\Lambda$CDM. Notice that the $H_0$ tension is also slightly relieved by these data, and the fit is significantly improved compared to $\Lambda$CDM ($\Delta\chi^2\simeq -8$). These features are in stark contrast with previous attempts to restore cosmological concordance (see~\cite{Hill:2020osr, Ivanov:2020ril} and Appendix~\ref{sec:numerics}). 

Interpreting the residual $2.8\sigma$ tension on $H_0$ as a moderate statistical fluctuation, it therefore seems justified to also combine the previous dataset with the local measurements from SH$_0$ES and Pantheon. This leads to the third important result of this work, a significant improvement of the fit to data as compared to $\Lambda$CDM, $\Delta\chi^2\simeq -18$ (with three extra parameters and $z_a$ fixed), driven mainly by a dramatically better fit to SH$_0$ES and the $S_8$ prior. The DDE component is now detected at $\simeq4\sigma$ (defined as $4\times$ the $1\sigma$ interval), with a preference for $\fpt\simeq 12\%$ at $z\sim 5,000$. The preference for the ULA component is also increased, with a vanishing relic abundance excluded at $\simeq 2\sigma$ in the posterior distributions, and its best-fit value being $r_a\sim 5\%$, which corresponds to $f_a\sim 10^{17}~\text{GeV}$, as expected from the earlier discussion of the model. These detections are the fourth important result of this work.
With this combined dataset, both the $S_8$ and $H_0$ tensions are essentially resolved in our DS model, again in stark contrast with the $\Lambda$CDM and EDE/NEDE models (see Appendix~\ref{sec:numerics} and~\cite{Hill:2020osr, Ivanov:2020ril, DAmico:2020ods}). In order to further assess the improvement of the fit to this dataset, we use the Akaike Information Criterium (AIC)~\cite{1100705} (see also~\cite{Liddle:2007fy}): $\Delta\text{AIC} = \Delta\chi^2 + 2\Delta N$, where $\Delta N$ is the number of additonal parameters compared to the $\Lambda$CDM model. We find $\Delta\text{AIC}_{\text{DS}}= -11.9$ considering the three parameters of the DS model that we scan over in our MCMC analysis, and $\Delta\text{AIC}_{\text{DS}}= -9.9$ if we also count $z_a$. Using $p=\text{exp}(-\Delta\text{AIC}/2)$, we find that the DS model has strong evidence over $\Lambda$CDM according to the revised Jeffreys' scale of~\cite{Trotta:2017wnx}.

Overall, we conclude that our DS model can restore cosmological concordance when a wide combination of early and late time datasets is considered. Importantly, both the $S_8$ and $H_0$ tensions remain below the $\simeq 3\sigma$ level even when the model is confronted with early time datasets only.\footnote{We have checked that the DS model qualifies for a ``gold medal'' according to the criteria put forth by~\cite{Schoneberg:2021qvd}, which appeared after the first version of our paper was posted on the arXiv. In particular, we find that the Gaussian Tension (GT) with respect to the measurement of $M_b$ by SH$_0$ES is $2.4\sigma$, when the same dataset as in~\cite{Schoneberg:2021qvd} is used (Planck18+BAO w/o FS+Pantheon). We find that NEDE with $w=c_s^2=2/3$ also qualifies for a gold medal, with a slightly worse GT of $2.7\sigma$. On the other hand,~\cite{Schoneberg:2021qvd} considers NEDE to be defined with $w=c_s^2$ free to vary (in contrast, EDE is taken to be defined as having a fixed effective $w>1/3$ in~\cite{Schoneberg:2021qvd}). Using this same choice, we have checked that the GT is $3.1~\sigma$ for the DS model, similar to NEDE. These differences in the conclusions may signal the appearance of volume-sampling effects when $w=c_s^2$ is let free to vary, see also the discussion in~\cite{Murgia:2020ryi}. We thank V.~Poulin for discussions on this point.}


{\bf Discussion}---We would now like to explore whether the features of our phenomenological DS model can arise within a plausible particle physics scenario.
The presence of ULAs with standard potentials is natural and appears to be a generic prediction of extra-dimensional UV theories such as String Theory, where the required $f_a\sim 10^{17}~\text{GeV}$ is a reasonable value~\cite{Svrcek:2006yi, Arvanitaki:2009fg}. These would-be massless particles get their potential from non-perturbative physics, e.g., from instantons of a gauge theory that confines at some scale $\Lambda_c$.\footnote{Non-perturbative UV contributions may also be present (see e.g.~\cite{Svrcek:2006yi} and~\cite{Hebecker:2018ofv}).} In this case, the natural expectation is $m_a\simeq \Lambda^2_{\text{c}}/f_a$.
The values of $m_a$ and $f_a$ obtained in our analysis then suggest the existence of a confining dark gauge theory with $\Lambda_c\sim 1$\,eV. For a unified model, can this gauge theory play the role of the DDE fluid? To answer this question, we need to address two separate aspects: (I) Can a confining gauge sector behave as dark energy at early times, at least sufficiently before matter-radiation equality? (II) Can it then behave as a fluid with $w>1/3$ below its confinement scale, at least for a sufficient amount of time after equality? 

First, gauge theory sectors do indeed generically feature two very distinct behaviors in their cosmological history: On the one hand, for $T_{\text{DS}}\gg \Lambda_c$ they are in a deconfined phase and their elementary constituents (``quarks'' and ``gluons'') behave as relativistic components.\footnote{Here we use $T_{\text{DS}}$ to denote the temperature of the confining DS, which can in general differ from that of the SM plasma.} 
On the other hand, for $T_{\text{DS}}\ll \Lambda_c$ they are in the confined phase, where massive bound states (``hadrons'') form. A PT normally occurs around the critical temperature $T_{\text{DS}, c} \lesssim \Lambda_c$. Close to the PT, gauge theories can exhibit a wide range of phenomena and behaviors, depending on the gauge group and the matter content.

Very interestingly, confinement PTs can be of first order kind in several simple examples~(see e.g.~\cite{Pisarski:1983ms, Creminelli:2001th}), in which case they may also naturally exhibit the phenomenon of strong {\it supercooling}, where the PT is delayed to $T_{\text{ds, n}}\ll T_{\text{DS}, c}$.  At temperatures $T_\text{ds, n}\lesssim T_\text{DS}\lesssim T_{\text{DS}, c}$, the confining sector is dominated by the vacuum energy gap between the two phases (see~\cite{Creminelli:2001th} and~\cite{vonHarling:2017yew, Baratella:2018pxi, Agashe:2019lhy} for discussions in the context of strongly-coupled solutions to the hierarchy problem). 
This can reproduce the required dark energy behavior of the DDE fluid up to sufficiently high redshifts. 
Further details on this possibility are provided in Appendix~\ref{sec:supercooledDS}.

Having established that dark energy behavior is feasible at early times, we now turn to the required $w>1/3$ behavior at late times. The generic expectation after a first order PT is that bubble collisions lead to an initially relativistic bath of DS states. Nonetheless, the authors of~\cite{Niedermann:2020dwg} have argued in favor of $w>1/3$ after a first order PT as a consequence of subhorizon anisotropies and nonlinearities. This is an interesting possibility which requires further investigation.

Here, we would like to suggest an alternative, albeit speculative, possibility. The equation of state (EoS) of a confining gauge theory can be affected by parameters beyond temperature; for instance, general arguments suggest that at very large ``baryon'' densities, the EoS can indeed be stiff~\cite{Zeldovich:1962emp}, i.e. $c_s^2 > 1/3$ (see also~\cite{Alford:2013aca, Bedaque:2014sqa} for a discussion in the context of neutron star cores, and e.g.~\cite{Hoyos:2016cob, Ecker:2017fyh} for holographic models). Furthermore, a recent holographic model with a cosmological first order confinement PT and a stiff equation of state below the nucleation temperature was presented in~\cite{Ares:2020lbt}, where it was found that the stiffness increases as the PT becomes strongly supercooled.

Yet another strategy to realize the DDE fluid, abandoning the connection with the ULA potential, is to make use of one or more homogeneous scalar fields that approach a near vanishing potential, thus becoming kinetic dominated, leading to $w>1/3$ at late times.

Whether these possibilities for $w>1/3$ are viable is a very interesting question for future exploration.

Finally, let us discuss further the possible constraints/signatures of our DS model. For $m_a \sim 10^{-26}~\text{eV}$, the state-of-the-art constraint on $\Faxdm$ from the Lyman-$\alpha$ forest is $\Faxdm\leq 0.18$ at $95\%$ C.L.~\cite{Kobayashi:2017jcf}, which is far from the $2\sigma$ upper value obtained in our MCMC analysis. The presence of an ULA in our mass range can also affect halo formation. However, existing analyses of high-$z$ galaxies do not constrain the axion DM fraction considered in this work~\cite{Bozek:2014uqa} (see also~\cite{Marsh:2015xka}). Nonetheless, these constraints may soon improve~(see e.g.~\cite{Schwabe:2020eac}).

However, it is anticipated that future CMB-S4 can detect a fraction of DM in an ULA with $m_a \sim 10^{-26}~\text{eV}$ at the percent level~\cite{Abazajian:2016yjj} and that further improvements may be possible with intensity mapping of neutral hydrogen~\cite{Bauer:2020zsj}. Future LSS surveys, such as {\it Euclid}~\cite{Amendola:2016saw}, DESI~\cite{Levi:2019ggs},  WFIRST/Roman~\cite{2019arXiv190205569A} and the Vera Rubin Observatory~\cite{Ivezic:2008fe} will also further probe the existence of a DDE component. Hence upcoming observations should be able to confirm or rule out our DS model.

{\bf Acknowledgments}---We would like to thank Vivian Poulin for help in setting up {\tt AxiCLASS}, Guido D'Amico for support with {\tt PyBird} and Evan McDonough for discussions on the $S_8$ prior. We also thank Mark Alford for discussions on the equation of state of QCD matter in neutron stars. We thank Florian Niedermann and Martin Sloth for useful comments on a first version of this paper. We acknowledge use of Tufts HPC research cluster. The work of MPH and FR is supported in part by National Science Foundation Grant No. PHY-2013953.

\bibliography{biblio}

\providecommand{\href}[2]{#2}\begingroup\raggedright\begin{thebibliography}{10}

\bibitem{Aghanim:2019ame}
{\bf Planck} Collaboration, N.~Aghanim {\em et.~al.}, {\it {Planck 2018
  results. V. CMB power spectra and likelihoods}},  Astron. Astrophys. {\bf
  641} (2020) A5, [\href{http://arxiv.org/abs/1907.12875}{{\tt
  arXiv:1907.12875}}].

\bibitem{Riess:2019cxk}
A.~G. Riess, S.~Casertano, W.~Yuan, L.~M. Macri, and D.~Scolnic, {\it {Large
  Magellanic Cloud Cepheid Standards Provide a 1\% Foundation for the
  Determination of the Hubble Constant and Stronger Evidence for Physics beyond
  $\Lambda$CDM}},  Astrophys. J. {\bf 876} (2019), no.~1 85,
  [\href{http://arxiv.org/abs/1903.07603}{{\tt arXiv:1903.07603}}].

\bibitem{Riess:2020fzl}
A.~G. Riess, S.~Casertano, W.~Yuan, J.~B. Bowers, L.~Macri, J.~C. Zinn, and
  D.~Scolnic, {\it {Cosmic Distances Calibrated to 1\% Precision with Gaia EDR3
  Parallaxes and Hubble Space Telescope Photometry of 75 Milky Way Cepheids
  Confirm Tension with $\Lambda$CDM}},  Astrophys. J. Lett. {\bf 908} (2021),
  no.~1 L6, [\href{http://arxiv.org/abs/2012.08534}{{\tt arXiv:2012.08534}}].

\bibitem{Wong:2019kwg}
K.~C. Wong {\em et.~al.}, {\it {H0LiCOW \textendash{} XIII. A 2.4 per cent
  measurement of H0 from lensed quasars: 5.3\ensuremath{\sigma} tension between
  early- and late-Universe probes}},  Mon. Not. Roy. Astron. Soc. {\bf 498}
  (2020), no.~1 1420--1439, [\href{http://arxiv.org/abs/1907.04869}{{\tt
  arXiv:1907.04869}}].

\bibitem{Baumann:2010tm}
D.~Baumann, A.~Nicolis, L.~Senatore, and M.~Zaldarriaga, {\it {Cosmological
  Non-Linearities as an Effective Fluid}},  JCAP {\bf 07} (2012) 051,
  [\href{http://arxiv.org/abs/1004.2488}{{\tt arXiv:1004.2488}}].

\bibitem{Carrasco:2012cv}
J.~J.~M. Carrasco, M.~P. Hertzberg, and L.~Senatore, {\it {The Effective Field
  Theory of Cosmological Large Scale Structures}},  JHEP {\bf 09} (2012) 082,
  [\href{http://arxiv.org/abs/1206.2926}{{\tt arXiv:1206.2926}}].

\bibitem{Hertzberg:2012qn}
M.~P. Hertzberg, {\it {Effective field theory of dark matter and structure
  formation: Semianalytical results}},  Phys. Rev. D {\bf 89} (2014), no.~4
  043521, [\href{http://arxiv.org/abs/1208.0839}{{\tt arXiv:1208.0839}}].

\bibitem{Colas:2019ret}
T.~Colas, G.~D'amico, L.~Senatore, P.~Zhang, and F.~Beutler, {\it {Efficient
  Cosmological Analysis of the SDSS/BOSS data from the Effective Field Theory
  of Large-Scale Structure}},  JCAP {\bf 06} (2020) 001,
  [\href{http://arxiv.org/abs/1909.07951}{{\tt arXiv:1909.07951}}].

\bibitem{DAmico:2019fhj}
G.~D'Amico, J.~Gleyzes, N.~Kokron, K.~Markovic, L.~Senatore, P.~Zhang,
  F.~Beutler, and H.~Gil-Mar\'\i{}n, {\it {The Cosmological Analysis of the
  SDSS/BOSS data from the Effective Field Theory of Large-Scale Structure}},
  JCAP {\bf 05} (2020) 005, [\href{http://arxiv.org/abs/1909.05271}{{\tt
  arXiv:1909.05271}}].

\bibitem{Ivanov:2019pdj}
M.~M. Ivanov, M.~Simonovi\'c, and M.~Zaldarriaga, {\it {Cosmological Parameters
  from the BOSS Galaxy Power Spectrum}},  JCAP {\bf 05} (2020) 042,
  [\href{http://arxiv.org/abs/1909.05277}{{\tt arXiv:1909.05277}}].

\bibitem{Asgari:2019fkq}
M.~Asgari {\em et.~al.}, {\it {KiDS+VIKING-450 and DES-Y1 combined: Mitigating
  baryon feedback uncertainty with COSEBIs}},  Astron. Astrophys. {\bf 634}
  (2020) A127, [\href{http://arxiv.org/abs/1910.05336}{{\tt
  arXiv:1910.05336}}].

\bibitem{Joudaki:2019pmv}
S.~Joudaki {\em et.~al.}, {\it {KiDS+VIKING-450 and DES-Y1 combined: Cosmology
  with cosmic shear}},  Astron. Astrophys. {\bf 638} (2020) L1,
  [\href{http://arxiv.org/abs/1906.09262}{{\tt arXiv:1906.09262}}].

\bibitem{Heymans:2013fya}
C.~Heymans {\em et.~al.}, {\it {CFHTLenS tomographic weak lensing cosmological
  parameter constraints: Mitigating the impact of intrinsic galaxy
  alignments}},  Mon. Not. Roy. Astron. Soc. {\bf 432} (2013) 2433,
  [\href{http://arxiv.org/abs/1303.1808}{{\tt arXiv:1303.1808}}].

\bibitem{Hildebrandt:2018yau}
H.~Hildebrandt {\em et.~al.}, {\it {KiDS+VIKING-450: Cosmic shear tomography
  with optical and infrared data}},  Astron. Astrophys. {\bf 633} (2020) A69,
  [\href{http://arxiv.org/abs/1812.06076}{{\tt arXiv:1812.06076}}].

\bibitem{Abbott:2017wau}
{\bf DES} Collaboration, T.~M.~C. Abbott {\em et.~al.}, {\it {Dark Energy
  Survey year 1 results: Cosmological constraints from galaxy clustering and
  weak lensing}},  Phys. Rev. D {\bf 98} (2018), no.~4 043526,
  [\href{http://arxiv.org/abs/1708.01530}{{\tt arXiv:1708.01530}}].

\bibitem{Hikage:2018qbn}
{\bf HSC} Collaboration, C.~Hikage {\em et.~al.}, {\it {Cosmology from cosmic
  shear power spectra with Subaru Hyper Suprime-Cam first-year data}},  Publ.
  Astron. Soc. Jap. {\bf 71} (2019), no.~2 Publications of the Astronomical
  Society of Japan, Volume 71, Issue 2, April 2019, 43,
  https://doi.org/10.1093/pasj/psz010,
  [\href{http://arxiv.org/abs/1809.09148}{{\tt arXiv:1809.09148}}].

\bibitem{Heymans:2020gsg}
C.~Heymans {\em et.~al.}, {\it {KiDS-1000 Cosmology: Multi-probe weak
  gravitational lensing and spectroscopic galaxy clustering constraints}},
  Astron. Astrophys. {\bf 646} (2021) A140,
  [\href{http://arxiv.org/abs/2007.15632}{{\tt arXiv:2007.15632}}].

\bibitem{Poulin:2018cxd}
V.~Poulin, T.~L. Smith, T.~Karwal, and M.~Kamionkowski, {\it {Early Dark Energy
  Can Resolve The Hubble Tension}},  Phys. Rev. Lett. {\bf 122} (2019), no.~22
  221301, [\href{http://arxiv.org/abs/1811.04083}{{\tt arXiv:1811.04083}}].

\bibitem{Agrawal:2019lmo}
P.~Agrawal, F.-Y. Cyr-Racine, D.~Pinner, and L.~Randall, {\it {Rock 'n' Roll
  Solutions to the Hubble Tension}},
  \href{http://arxiv.org/abs/1904.01016}{{\tt arXiv:1904.01016}}.

\bibitem{Lin:2019qug}
M.-X. Lin, G.~Benevento, W.~Hu, and M.~Raveri, {\it {Acoustic Dark Energy:
  Potential Conversion of the Hubble Tension}},  Phys. Rev. D {\bf 100} (2019),
  no.~6 063542, [\href{http://arxiv.org/abs/1905.12618}{{\tt
  arXiv:1905.12618}}].

\bibitem{Hill:2020osr}
J.~C. Hill, E.~McDonough, M.~W. Toomey, and S.~Alexander, {\it {Early dark
  energy does not restore cosmological concordance}},  Phys. Rev. D {\bf 102}
  (2020), no.~4 043507, [\href{http://arxiv.org/abs/2003.07355}{{\tt
  arXiv:2003.07355}}].

\bibitem{Ivanov:2020ril}
M.~M. Ivanov, E.~McDonough, J.~C. Hill, M.~Simonovi\'c, M.~W. Toomey,
  S.~Alexander, and M.~Zaldarriaga, {\it {Constraining Early Dark Energy with
  Large-Scale Structure}},  Phys. Rev. D {\bf 102} (2020), no.~10 103502,
  [\href{http://arxiv.org/abs/2006.11235}{{\tt arXiv:2006.11235}}].

\bibitem{DAmico:2020ods}
G.~D'Amico, L.~Senatore, P.~Zhang, and H.~Zheng, {\it {The Hubble Tension in
  Light of the Full-Shape Analysis of Large-Scale Structure Data}},
  \href{http://arxiv.org/abs/2006.12420}{{\tt arXiv:2006.12420}}.

\bibitem{Smith:2020rxx}
T.~L. Smith, V.~Poulin, J.~L. Bernal, K.~K. Boddy, M.~Kamionkowski, and
  R.~Murgia, {\it {Early dark energy is not excluded by current large-scale
  structure data}},  \href{http://arxiv.org/abs/2009.10740}{{\tt
  arXiv:2009.10740}}.

\bibitem{Murgia:2020ryi}
R.~Murgia, G.~F. Abell\'an, and V.~Poulin, {\it {Early dark energy resolution
  to the Hubble tension in light of weak lensing surveys and lensing
  anomalies}},  Phys. Rev. D {\bf 103} (2021), no.~6 063502,
  [\href{http://arxiv.org/abs/2009.10733}{{\tt arXiv:2009.10733}}].

\bibitem{Gonzalez:2020fdy}
M.~Gonzalez, M.~P. Hertzberg, and F.~Rompineve, {\it {Ultralight Scalar Decay
  and the Hubble Tension}},  JCAP {\bf 10} (2020) 028,
  [\href{http://arxiv.org/abs/2006.13959}{{\tt arXiv:2006.13959}}].

\bibitem{Niedermann:2019olb}
F.~Niedermann and M.~S. Sloth, {\it {New Early Dark Energy}},
  \href{http://arxiv.org/abs/1910.10739}{{\tt arXiv:1910.10739}}.

\bibitem{Niedermann:2020dwg}
F.~Niedermann and M.~S. Sloth, {\it {Resolving the Hubble tension with new
  early dark energy}},  Phys. Rev. D {\bf 102} (2020), no.~6 063527,
  [\href{http://arxiv.org/abs/2006.06686}{{\tt arXiv:2006.06686}}].

\bibitem{Niedermann:2020qbw}
F.~Niedermann and M.~S. Sloth, {\it {New Early Dark Energy is compatible with
  current LSS data}},  \href{http://arxiv.org/abs/2009.00006}{{\tt
  arXiv:2009.00006}}.

\bibitem{Kaloper:2019lpl}
N.~Kaloper, {\it {Dark energy, $H_0$ and weak gravity conjecture}},  Int. J.
  Mod. Phys. D {\bf 28} (2019), no.~14 1944017,
  [\href{http://arxiv.org/abs/1903.11676}{{\tt arXiv:1903.11676}}].

\bibitem{Lin:2018nxe}
M.-X. Lin, M.~Raveri, and W.~Hu, {\it {Phenomenology of Modified Gravity at
  Recombination}},  Phys. Rev. D {\bf 99} (2019), no.~4 043514,
  [\href{http://arxiv.org/abs/1810.02333}{{\tt arXiv:1810.02333}}].

\bibitem{Sola:2019jek}
J.~Sol\`a~Peracaula, A.~Gomez-Valent, J.~de~Cruz~P\'erez, and C.~Moreno-Pulido,
  {\it {Brans\textendash{}Dicke Gravity with a Cosmological Constant Smoothes
  Out $\Lambda$CDM Tensions}},  Astrophys. J. Lett. {\bf 886} (2019), no.~1 L6,
  [\href{http://arxiv.org/abs/1909.02554}{{\tt arXiv:1909.02554}}].

\bibitem{Sakstein:2019fmf}
J.~Sakstein and M.~Trodden, {\it {Early Dark Energy from Massive Neutrinos as a
  Natural Resolution of the Hubble Tension}},  Phys. Rev. Lett. {\bf 124}
  (2020), no.~16 161301, [\href{http://arxiv.org/abs/1911.11760}{{\tt
  arXiv:1911.11760}}].

\bibitem{Zumalacarregui:2020cjh}
M.~Zumalacarregui, {\it {Gravity in the Era of Equality: Towards solutions to
  the Hubble problem without fine-tuned initial conditions}},  Phys. Rev. D
  {\bf 102} (2020), no.~2 023523, [\href{http://arxiv.org/abs/2003.06396}{{\tt
  arXiv:2003.06396}}].

\bibitem{Ballesteros:2020sik}
G.~Ballesteros, A.~Notari, and F.~Rompineve, {\it {The $H_0$ tension: $\Delta
  G_N$ vs. $\Delta N_{\rm eff}$}},  JCAP {\bf 11} (2020) 024,
  [\href{http://arxiv.org/abs/2004.05049}{{\tt arXiv:2004.05049}}].

\bibitem{Braglia:2020iik}
M.~Braglia, M.~Ballardini, W.~T. Emond, F.~Finelli, A.~E. Gumrukcuoglu,
  K.~Koyama, and D.~Paoletti, {\it {Larger value for $H_0$ by an evolving
  gravitational constant}},  Phys. Rev. D {\bf 102} (2020), no.~2 023529,
  [\href{http://arxiv.org/abs/2004.11161}{{\tt arXiv:2004.11161}}].

\bibitem{Braglia:2020auw}
M.~Braglia, M.~Ballardini, F.~Finelli, and K.~Koyama, {\it {Early modified
  gravity in light of the $H_0$ tension and LSS data}},  Phys. Rev. D {\bf 103}
  (2021), no.~4 043528, [\href{http://arxiv.org/abs/2011.12934}{{\tt
  arXiv:2011.12934}}].

\bibitem{Khlopov:1985jw}
M.~Khlopov, B.~A. Malomed, and I.~B. Zeldovich, {\it {Gravitational instability
  of scalar fields and formation of primordial black holes}},  Mon. Not. Roy.
  Astron. Soc. {\bf 215} (1985) 575--589.

\bibitem{Hu:1998kj}
W.~Hu, {\it {Structure formation with generalized dark matter}},  Astrophys. J.
  {\bf 506} (1998) 485--494, [\href{http://arxiv.org/abs/astro-ph/9801234}{{\tt
  astro-ph/9801234}}].

\bibitem{Hwang:2009js}
J.-c. Hwang and H.~Noh, {\it {Axion as a Cold Dark Matter candidate}},  Phys.
  Lett. B {\bf 680} (2009) 1--3, [\href{http://arxiv.org/abs/0902.4738}{{\tt
  arXiv:0902.4738}}].

\bibitem{Fung:2021wbz}
L.~W. Fung, L.~Li, T.~Liu, H.~N. Luu, Y.-C. Qiu, and S.~H.~H. Tye, {\it
  {Axi-Higgs Cosmology}},  \href{http://arxiv.org/abs/2102.11257}{{\tt
  arXiv:2102.11257}}.

\bibitem{Kobayashi:2017jcf}
T.~Kobayashi, R.~Murgia, A.~De~Simone, V.~Ir\v{s}i\v{c}, and M.~Viel, {\it
  {Lyman-$\alpha$ constraints on ultralight scalar dark matter: Implications
  for the early and late universe}},  Phys. Rev. D {\bf 96} (2017), no.~12
  123514, [\href{http://arxiv.org/abs/1708.00015}{{\tt arXiv:1708.00015}}].

\bibitem{Blas:2011rf}
D.~Blas, J.~Lesgourgues, and T.~Tram, {\it {The Cosmic Linear Anisotropy
  Solving System (CLASS) II: Approximation schemes}},  JCAP {\bf 07} (2011)
  034, [\href{http://arxiv.org/abs/1104.2933}{{\tt arXiv:1104.2933}}].

\bibitem{Poulin:2018dzj}
V.~Poulin, T.~L. Smith, D.~Grin, T.~Karwal, and M.~Kamionkowski, {\it
  {Cosmological implications of ultralight axionlike fields}},  Phys. Rev. D
  {\bf 98} (2018), no.~8 083525, [\href{http://arxiv.org/abs/1806.10608}{{\tt
  arXiv:1806.10608}}].

\bibitem{Smith:2019ihp}
T.~L. Smith, V.~Poulin, and M.~A. Amin, {\it {Oscillating scalar fields and the
  Hubble tension: a resolution with novel signatures}},  Phys. Rev. D {\bf 101}
  (2020), no.~6 063523, [\href{http://arxiv.org/abs/1908.06995}{{\tt
  arXiv:1908.06995}}].

\bibitem{Audren:2012wb}
B.~Audren, J.~Lesgourgues, K.~Benabed, and S.~Prunet, {\it {Conservative
  Constraints on Early Cosmology: an illustration of the Monte Python
  cosmological parameter inference code}},  JCAP {\bf 02} (2013) 001,
  [\href{http://arxiv.org/abs/1210.7183}{{\tt arXiv:1210.7183}}].

\bibitem{Brinckmann:2018cvx}
T.~Brinckmann and J.~Lesgourgues, {\it {MontePython 3: boosted MCMC sampler and
  other features}},  Phys. Dark Univ. {\bf 24} (2019) 100260,
  [\href{http://arxiv.org/abs/1804.07261}{{\tt arXiv:1804.07261}}].

\bibitem{Lewis:2019xzd}
A.~Lewis, {\it {GetDist: a Python package for analysing Monte Carlo samples}},
  \href{http://arxiv.org/abs/1910.13970}{{\tt arXiv:1910.13970}}.

\bibitem{Hlozek:2014lca}
R.~Hlozek, D.~Grin, D.~J.~E. Marsh, and P.~G. Ferreira, {\it {A search for
  ultralight axions using precision cosmological data}},  Phys. Rev. D {\bf 91}
  (2015), no.~10 103512, [\href{http://arxiv.org/abs/1410.2896}{{\tt
  arXiv:1410.2896}}].

\bibitem{Lague:2021frh}
A.~Lagu\"e, J.~R. Bond, R.~Hlo\v{z}ek, K.~K. Rogers, D.~J.~E. Marsh, and
  D.~Grin, {\it {Constraining Ultralight Axions with Galaxy Surveys}},
  \href{http://arxiv.org/abs/2104.07802}{{\tt arXiv:2104.07802}}.

\bibitem{Beutler:2011hx}
F.~Beutler, C.~Blake, M.~Colless, D.~H. Jones, L.~Staveley-Smith, L.~Campbell,
  Q.~Parker, W.~Saunders, and F.~Watson, {\it {The 6dF Galaxy Survey: Baryon
  Acoustic Oscillations and the Local Hubble Constant}},  Mon. Not. Roy.
  Astron. Soc. {\bf 416} (2011) 3017--3032,
  [\href{http://arxiv.org/abs/1106.3366}{{\tt arXiv:1106.3366}}].

\bibitem{Ross:2014qpa}
A.~J. Ross, L.~Samushia, C.~Howlett, W.~J. Percival, A.~Burden, and M.~Manera,
  {\it {The clustering of the SDSS DR7 main Galaxy sample \textendash{} I. A 4
  per cent distance measure at $z = 0.15$}},  Mon. Not. Roy. Astron. Soc. {\bf
  449} (2015), no.~1 835--847, [\href{http://arxiv.org/abs/1409.3242}{{\tt
  arXiv:1409.3242}}].

\bibitem{Alam:2016hwk}
{\bf BOSS} Collaboration, S.~Alam {\em et.~al.}, {\it {The clustering of
  galaxies in the completed SDSS-III Baryon Oscillation Spectroscopic Survey:
  cosmological analysis of the DR12 galaxy sample}},  Mon. Not. Roy. Astron.
  Soc. {\bf 470} (2017), no.~3 2617--2652,
  [\href{http://arxiv.org/abs/1607.03155}{{\tt arXiv:1607.03155}}].

\bibitem{DAmico:2020kxu}
G.~D'Amico, L.~Senatore, and P.~Zhang, {\it {Limits on $w$CDM from the EFTofLSS
  with the PyBird code}},  \href{http://arxiv.org/abs/2003.07956}{{\tt
  arXiv:2003.07956}}.

\bibitem{Scolnic:2017caz}
D.~M. Scolnic {\em et.~al.}, {\it {The Complete Light-curve Sample of
  Spectroscopically Confirmed SNe Ia from Pan-STARRS1 and Cosmological
  Constraints from the Combined Pantheon Sample}},  Astrophys. J. {\bf 859}
  (2018), no.~2 101, [\href{http://arxiv.org/abs/1710.00845}{{\tt
  arXiv:1710.00845}}].

\bibitem{Hlozek:2016lzm}
R.~Hlo\v{z}ek, D.~J.~E. Marsh, D.~Grin, R.~Allison, J.~Dunkley, and
  E.~Calabrese, {\it {Future CMB tests of dark matter: Ultralight axions and
  massive neutrinos}},  Phys. Rev. D {\bf 95} (2017), no.~12 123511,
  [\href{http://arxiv.org/abs/1607.08208}{{\tt arXiv:1607.08208}}].

\bibitem{10.1214/ss/1177011136}
A.~Gelman and D.~B. Rubin, {\it {Inference from Iterative Simulation Using
  Multiple Sequences}},  Statistical Science {\bf 7} (1992), no.~4 457 -- 472.

\bibitem{1100705}
H.~Akaike, {\it A new look at the statistical model identification},  IEEE
  Transactions on Automatic Control {\bf 19} (1974), no.~6 716--723.

\bibitem{Liddle:2007fy}
A.~R. Liddle, {\it {Information criteria for astrophysical model selection}},
  Mon. Not. Roy. Astron. Soc. {\bf 377} (2007) L74--L78,
  [\href{http://arxiv.org/abs/astro-ph/0701113}{{\tt astro-ph/0701113}}].

\bibitem{Trotta:2017wnx}
R.~Trotta, {\it {Bayesian Methods in Cosmology}},  1, 2017.
\newblock \href{http://arxiv.org/abs/1701.01467}{{\tt arXiv:1701.01467}}.

\bibitem{Schoneberg:2021qvd}
N.~Sch\"oneberg, G.~Franco~Abell\'an, A.~P\'erez~S\'anchez, S.~J. Witte,
  V.~Poulin, and J.~Lesgourgues, {\it {The $H_0$ Olympics: A fair ranking of
  proposed models}},  \href{http://arxiv.org/abs/2107.10291}{{\tt
  arXiv:2107.10291}}.

\bibitem{Svrcek:2006yi}
P.~Svrcek and E.~Witten, {\it {Axions In String Theory}},  JHEP {\bf 06} (2006)
  051, [\href{http://arxiv.org/abs/hep-th/0605206}{{\tt hep-th/0605206}}].

\bibitem{Arvanitaki:2009fg}
A.~Arvanitaki, S.~Dimopoulos, S.~Dubovsky, N.~Kaloper, and J.~March-Russell,
  {\it {String Axiverse}},  Phys. Rev. D {\bf 81} (2010) 123530,
  [\href{http://arxiv.org/abs/0905.4720}{{\tt arXiv:0905.4720}}].

\bibitem{Hebecker:2018ofv}
A.~Hebecker, T.~Mikhail, and P.~Soler, {\it {Euclidean wormholes, baby
  universes, and their impact on particle physics and cosmology}},  Front.
  Astron. Space Sci. {\bf 5} (2018) 35,
  [\href{http://arxiv.org/abs/1807.00824}{{\tt arXiv:1807.00824}}].

\bibitem{Pisarski:1983ms}
R.~D. Pisarski and F.~Wilczek, {\it {Remarks on the Chiral Phase Transition in
  Chromodynamics}},  Phys. Rev. D {\bf 29} (1984) 338--341.

\bibitem{Creminelli:2001th}
P.~Creminelli, A.~Nicolis, and R.~Rattazzi, {\it {Holography and the
  electroweak phase transition}},  JHEP {\bf 03} (2002) 051,
  [\href{http://arxiv.org/abs/hep-th/0107141}{{\tt hep-th/0107141}}].

\bibitem{vonHarling:2017yew}
B.~von Harling and G.~Servant, {\it {QCD-induced Electroweak Phase
  Transition}},  JHEP {\bf 01} (2018) 159,
  [\href{http://arxiv.org/abs/1711.11554}{{\tt arXiv:1711.11554}}].

\bibitem{Baratella:2018pxi}
P.~Baratella, A.~Pomarol, and F.~Rompineve, {\it {The Supercooled Universe}},
  JHEP {\bf 03} (2019) 100, [\href{http://arxiv.org/abs/1812.06996}{{\tt
  arXiv:1812.06996}}].

\bibitem{Agashe:2019lhy}
K.~Agashe, P.~Du, M.~Ekhterachian, S.~Kumar, and R.~Sundrum, {\it {Cosmological
  Phase Transition of Spontaneous Confinement}},  JHEP {\bf 05} (2020) 086,
  [\href{http://arxiv.org/abs/1910.06238}{{\tt arXiv:1910.06238}}].

\bibitem{Zeldovich:1962emp}
Y.~B. Zel'dovich, {\it {The equation of state at ultrahigh densities and its
  relativistic limitations}},  Sov. Phys. JETP {\bf 14} (1962) 1143--1147.

\bibitem{Alford:2013aca}
M.~G. Alford, S.~Han, and M.~Prakash, {\it {Generic conditions for stable
  hybrid stars}},  Phys. Rev. D {\bf 88} (2013), no.~8 083013,
  [\href{http://arxiv.org/abs/1302.4732}{{\tt arXiv:1302.4732}}].

\bibitem{Bedaque:2014sqa}
P.~Bedaque and A.~W. Steiner, {\it {Sound velocity bound and neutron stars}},
  Phys. Rev. Lett. {\bf 114} (2015), no.~3 031103,
  [\href{http://arxiv.org/abs/1408.5116}{{\tt arXiv:1408.5116}}].

\bibitem{Hoyos:2016cob}
C.~Hoyos, N.~Jokela, D.~Rodr\'\i{}guez~Fern\'andez, and A.~Vuorinen, {\it
  {Breaking the sound barrier in AdS/CFT}},  Phys. Rev. D {\bf 94} (2016),
  no.~10 106008, [\href{http://arxiv.org/abs/1609.03480}{{\tt
  arXiv:1609.03480}}].

\bibitem{Ecker:2017fyh}
C.~Ecker, C.~Hoyos, N.~Jokela, D.~Rodr\'\i{}guez~Fern\'andez, and A.~Vuorinen,
  {\it {Stiff phases in strongly coupled gauge theories with holographic
  duals}},  JHEP {\bf 11} (2017) 031,
  [\href{http://arxiv.org/abs/1707.00521}{{\tt arXiv:1707.00521}}].

\bibitem{Ares:2020lbt}
F.~R. Ares, M.~Hindmarsh, C.~Hoyos, and N.~Jokela, {\it {Gravitational waves
  from a holographic phase transition}},
  \href{http://arxiv.org/abs/2011.12878}{{\tt arXiv:2011.12878}}.

\bibitem{Bozek:2014uqa}
B.~Bozek, D.~J.~E. Marsh, J.~Silk, and R.~F.~G. Wyse, {\it {Galaxy
  UV-luminosity function and reionization constraints on axion dark matter}},
  Mon. Not. Roy. Astron. Soc. {\bf 450} (2015), no.~1 209--222,
  [\href{http://arxiv.org/abs/1409.3544}{{\tt arXiv:1409.3544}}].

\bibitem{Marsh:2015xka}
D.~J.~E. Marsh, {\it {Axion Cosmology}},  Phys. Rept. {\bf 643} (2016) 1--79,
  [\href{http://arxiv.org/abs/1510.07633}{{\tt arXiv:1510.07633}}].

\bibitem{Schwabe:2020eac}
B.~Schwabe, M.~Gosenca, C.~Behrens, J.~C. Niemeyer, and R.~Easther, {\it
  {Simulating mixed fuzzy and cold dark matter}},  Phys. Rev. D {\bf 102}
  (2020), no.~8 083518, [\href{http://arxiv.org/abs/2007.08256}{{\tt
  arXiv:2007.08256}}].

\bibitem{Abazajian:2016yjj}
{\bf CMB-S4} Collaboration, K.~N. Abazajian {\em et.~al.}, {\it {CMB-S4 Science
  Book, First Edition}},  \href{http://arxiv.org/abs/1610.02743}{{\tt
  arXiv:1610.02743}}.

\bibitem{Bauer:2020zsj}
J.~B. Bauer, D.~J.~E. Marsh, R.~Hlo\v{z}ek, H.~Padmanabhan, and A.~Lagu\"e,
  {\it {Intensity Mapping as a Probe of Axion Dark Matter}},  Mon. Not. Roy.
  Astron. Soc. {\bf 500} (2020), no.~3 3162--3177,
  [\href{http://arxiv.org/abs/2003.09655}{{\tt arXiv:2003.09655}}].

\bibitem{Amendola:2016saw}
L.~Amendola {\em et.~al.}, {\it {Cosmology and fundamental physics with the
  Euclid satellite}},  Living Rev. Rel. {\bf 21} (2018), no.~1 2,
  [\href{http://arxiv.org/abs/1606.00180}{{\tt arXiv:1606.00180}}].

\bibitem{Levi:2019ggs}
{\bf DESI} Collaboration, M.~E. Levi {\em et.~al.}, {\it {The Dark Energy
  Spectroscopic Instrument (DESI)}},
  \href{http://arxiv.org/abs/1907.10688}{{\tt arXiv:1907.10688}}.

\bibitem{2019arXiv190205569A}
R.~Akeson {\em et.~al.}, {\it {The Wide Field Infrared Survey Telescope: 100
  Hubbles for the 2020s}},  arXiv e-prints (Feb., 2019) arXiv:1902.05569,
  [\href{http://arxiv.org/abs/1902.05569}{{\tt arXiv:1902.05569}}].

\bibitem{Ivezic:2008fe}
{\bf LSST} Collaboration, v.~Ivezi\'c {\em et.~al.}, {\it {LSST: from Science
  Drivers to Reference Design and Anticipated Data Products}},  Astrophys. J.
  {\bf 873} (2019), no.~2 111, [\href{http://arxiv.org/abs/0805.2366}{{\tt
  arXiv:0805.2366}}].

\bibitem{Witten:1980ez}
E.~Witten, {\it {Cosmological Consequences of a Light Higgs Boson}},  Nucl.
  Phys. B {\bf 177} (1981) 477--488.

\bibitem{Iso:2017uuu}
S.~Iso, P.~D. Serpico, and K.~Shimada, {\it {QCD-Electroweak First-Order Phase
  Transition in a Supercooled Universe}},  Phys. Rev. Lett. {\bf 119} (2017),
  no.~14 141301, [\href{http://arxiv.org/abs/1704.04955}{{\tt
  arXiv:1704.04955}}].

\end{thebibliography}\endgroup
\bibliographystyle{BiblioStyle}

\appendix

\section{Dark energy from a confinement PT}
\label{sec:supercooledDS}

In this Section we further discuss the possibility of a period of dark energy behavior from a confinement PT. It is known that confinement PTs are of the first-order kind in several simple cases. The most studied scenario is perhaps that of $SU(N)$ in the large $N$ limit (see~\cite{Creminelli:2001th} for an holographic study, and e.g.~\cite{Baratella:2018pxi, Agashe:2019lhy} for recent updates). This is a conformal field theory (CFT) in the UV. So the considerations of this Section should apply to any strongly-interacting approximate CFT.

In these theories, a confinement scale $\Lambda_c$ can arise due to small marginally relevant deformations. Such a scale will naturally be much smaller than any UV scale of reference. The cosmological evolution of such a dark gauge sector then is (here we follow the discussion in~\cite{Baratella:2018pxi}): At high temperatures, $T_{\text{ds}}\gg \Lambda_c$, the theory is in the deconfined phase and behaves as a hot plasma of dark quarks and dark gluons. Its free energy is then given by $\mathcal{F}\sim -N^2~T_{\text{ds}}^4$. At low temperatures $\Tds \lesssim \Lambda_c$ the theory develops a mass gap $\sim \Lambda_c$ and its free energy then has a contribution from vacuum energy, $\mathcal{F}\sim V_{\text{ds}}\sim N^2~\Lambda_c^4$. The critical temperature is then determined by equating these two contributions, $T_{\text{ds}, c}\lesssim \Lambda_c$. At this temperature, the free energies of the confined and deconfined phase are degenerate, and below it, the confined phase is energetically favored.
In other words, the deconfined and confined phases of the gauge theory can be seen as two minima of the free energy, separated by a thermal potential barrier which is non vanishing at any finite temperature. The confinement PT from one minimum to the other can then be studied as a tunneling phenomenon, and proceeds via the nucleation of cosmic bubbles, with a characteristic temperature-dependent rate $\Gamma(T)$. As is well known in this context, the PT can complete only if $\Gamma(T)> H(T)$ at some $T$, corresponding to efficient bubble percolation. Whether or not this occurs should be assessed by computing the instanton action which connects the two minima.

This is in general a complicated task. However, in some cases the confined phase can be described by the effective field theory of a light {\it dilaton} field, which is the Goldstone boson of spontaneously broken scale invariance. The PT can then be studied as tunneling in a single field potential, as done e.g., in \cite{Baratella:2018pxi}. Very interestingly, the generic expectation is that tunneling in these theories is not efficient at $T_{\text{ds}}\lesssim T_{\text{ds}, c}$, thus the dark sector remains stuck for a while in the deconfined phase, whose free energy at low temperatures is dominated by vacuum energy. The dark confining sector then behaves effectively as a cosmological constant for temperatures below $T_{\text{ds}, c}$ and can thus act as the dark energy component of the effective DS model put forth in this work. This phenomenon is also known as {\it supercooling} (see ~\cite{Witten:1980ez} for seminal work in the context of a weakly coupled Coleman-Weinberg model).

In this regime the DS plasma keeps cooling because of Hubble friction. Tunneling may eventually become efficient at some $T_{\text{ds, n}}\ll T_{\text{ds, c}}$, at which point the vacuum energy is released via the collision of cosmic bubbles. Let us first assume that such an exit from the supercooling phase is indeed possible; we discuss the feasibility of this assumption below. Given the results of our MCMC analysis above, we can then estimate the relation of the DS temperature $T_{\text{ds}}$ to that of the SM bath, using 
\begin{align}
\frac{\rho_{\text{ds,c}}}{\rho_{SM}(z_c)} = \left(\frac{g_{\text{ds,c}}}{g_{\text{SM}}(z_c)}\right)\left(\frac{T_{ds,c}}{T_{\text{SM}}(z_c)}\right)^4\simeq\fpt\left(\frac{\zpt}{z_{c}}\right)^{4},
\end{align}
where $g_{\text{ds}/\text{SM}}$ is the number of relativistic degrees of freedom in the DS sector and in the SM respectively. Let us then identify $z_c$ with the axion parameter $z_a$, since in general we expect axion oscillations to start shortly before confinement of the gauge theory. Using the best-fit results of the previous section, we find
\begin{equation}
\label{eq:dst}
T_{\text{ds}}\simeq 0.2~g_{\text{ds}}^{-1/4}(z_c)~T_{\text{SM}},
\end{equation}
where we have used the fact that the DS sector behaves as radiation at $z>z_c$ and thus the ratio of its entropy and SM entropy remains constant, except for changes in the number of degrees of freedom. The contribution of our DS to dark radiation at BBN is too small to be observable, i.e. $\Delta N_{\text{eff}}= \frac{4}{7}(11/4)^{4/3} g_{\text{DS}} (T_{\text{ds}}/T_{\gamma})^4 \simeq 0.003$. Furthermore, since our dark gauge theory sector may not interact with the SM other than gravitationally, condition \eqref{eq:dst} appears relatively easy to obtain in a realistic model, for instance, by having moderately smaller couplings to the DS at reheating after inflation. 

Note that the confining DS differs from the DDE fluid at very high redshift $z > z_c\sim 16,000$, since it behaves as radiation, rather than dark energy. However, given that the CMB and the sound horizon at recombination are most sensitive to physics at $z\lesssim 10^4$, this difference should not dramatically alter the quality of the fit presented in the previous section. A detailed assessment is left for future work.

Let us now return to tunneling and show more quantitatively how an epoch of supercooling can arise in our scenario. We do so by adapting the results of~\cite{Creminelli:2001th, Baratella:2018pxi, Agashe:2019lhy} to our case. As discussed above, tunneling is expected to occur once the condition $\Gamma\sim \mathcal{A}(T_{\text{ds}}) e^{-S_B(\Tds)}\gtrsim H^4$ is met, where $H\sim T^2/\mpl$ in the radiation dominated Universe and $T>\Tds$ is the temperature of the SM bath.\footnote{Our setup differs in this respect from that of strongly coupled extensions of the SM, since the dark sector never comes to dominate the energy density of the Universe and no phase of inflation occurs. Furthermore, the temperature of the dark sector is smaller than that of the SM bath.} The prefactor $\mathcal{A}(T_{\text{ds}})$ is difficult to estimate; nonetheless, on dimensional grounds we expect $\mathcal{A} (T_{\text{ds}})\sim T_{\text{ds}}^4$. We then find the required size of the tunneling action to be
\begin{equation}
\label{eq:critical}
S_{B}\lesssim 4\ln\left[\frac{\mpl}{T}\frac{T_{\text{ds}}}{T}\right].
\end{equation}
By using the results of the discussion, we have $\Tds\sim 0.1/g_{\text{ds}}^{1/4}~T\sim 0.1/\sqrt{N}~T$. Close to the critical temperature, $T_{\text{ds, c}}\sim \Lambda_c\lesssim 1~\text{eV}$, the tunneling action is then $S_B\sim 230$.
Tunneling can occur either close to the critical temperature, in the thin wall regime, or at lower temperatures in the thick wall regime. In both cases the bounce action is proportional to $N^2$ and tunneling is thus suppressed. More precisely, in the thin wall regime tunneling is expected to be dominated by $O(3)$ bounces, which give~\cite{Creminelli:2001th, Agashe:2019lhy}
\begin{equation}
\label{eq:thinwall}
S_{B, \text{thin}}=\frac{S_3(T_{\text{ds}})}{T_{\text{ds}}}\sim N^2(\beta_{\lambda}\lambda_0)^{-3/4}\frac{T_{\text{ds,c}}/T_\text{ds}}{\left(1-T_\text{ds}^4/T_{\text{ds,c}}^4\right)^2}.
\end{equation}
Here $\lambda$ is the running quartic coupling of the dilaton field, $\lambda_0$ being its value at high scales, and  $\beta_{\lambda}=d\lambda/d\ln\mu$. In this notation, the scale $\Lambda_c$ arises naturally via dimensional transmutation as $\Lambda_c\simeq e^{-1/\beta_{\lambda}}\Lambda_{\text{UV}}$. For $\Lambda_{\text{UV}}\simeq \mpl$ and $\Lambda_c\simeq \text{eV}$ one finds $\beta_{\lambda}\simeq 1/60\ll 1$. Then the action \eqref{eq:thinwall} is always much larger than \eqref{eq:critical} when $T_{\text{ds}}\lesssim T_{c, \text{ds}}$ for any $N$ and tunneling cannot occur in the thin wall regime.
In the thick wall regime, tunneling is dominated by $O(4)$ bounces, which give~\cite{Baratella:2018pxi}
\begin{equation}
\label{eq:thickwall}
S_{B, \text{thick}}\sim \frac{24 N^2}{\beta_{\lambda} \ln\left(\frac{T_{\text{ds, c}}}{\Tds}\right)}.
\end{equation}
We see that it is possible to tunnel only after a long supercooling epoch, i.e., when $T_{c, \text{ds}}/T_{\text{ds}}\gg 1$. However, for our values of parameters we find that thick wall tunneling is not observationally feasible for $N>1$, unless $\beta_{\lambda}\gtrsim 0.1 $. In certain constructions, larger values of $\beta_{\lambda}$ may be allowed and thus tunneling might be faster, see e.g., \cite{Baratella:2018pxi, Agashe:2019lhy}. In this case it might be possible to complete the PT when the vacuum energy of the dark sector is still subdominant with respect to the SM bath (our DS model requires it to be $\sim 15\%$ of the total energy close to matter-radiation equality). 

The discussion above shows that in simple theories with a confinement PT, an epoch of dark energy behavior is expected, and can actually constitute an observationally dangerous prediction of these models, which closely resembles the graceful exit problem of old inflation. A possible way out has been first pointed out in~\cite{Witten:1980ez} and more recently discussed by~\cite{ Iso:2017uuu} and \cite{vonHarling:2017yew} for strongly-coupled models (see also~\cite{Baratella:2018pxi}): the introduction of an explicit breaking of scale invariance, which corresponds to some massive states coupled to the approximately conformal dark sector.\footnote{For scenarios with scale invariance in the electroweak sector, such as those considered in~\cite{Witten:1980ez, vonHarling:2017yew, Iso:2017uuu, Baratella:2018pxi}, such a breaking naturally comes from QCD confinement. A similar process may occur in our scenario, if the dark sector is actually made of two theories with similar confinement scales.} These states would then play a role similar to that of the trigger field in NEDE. In this case, the tunneling action is drastically modified once the temperature of the dark sector has dropped to the mass scale at which scale invariance is explicitly broken and tunneling can then proceed efficiently, on a time scale which is much smaller than the corresponding Hubble time. In our setup, such massive states would have to be relevant at a scale $\Tds\lesssim 1~\text{eV}$. Notice that the abundance of these states at any time can be very small and thus should not introduce further cosmological problems. Furthermore, this possibility also evades the generation of large-scale observable anisotropies from bubbles: First, bubbles which would be nucleated around $z_a\simeq 16,000$ have a very small probability of being nucleated at all, since $S_{\text{B}}\gg 240$ (see also the discussion in \cite{Niedermann:2020dwg}). Second, bubbles nucleated at $\zpt\simeq 5,000$ never grow to horizon size, since the time scale of the PT $\beta^{-1}\sim (\dot{\Gamma}/\Gamma)^{-1}$ can be much smaller than one Hubble time, i.e. typically $\beta/H\sim 10^{3}-10^{4}$.

While we leave a more detailed investigation of the aspects above for future work, we would like to remark that the problem of obtaining a sharp PT around equality is perhaps analogous to the coincidence problem of popular scenarios which address the Hubble tension only. In EDE, the mass of the scalar field and its initial value are chosen ad-hoc to obtain a sharp transition around equality. In NEDE scenarios, two mass scales have to be chosen ad-hoc: the vacuum energy released by the dark sector and the mass of the scalar field which triggers the PT. In our scenario, the first mass scale arises naturally because of dimensional transmutation, whereas the second scale likely requires additional ingredients to the dark sector, which would play a role similar to that of the trigger field in the NEDE scenario. 

The details of the generation of the axion mass from non-perturbative effects during the supercooled stage in the deconfined phase is another interesting task for future work.

Concerning the evolution of the DS after the PT, we remark that a detailed investigation of the plausibility of achieving $w\approx 2/3$ in this era is needed. As we have mentioned above, one possibility is that the EoS may be stiff in the presence of a large chemical potential, as appears to be hinted by studies of QCD matter inside neutron stars. In contrast to QCD, a DS can exhibit a large dark baryon asymmetry $\eta_{dB}\sim \mathcal{O}(1)$ in the early Universe, thus dense dark QCD matter may be achieved even at finite temperature. 
We leave it to future work to examine this, and to present concrete numerical results that implement this. Or, alternatively, to develop theoretical reasons that constrain this possibility completely.

\section{Model Comparison}
\label{sec:numerics}

We report in this section further detailed numerical results of our MCMC analysis, for our DS model as well as for the $\Lambda$CDM model, including comparisons of the matter power spectra and temperature anisotropies obtained with {\tt CLASS}. As a means to compare with existing models which aim at addressing the Hubble tension, we also presents results for the 2-parameters NEDE model of~\cite{Niedermann:2019olb, Niedermann:2020dwg}. Detailed results for the EDE model of~\cite{Poulin:2018dzj} can be found in~\cite{Hill:2020osr, Ivanov:2020ril, DAmico:2020ods, Smith:2020rxx, Murgia:2020ryi}.

Let us first make a remark on our numerical strategy for the parameter $\zpt$, as compared to that adopted for the NEDE model in~\cite{Niedermann:2019olb} and for the EDE model in \cite{Smith:2020rxx}: We have chosen to keep the DDE parameters $\fpt$ and $\zpt$ (or more precisely, the ``trigger'' field mass $m_t$) free to vary in our MCMC analysis to fully explore the parameter space of this component, for all the four combinations of datasets used in this work. The authors of~\cite{Niedermann:2020dwg} and \cite{Smith:2020rxx} have argued that such a choice can lead to volume sampling effects when the SH$_0$ES prior is not included, which could then negatively impact the inferred values of the NEDE or EDE parameters. Therefore, they have chosen to consider 1-parameter submodels of the NEDE and EDE scenarios, by fixing respectively $m_t$ for the NEDE model and $z_a, \theta_i$ for the EDE model (we use the same notation as for our DS model; however notice that in the EDE model the axion field has a non-standard $\sim \cos(a/f)^3$ potential, rather than a standard $\cos(a/f)$ potential as in our case). Using this strategy, they found values of $H_0$ that are significantly larger than what would be inferred from the full 2- and 3-parameters NEDE and EDE models respectively. On the other hand, the authors of~\cite{Ivanov:2020ril, DAmico:2020ods} have argued that the 3-parameter MCMC analysis of the EDE model is not significantly affected by such volume effects, even in the absence of the SH$_0$ES prior, and thus that one should draw conclusions on the $H_0$ tension from the 3-parameter MCMC analysis. This would then justify our choice. However, it is important to keep in mind that the results reported in this work for our DS model should not be compared to the results of~\cite{Niedermann:2019olb} and \cite{Smith:2020rxx} for the 1-parameter NEDE and EDE submodels. 

\begin{figure*}[hbt!]

	\centering
	\includegraphics[width=0.45\linewidth]{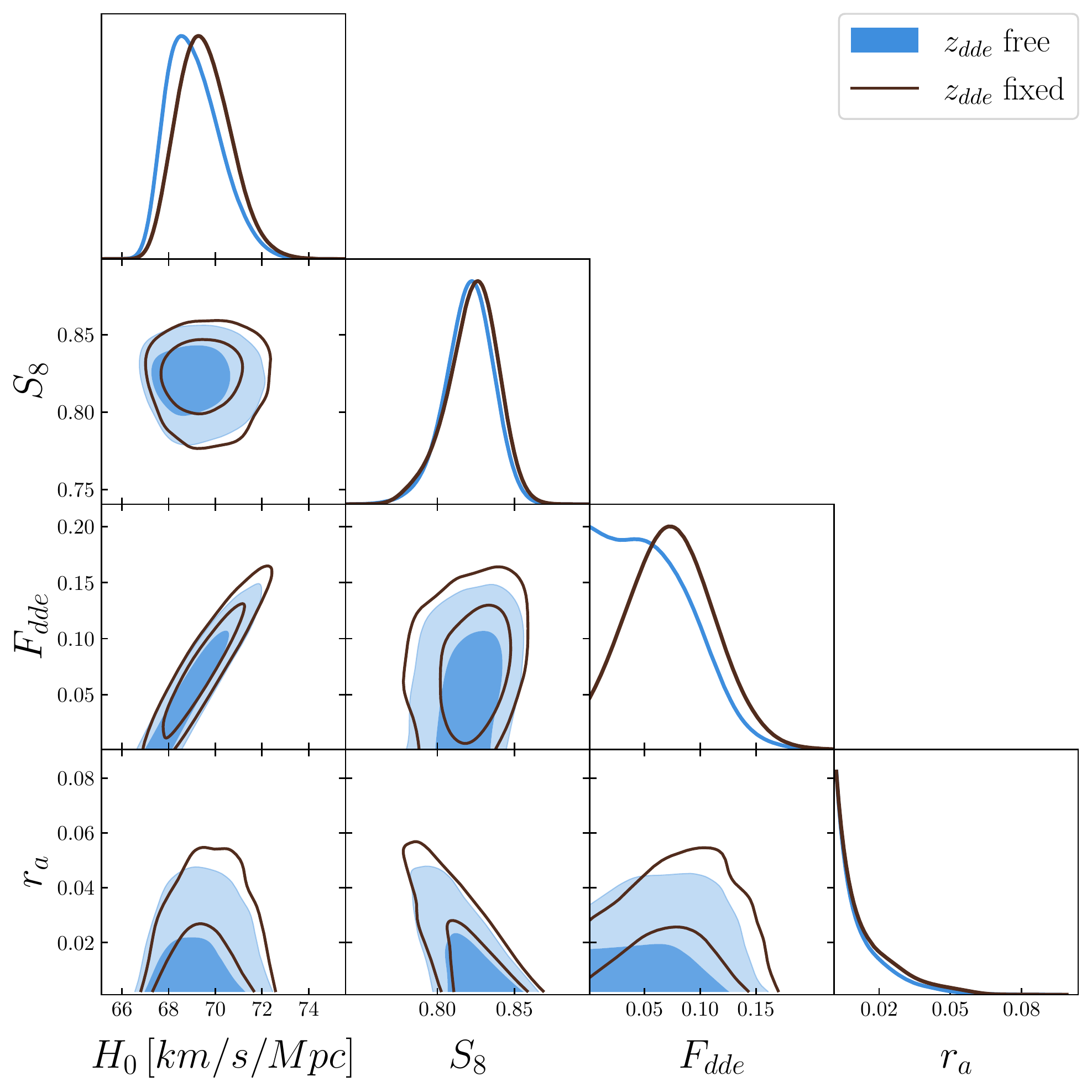}
	\caption{{\small Comparison of 1-D and 2-D posteriors for $H_0$, $S_8$, $\fpt$, and $r_a$ between the DS model presented in this work and the 2-parameter submodel wherein $m_t$ (or almost equivalently $\zpt$) is fixed. These results were obtained with the P18+BAO+EFT dataset.}} 
\label{fig:3pv2pEFT}
\end{figure*}

Rather, to perform a fair comparison of the values of $H_0$ inferred without the SH$_0$ES prior by~\cite{Niedermann:2020dwg} and~\cite{Smith:2020rxx}, we have performed a MCMC analysis of a 2-parameter submodel of our DS model, by keeping $m_t$ (or almost equivalently, $\zpt$, see footnote 1 in the main text) fixed, and only $\fpt$ and $\Faxdm$ free to vary (in our MCMC analysis we are forced to fix $\zax$ to obtain reliable results, due to strong parameter degeneracies). The results of this choice are shown for the dataset P18+BAO+EFT in Figure~\ref{fig:3pv2pEFT} and Table~\ref{tab:3pv2pEFT}.
Overall, the analysis of this 2-parameter submodel shows an increase in $H_0$ compared to what we presented in this work for the same datasets, as well as a detection of $\fpt>0$ at greater than 1$\sigma$. In addition, the fraction of axion dark matter $r_a$ allowed at 95\% CL is increased to $r_a<0.043$.

In this sense, our choice in this work can be considered as a conservative one. Since results for the 2-parameters NEDE model without the SH$_0$ES prior have not yet been presented in the literature, we have obtained them by using the publicly released {\tt TriggerCLASS} code developed in ~\cite{Niedermann:2019olb, Niedermann:2020dwg}.

\begin{table*}[hbt!]
\begin{tabular} {| l | c| c|}
\hline\hline
 \multicolumn{1}{|c|}{ Parameter} &  \multicolumn{1}{|c|}{~~~$z_{dde}$ free~~~} &  \multicolumn{1}{|c|}{~~~~~$z_{dde}$ fixed~~~~~}\\
\hline\hline
$r_a\equiv\Omega_a/\Omega_{\text{dm}}                       $ & $< 0.039~[95\%]~(0.014)$ & $< 0.043~[95\%]~(0.007)$\\
$F_{dde}                   $ & $< 0.124~[95\%]~(0.11)$ & $0.075~(0.082)^{+0.035}_{-0.041}   $\\
$H_0\,[\text{km/s/Mpc}]                       $ & $69.09~(70.58)^{+0.86}_{-1.4}      $ & $69.53~(69.65)^{+0.99}_{-1.3}      $\\
$S_8                       $ & $0.820~(0.826)^{+0.017}_{-0.014}   $ & $0.823~(0.835)^{+0.019}_{-0.014}   $\\
\hline
Tension with SH$_0$ES & $3.0\, \sigma $ & $2.6\, \sigma $\\
Tension with S$_8$ & $2.8\, \sigma $ & $2.9\, \sigma $\\
\hline
$\chi^2_{DS}-\chi^2_{\Lambda CDM}$ & -1.6  & -1.8\\
\hline
\end{tabular}
\caption{The mean (best-fit in parenthesis) $\pm1\sigma$ error of $r_a$, $\fpt$, $H_0$, and $S_8$ obtained by fitting the DS model presented in this work and the 2-parameter submodel wherein $m_t$ (or almost equivalently $\zpt$) is fixed to the cosmological datasets: Planck18+BAO+EFT. The tensions with SH$_0$ES and the $S_8$ prior are shown, demonstrating the effect of fixing this parameter.}
\label{tab:3pv2pEFT}
\end{table*}
 
Finally, before moving to the numerical results, let us discuss how our results would be affected by a different choice of the axion redshift $\zax$. In this work, we have fixed $\log_{10}z_a=4.2$, which approximately corresponds to $m_a\simeq 10^{-26}~\text{eV}$. This choice was guided by the best-fit values of 4-parameters MCMC analyses, which however fail to achieve reliable results for central values and corresponding error bars because of known degeneracies between ULA and cold DM (see also~\cite{Poulin:2018dzj} for the same choice of fixing $z_a$, as well as the discussion in Appendix F of~\cite{Marsh:2015xka}) when $\log_{10}z_a\gg 4.5$. This is the reason why we kept $\zax$ fixed, rather than free to vary as $\zpt$.
Nonetheless, we find that other choices for $\zax$ are possible, without strongly altering our conclusions. In general, the larger $\zax$, the looser CMB constraints on the axion DM fraction $r_a$ become. Furthermore, for larger $\zax$, a larger $r_a$ is required to address the $S_8$ tension (see the section describing the phenomenological model above and~\cite{Kobayashi:2017jcf}).

\begin{figure*}[hbt!]

	\centering
	\includegraphics[width=0.45\linewidth]{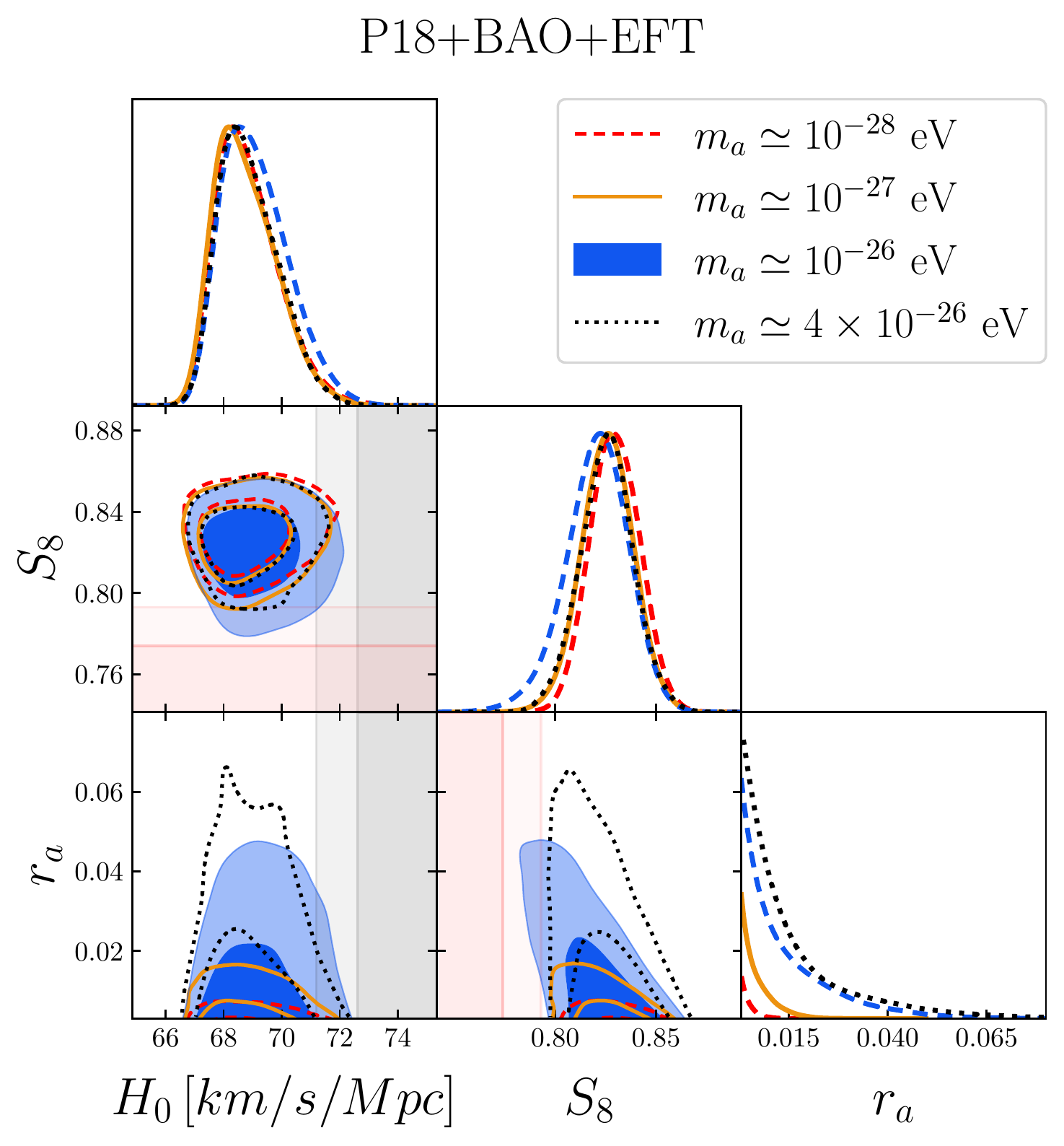}
	\caption{{\small Comparison of 1-D and 2-D posteriors of the DS model for $H_0$, $S_8$, and $r_a$, using four different values of the parameter $\log_{10}z_a = 3.2,\,3.7,\,4.2,\,4.5$. These four choices correspond to axion masses $m_a\simeq 10^{-28},\,10^{-27},\,10^{-26},$ and $4\times 10^{-26}\,\mbox{eV}$, respectively. These results were obtained with the P18+BAO+EFT dataset. In grey are shown the 1-$\sigma$ (darker) and 2-$\sigma$ (lighter) ranges for $H_0$ from SH$_{0}$ES, and similarly the $S_8$ prior is shown in pink.}} 
\label{fig:ac_comp_EFT}
\end{figure*}

\begin{figure*}[hbt!]

	\centering
	\includegraphics[width=0.45\linewidth]{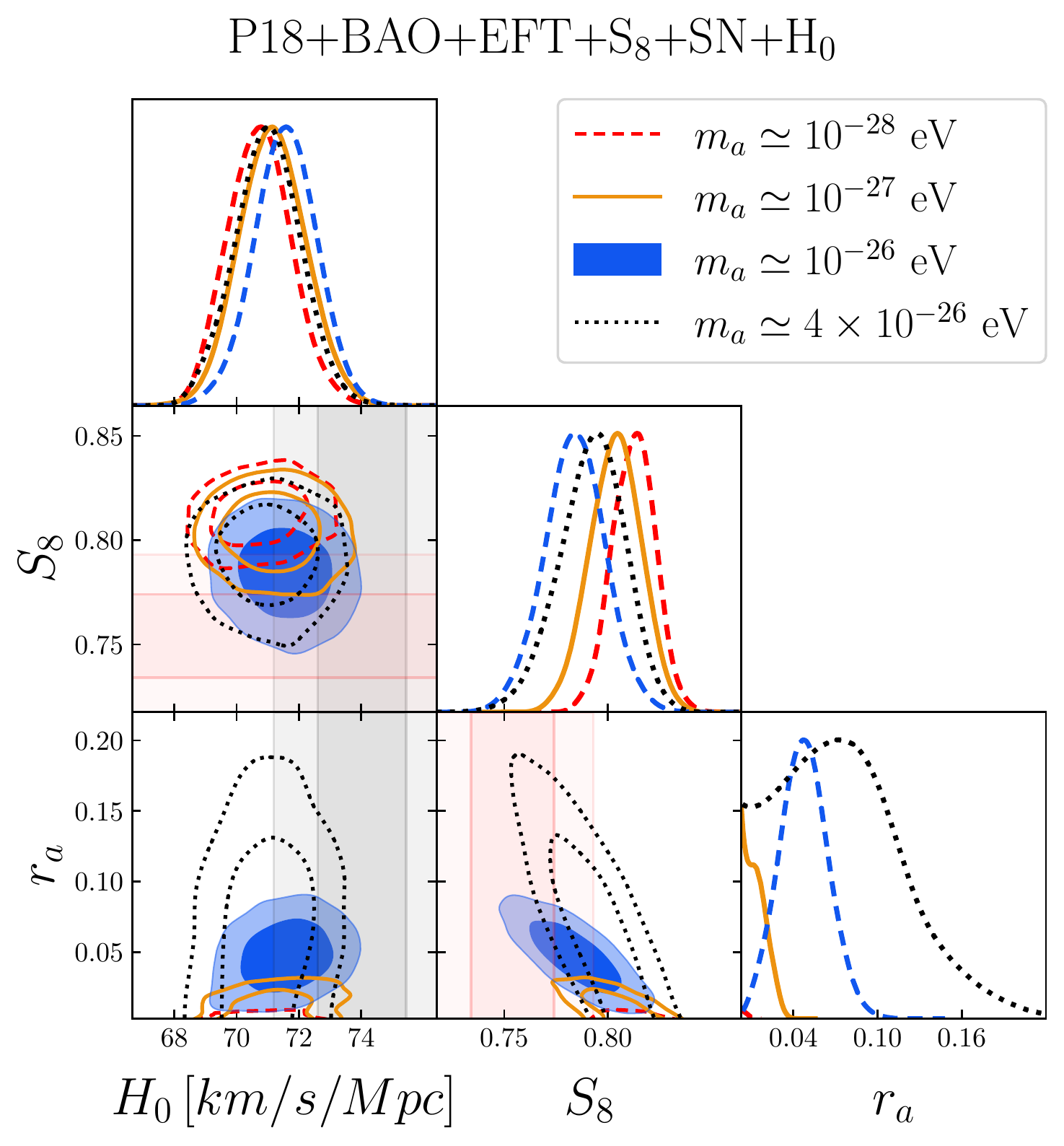}
	\caption{{\small Comparison of 1-D and 2-D posteriors of the DS model for $H_0$, $S_8$, and $r_a$, using four different values of the parameter $\log_{10}z_a = 3.2,\,3.7,\,4.2,\,4.5$. These four choices correspond to axion masses $m_a\simeq 10^{-28},\,10^{-27},\,10^{-26},$ and $4\times 10^{-26}\,\mbox{eV}$, respectively. These results were obtained with the full P18+BAO+EFT+S$_8$+SN+H$_0$ dataset. In grey are shown the 1-$\sigma$ (darker) and 2-$\sigma$ (lighter) ranges for $H_0$ from SH$_{0}$ES, and similarly the $S_8$ prior is shown in pink.}} 
\label{fig:ac_comp_full}
\end{figure*}

We present the MCMC analysis for several choices for $\log_{10}z_a$ in Figures~\ref{fig:ac_comp_EFT} and~\ref{fig:ac_comp_full}, and Tables~\ref{tab:ac_comp_EFT} and~\ref{tab:ac_comp_full}. The values of $\log_{10}z_a$ are chosen to be $3.2$, $3.7$, $4.2$ (corresponding to the analysis presented in our main work), and $4.5$. These roughly correspond to axion masses $m_a\simeq 10^{-28},\,10^{-27},\,10^{-26},$ and $4\times 10^{-26}\,\mbox{eV}$, respectively. Comparing these four choices when applied to a fit to the dataset P18+BAO+EFT, we find that $10^{-27}~\text{eV}~\lesssim m_a \lesssim 10^{-25}~\text{eV}~$ perform almost equally well in their goodness-of-fit, giving similar $\chi^2$s. However, Figure~\ref{fig:ac_comp_EFT} clearly shows that the choice $\log_{10}z_a=4.2$ or $m_a\simeq10^{-26}\,\mbox{eV}$ allows for the least tension with both SH$_0$ES and the $S_8$ prior, indicated also in Table~\ref{tab:ac_comp_EFT}. Turning now to the use of the full dataset P18+BAO+EFT+S$_8$+SN+H$_0$, we see that the choice $\log_{10}z_a=3.2$ corresponding to $m_a\simeq10^{-28}\,\mbox{eV}$ is far too constrained to significantly affect the $S_8$ tension, while the other choices allow for high enough $r_a$ to relieve the tension. However, as indicated by conducting the MCMC analysis with $z_a$ kept free, the best choice is indeed $\log_{10}z_a =4.2$ as it allows for the most significant resolution of the tensions. This is clearly reflected in the values of the $\chi^2$ in Table~\ref{tab:ac_comp_full}.

\begin{table*}[hbt!]
\begin{tabular} {| l | c| c| c| c|}
\hline\hline
 \multicolumn{1}{|c|}{ Parameter} &  \multicolumn{1}{|c|}{$m_a\simeq 10^{-28}\,\text{eV}$} &  \multicolumn{1}{|c|}{$m_a\simeq 10^{-27}\,\text{eV}$} &  \multicolumn{1}{|c|}{$m_a\simeq 10^{-26}\,\text{eV}$} &  \multicolumn{1}{|c|}{$m_a\simeq 4\times 10^{-26}\,\text{eV}$}\\
\hline\hline
$H_0\,[\text{km/s/Mpc}]                       $ & $68.83~(68.08)^{+0.78}_{-1.3}      $ & $68.77~(69.9)^{+0.80}_{-1.3}      $ & $69.09~(70.58)^{+0.86}_{-1.4}      $ & $68.86~(68.75)^{+0.78}_{-1.2}      $\\
$S_8                       $ & $0.830~(0.831)^{+0.012}_{-0.012}   $ & $0.826~(0.824)^{+0.013}_{-0.013}   $ & $0.820~(0.826)^{+0.017}_{-0.014}   $ & $0.826~(0.834)^{+0.013}_{-0.013}   $\\
\hline
Tension with SH$_0$ES & $3.2\, \sigma $ & $3.2\, \sigma $ & $3.0\, \sigma $ & $3.2\, \sigma $\\
Tension with S$_8$ & $3.3\, \sigma $ & $3.1\, \sigma $ & $2.8\, \sigma $ & $3.1\, \sigma $\\
\hline
$\chi^2_{DS}-\chi^2_{\Lambda CDM}$ & -0.6 & -2.0 & -1.6 & -1.0\\
\hline
\end{tabular}
\caption{The mean (best-fit in parenthesis) $\pm1\sigma$ error of $H_0$ and $S_8$ obtained by fitting the DS model to the cosmological dataset P18+BAO+EFT, choosing four different values of the parameter $\log_{10}z_a = 3.2,\,3.7,\,4.2,\,4.5$. These four choices correspond to axion masses $m_a\simeq 10^{-28},\,10^{-27},\,10^{-26},$ and $4\times 10^{-26}\,\mbox{eV}$, respectively. The tensions with SH$_0$ES and the $S_8$ prior are shown, demonstrating the effect of choosing this parameter.}
\label{tab:ac_comp_EFT}
\end{table*}

We now report detailed tables of cosmological and derived parameters for all the four combination of datasets considered in this work. We also present complete posteriors for $\Lambda$CDM and for the DS model. Furthermore, for completeness, we also report results obtained without the EFTofLSS information, i.e. without the {\tt PyBird} likelihood. We denote datasets according to the notation outlined above. In order to assess tensions with the $H_0$ and $S_8$ measurements, we use the measure described in p. 4 of our paper. This measure of the $H_0$ tension may be an overestimate for models where the $H_0$ posterior is non-Gaussian, such is the case in both the DS and NEDE models when the $H_0$ prior is not included (this can also be understood from the results reported in Table~\ref{tab:3pv2pEFT}). While an alternative measure may further reduce the $H_0$ tension in both the DS and NEDE models than what we present here~\cite{Schoneberg:2021qvd}, the comparison between the DS and NEDE model would not be significantly affected, since they both feature non-Gaussian $H_0$ posteriors. Our results are obtained by merging two publicly available extensions of {\tt CLASS}~\cite{Blas:2011rf}, {\tt AxiCLASS}~\cite{Poulin:2018dzj, Smith:2019ihp} to model ULAs and {\tt TriggerCLASS}~\cite{Niedermann:2019olb, Niedermann:2020dwg} to model a NEDE fluid. In our MCMC scan, for the ULA component we fix $a_i/f_a = 2, \log_{10}z_a=4.2$ and keep the ULA fraction of the total energy density at the redshift $z_a$, $\log_{10}\Omega_a(z_{a})$, free to vary. In all our analyses, we set a lower bound on $\log_{10}\Omega_{z_{a}}\geq -3.5$ to avoid volume effects. We then present results in terms of the derived fraction of the DM abundance today: $r_a\equiv \frac{\Omega_a}{\Omega_{dm}}$. For the DDE/NEDE component, we fix $H/m_{t}=0.2$ and $w_f=2/3$. We keep the parameters $\fpt$ and $\log_{10} m_{t}$ free to vary, while presenting the derived redshift of the NEDE transition $z_{dde}$ rather than $\log_{10} m_{t}$.


\begin{widetext}

\begin{table*}[t!]
\begin{tabular} {| l | c| c| c| c|}
\hline\hline
 \multicolumn{1}{|c|}{ Parameter} &  \multicolumn{1}{|c|}{$m_a\simeq 10^{-28}\,\text{eV}$} &  \multicolumn{1}{|c|}{$m_a\simeq 10^{-27}\,\text{eV}$} &  \multicolumn{1}{|c|}{$m_a\simeq 10^{-26}\,\text{eV}$} &  \multicolumn{1}{|c|}{$m_a\simeq 4\times 10^{-26}\,\text{eV}$}\\
\hline\hline
$H_0\,[\text{km/s/Mpc}]                       $ & $70.7~(70.2)^{+1.0}_{-1.0}        $ & $71.2~(71.4)^{+1.1}_{-1.1}        $ & $71.56~(70.99)^{+0.99}_{-0.98}     $ & $71.0~(70.9)^{+1.1}_{-1.1}        $\\
$S_8                       $ & $0.812~(0.812)^{+0.010}_{-0.010}   $ & $0.804~(0.797)^{+0.013}_{-0.013}   $ & $0.784~(0.789)^{+0.014}_{-0.014}   $ & $0.792~(0.783)^{+0.017}_{-0.014}   $\\
\hline
Tension with SH$_0$ES & $1.9\, \sigma $ & $1.6\, \sigma $ & $1.4\, \sigma $ & $1.7\, \sigma $\\
Tension with S$_8$ & $2.6\, \sigma $ & $2.1\, \sigma $ & $1.2\, \sigma $ & $1.6\, \sigma $\\
\hline
$\chi^2_{DS}-\chi^2_{\Lambda CDM}$ & -6.9 & -9.7 & -17.9 & -11.7\\
\hline
\end{tabular}
\caption{The mean (best-fit in parenthesis) $\pm1\sigma$ error of $H_0$ and $S_8$ obtained by fitting the DS model to the cosmological dataset P18+BAO+EFT+S$_8$+SN+H$_0$, choosing four different values of the parameter $\log_{10}z_a = 3.2,\,3.7,\,4.2,\,4.5$. These four choices correspond to axion masses $m_a\simeq 10^{-28},\,10^{-27},\,10^{-26},$ and $4\times 10^{-26}\,\mbox{eV}$, respectively. The tensions with SH$_0$ES and the $S_8$ prior are shown, demonstrating the effect of choosing this parameter.}
\label{tab:ac_comp_full}
\end{table*}

\subsection*{P18+BAO+EFT}

\begin{figure*}[h]
	\centering
	\includegraphics[width=0.45\linewidth]{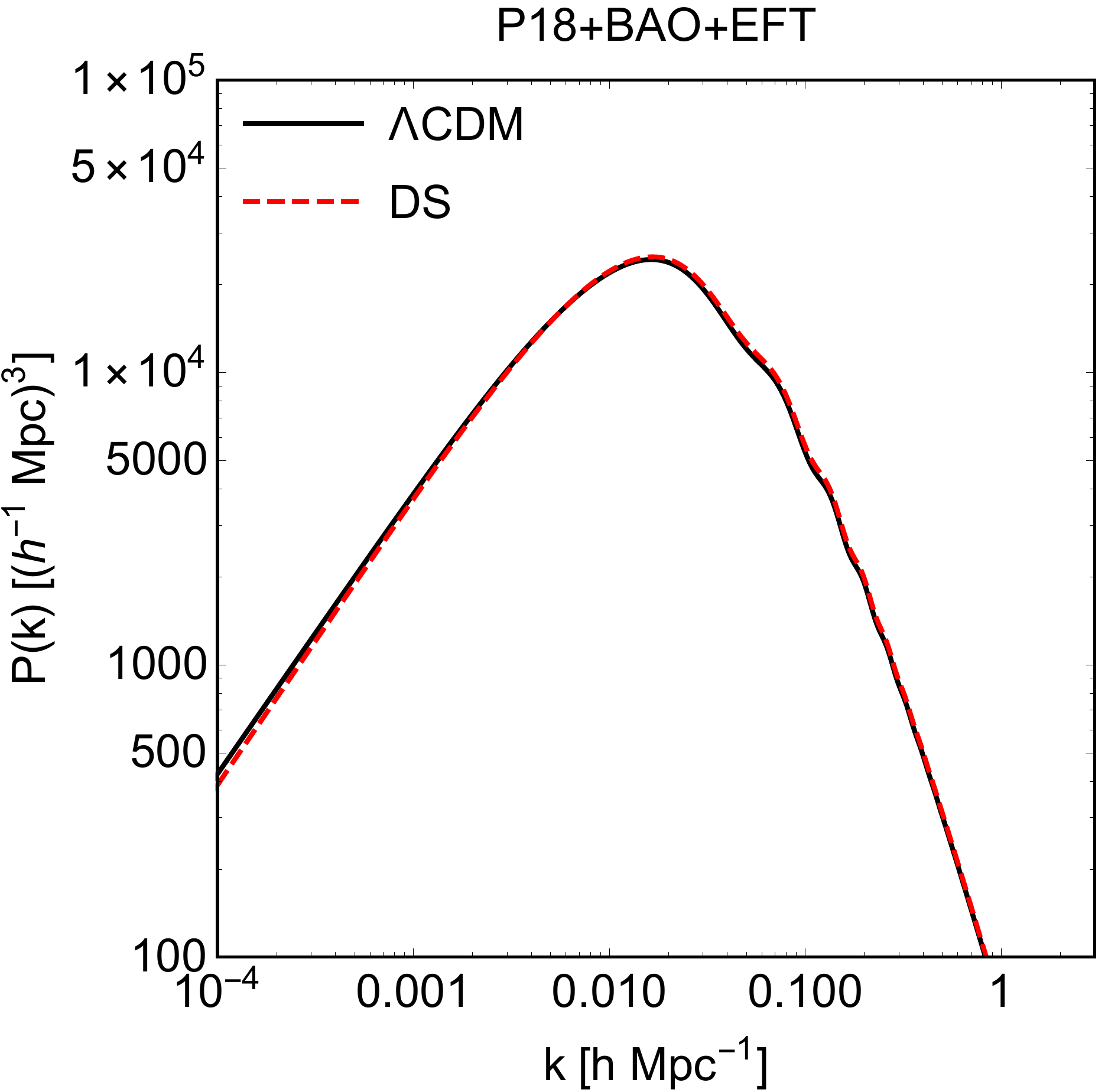}
	\caption{\small Linear matter power spectra at $z=0$ for $\Lambda$CDM (solid, black) and the DS model (dashed, red) for the best fit values of cosmological parameters obtained with the P18+BAO+EFT dataset.  \label{fig:PkPBE}}
\end{figure*}

The matter and temperature anisotropy power spectra for the best fit values of cosmological parameters in the $\Lambda$CDM and DS models are almost indistinguishable in this case.

The 2-parameter NEDE model has a similar $\chi^2$ as the 3-parameter DS model. However, the additional ULA component in the DS scenario leads to a significant reduction of the $S_8$ tension, from $3.5\sigma$ in NEDE to $2.8\sigma$ in the DS model and to a milder improvement compared to the $\Lambda$CDM model, in the absence of an $S_8$ prior. In contrast, in the NEDE model the $S_8$ tension is stronger than in the $\Lambda$CDM model. The $H_0$ tension is also slightly smaller in the DS model than in the NEDE model.

\begin{figure*}[b]
	\centering
	\includegraphics[width=0.45\linewidth]{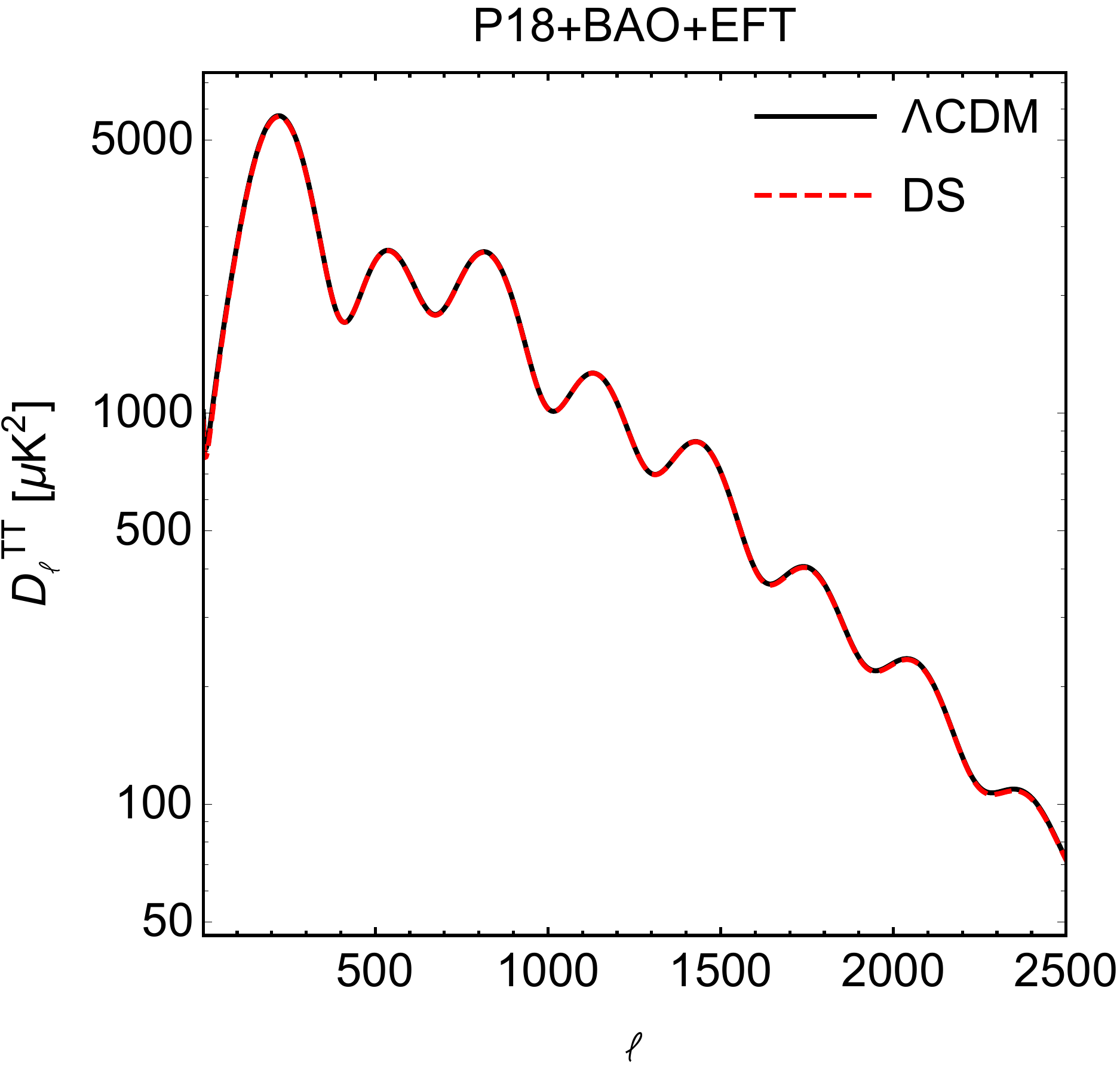}
	\, \,
	\includegraphics[width=0.44\linewidth]{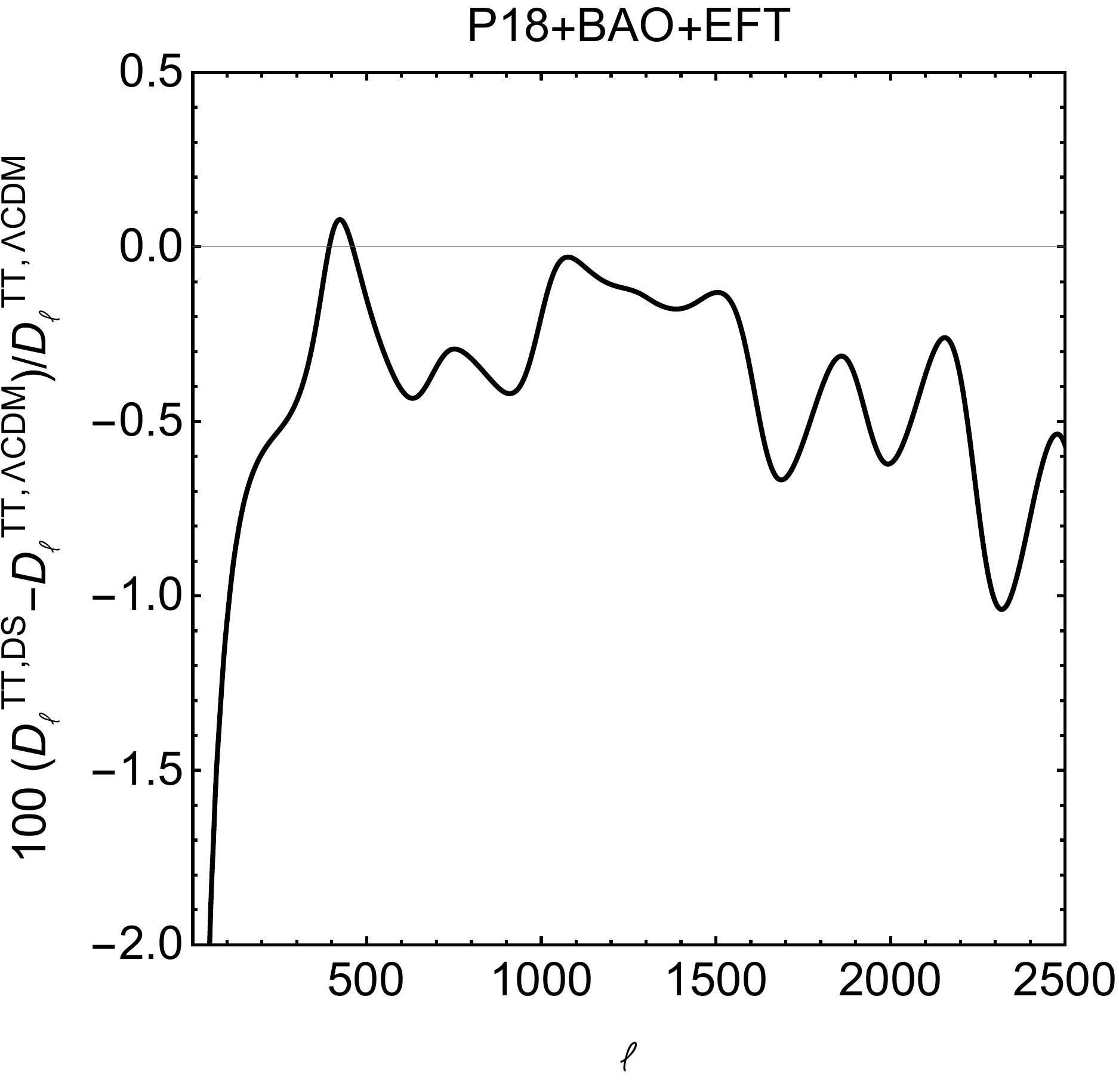}
	\caption{\small {\it Left}: CMB temperature anisotropy power spectra for $\Lambda$CDM (solid, black) and the DS model. {\it Right}: Residuals, multiplied by $10^2$. The curves have been obtained using the best fit values of cosmological parameters obtained with the P18+BAO+EFT dataset, reported in Table~\ref{table:resultsEPB}. \label{fig:DlPBE}}
\end{figure*}

\begin{table*}[hbt!]
\begin{tabular} {| l | c| c| c|}
\hline\hline
 \multicolumn{1}{|c|}{ Parameter} &  \multicolumn{1}{|c|}{~~~$\Lambda$CDM~~~} &  \multicolumn{1}{|c|}{~~~~~DS~~~~~} &  \multicolumn{1}{|c|}{~~~~NEDE~~~~}\\
\hline\hline
$100 \omega_b              $ & $2.239~(2.25)^{+0.013}_{-0.013}   $ & $2.265~(2.289)^{+0.020}_{-0.027}   $ & $2.259~(2.256)^{+0.019}_{-0.024}   $\\
$\omega_{cdm }             $ & $0.11949~(0.1192)^{+0.00094}_{-0.00096}$ & $0.1227~(0.127)^{+0.0027}_{-0.0040}$ & $0.1235~(0.1235)^{+0.0021}_{-0.0039}$\\
$\ln 10^{10}A_s            $ & $3.046~(3.029)^{+0.014}_{-0.014}   $ & $3.054~(3.058)^{+0.015}_{-0.015}   $ & $3.053~(3.052)^{+0.014}_{-0.016}   $\\
$n_{s }                    $ & $0.9656~(0.9661)^{+0.0038}_{-0.0038}$ & $0.9743~(0.9864)^{+0.0067}_{-0.0087}$ & $0.9734~(0.9745)^{+0.0062}_{-0.0084}$\\
$\tau_{reio }              $ & $0.0555~(0.0471)^{+0.0068}_{-0.0074}$ & $0.0561~(0.0551)^{+0.0068}_{-0.0075}$ & $0.0557~(0.0539)^{+0.0064}_{-0.0075}$\\
$H_0\,[\text{km/s/Mpc}]                       $ & $67.59~(67.78)^{+0.42}_{-0.42}     $ & $69.09~(70.58)^{+0.86}_{-1.4}      $ & $68.86~(68.34)^{+0.75}_{-1.3}      $\\
\hline
$F_{dde}                   $ &- & $< 0.124~[95\%]~(0.11)$ & $< 0.119~[95\%]~(0.043)$\\
$z_{dde}                   $ &- & $5193~(5352)^{+1300}_{-1600}  $ & $5330~(5604)^{+1500}_{-2000}  $\\
$r_a\equiv\Omega_a/\Omega_{\text{dm}}                       $ &- & $< 0.039~[95\%]~(0.014)$ &-\\
\hline
$\log_{10}z_a$ &- & fixed to: 4.2 &-\\
$S_8                       $ & $0.825~(0.814)^{+0.011}_{-0.011}   $ & $0.820~(0.826)^{+0.017}_{-0.014}   $ & $0.833~(0.841)^{+0.011}_{-0.012}   $\\
\hline
Tension with SH$_0$ES & $4.3\, \sigma $ & $3.0\, \sigma $ & $3.2\, \sigma $\\
Tension with S$_8$ & $3.2\, \sigma $ & $2.8\, \sigma $ & $3.5\, \sigma $\\
\hline
$\chi^2_{DS}-\chi^2_{\Lambda CDM}$ & - & $-1.6$ & $-1.0$\\
\hline
\end{tabular}
  \caption{The mean (best-fit in parenthesis) $\pm1\sigma$ error of the cosmological parameters obtained by fitting $\Lambda$CDM, our Dark Sector (DS) model, and the New Early Dark Energy (NEDE) model to the cosmological datasets: Planck18+BAO+EFT. Upper bounds are presented at 95\% CL. }
  \label{table:resultsEPB}
\end{table*}

\begin{table}[h!]
\centering
{\renewcommand{\arraystretch}{1.25} 
{\begin{tabular}{|c | c | c | c |}
\hline
Dataset & $\Lambda$CDM & DS & NEDE \\
\hline
 Planck highl TTTEEE  & 2353.91 & 2353.25 & 2351.26\\
\hline
  Planck lowl EE   & 395.82 & 396.09 & 397.45\\
\hline
  Planck lowl TT  & 23.05 & 20.95 & 22.20\\
\hline
 Planck lensing  & 9.80 & 9.36 &  8.88\\
\hline
   eft with bao highz NGC & 67.70 & 67.61 &  69.00\\
\hline
  eft with bao highz SGC & 61.42 & 61.79 & 61.60  \\
\hline
   eft with bao lowz NGC & 69.25 & 70.10 & 69.63   \\
   \hline
  bao smallz 2014 &  1.26 & 1.47 & 1.15  \\
\hline
Total  & 2982.20 & 2980.61 & 2981.16\\
\hline
\end{tabular}}
}
\caption{\small Contributions to the total $\chi^2_{\rm eff}$ for individual data sets, for the best-fits of the $\Lambda$CDM, DS, and NEDE models. These were obtained when performing a fit to the data sets P18+BAO+EFT.\label{chi2EPB}}
\end{table}

\begin{figure*}[t]
\includegraphics[width=0.87\textwidth]{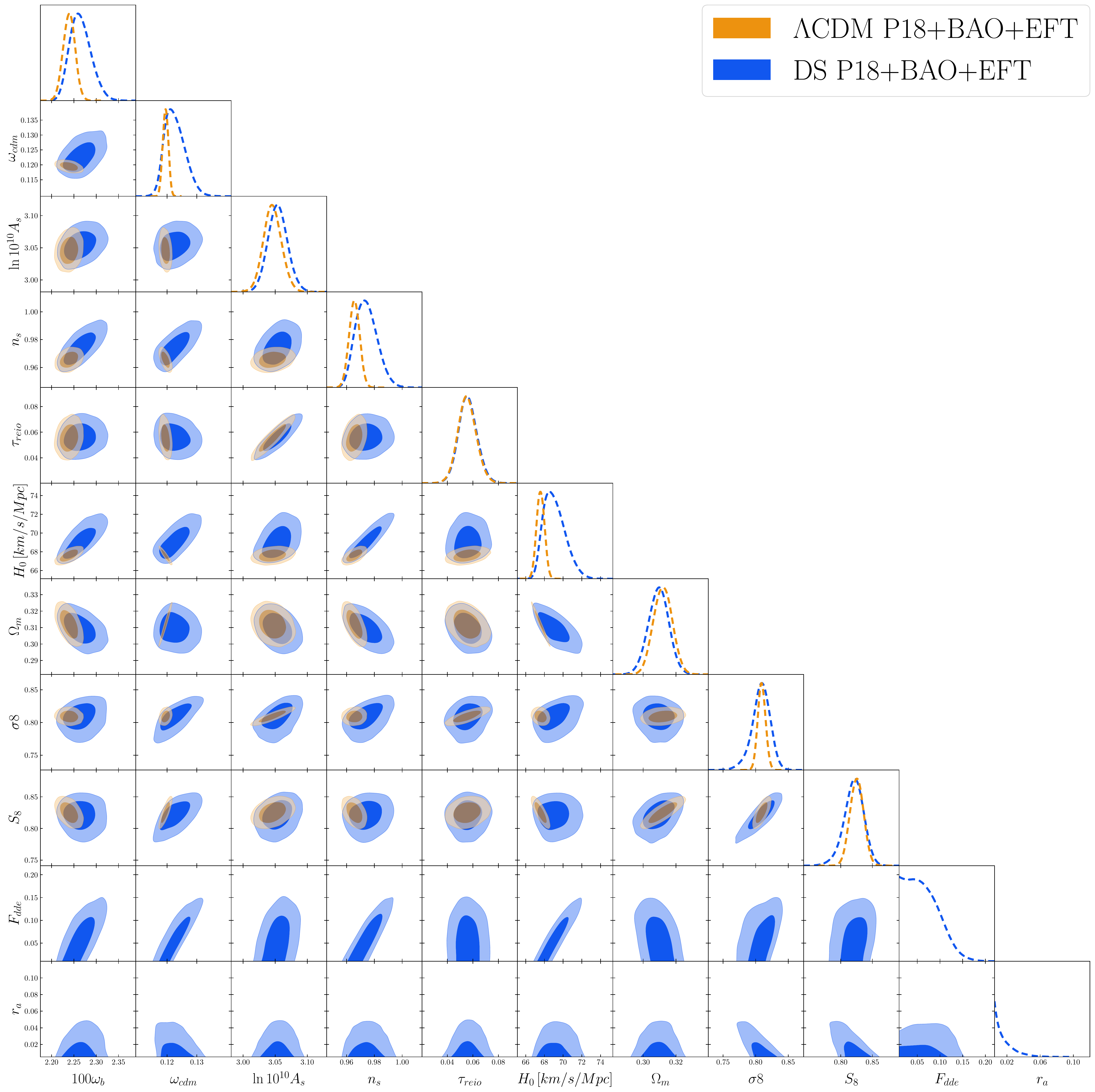}
\caption{Comparison of posterior distributions of $\Lambda$CDM parameters for $\Lambda$CDM and the Dark Sector model, fit to the Planck18+BAO+EFT data set.}
\label{fig:ELvPB}
\end{figure*}
\FloatBarrier

\subsection*{P18+BAO+EFT+S$_8$}

With the inclusion of the $S_8$ prior, the best fit value of the ULA DM fraction $r_a$ is raised to $~4\%$. As a consequence, the linear matter power spectrum is significantly suppressed at scales $0.1 h/\text{Mpc}\lesssim k\lesssim 1 h/\text{Mpc}$ in the DS model as compared to the $\Lambda$CDM model, see Fig.~\ref{fig:PkPBES}. 

This leads to a significant reduction of the $S_8$ tension in the DS model, while the temperature anistropy power spectrum is still almost indistinguishable from that obtained with the $\Lambda$CDM model, see Fig.~\ref{fig:DlPBES}. The inclusion of a prior on $S_8$ thus leads to a significant improvement of the fit for our DS model as compared to $\Lambda$CDM and NEDE, see Table~\ref{chi2EPBS8}. The latter in particular now performs slightly worse than $\Lambda$CDM, as expected from the previous results without the prior. The NEDE model exhibits a significant residual $2.7\sigma$ tension with the S$_8$ prior. Furthermore, the $H_0$ tension is slightly ameliorated in the DS model, whereas it is slightly exacerbated in the NEDE model. 

\begin{figure*}[t]
	\centering
	\includegraphics[width=0.45\linewidth]{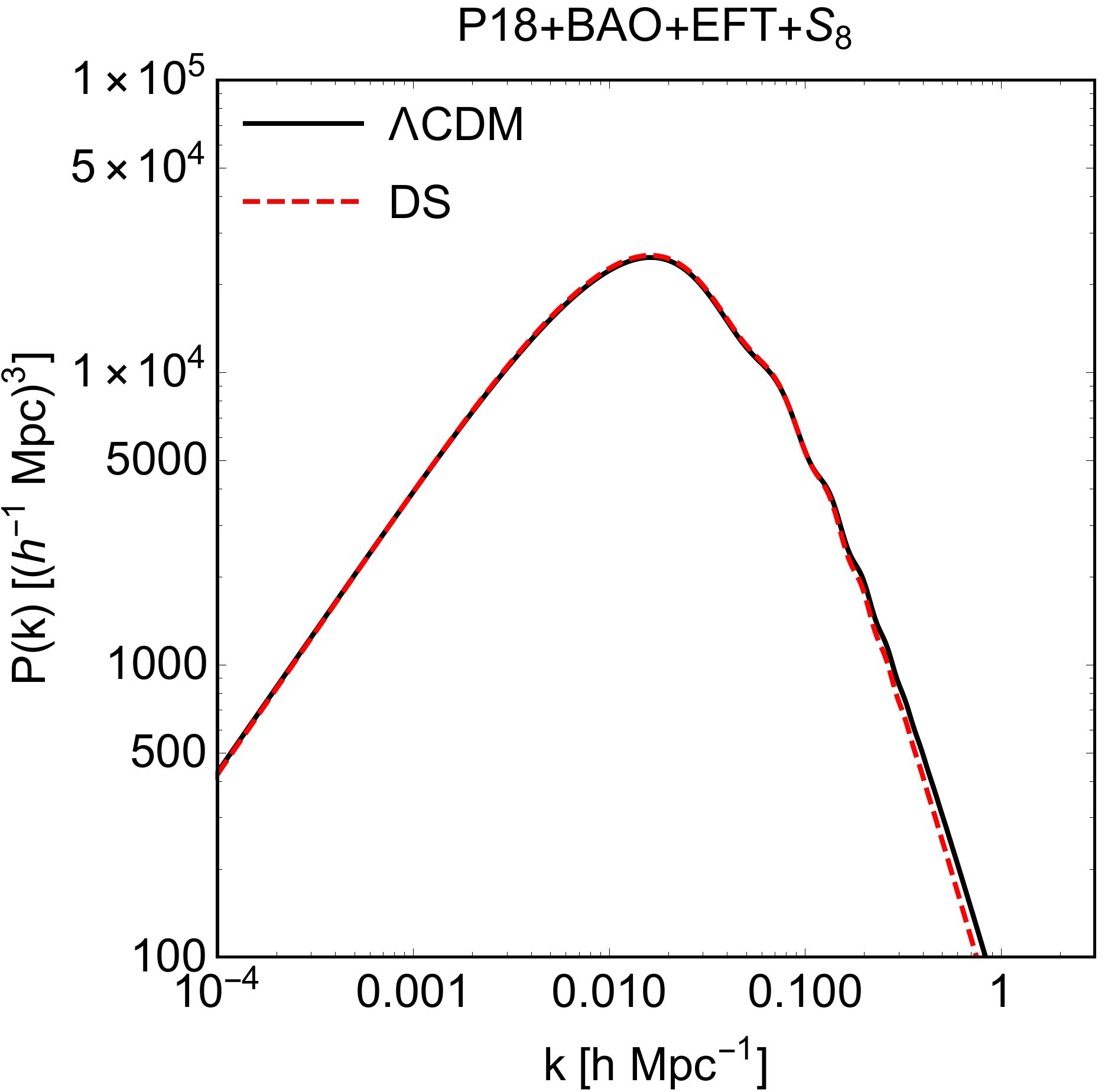}
	\caption{\small Linear matter power spectra at $z=0$ for $\Lambda$CDM (solid, black) and the DS model (dashed, red) for the best fit values of cosmological parameters obtained with the P18+BAO+EFT+S$_8$ dataset.  \label{fig:PkPBES}}
\end{figure*}

\begin{figure*}[h]
	\centering
	\includegraphics[width=0.45\linewidth]{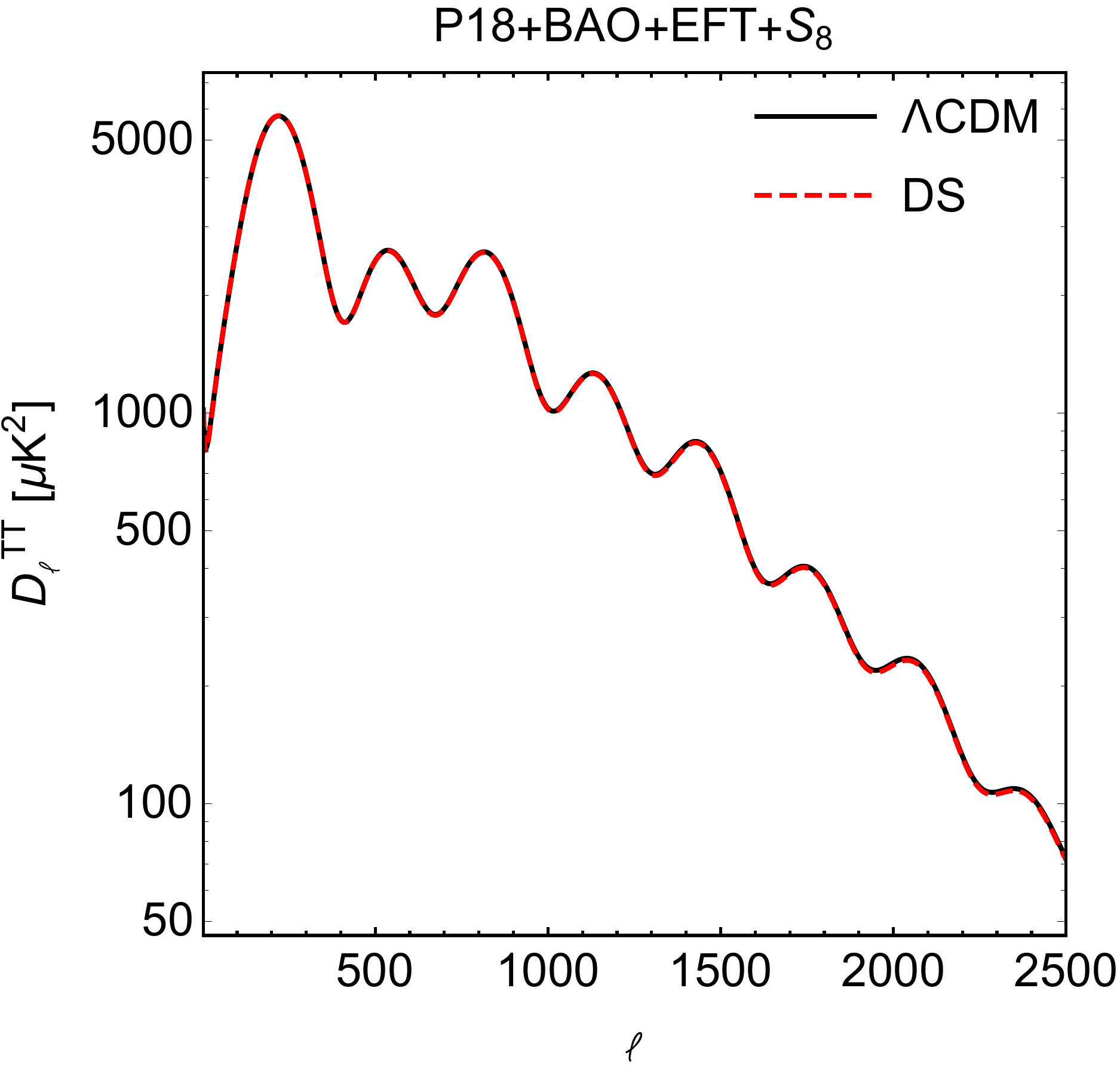}
	\, \,
	\includegraphics[width=0.435\linewidth]{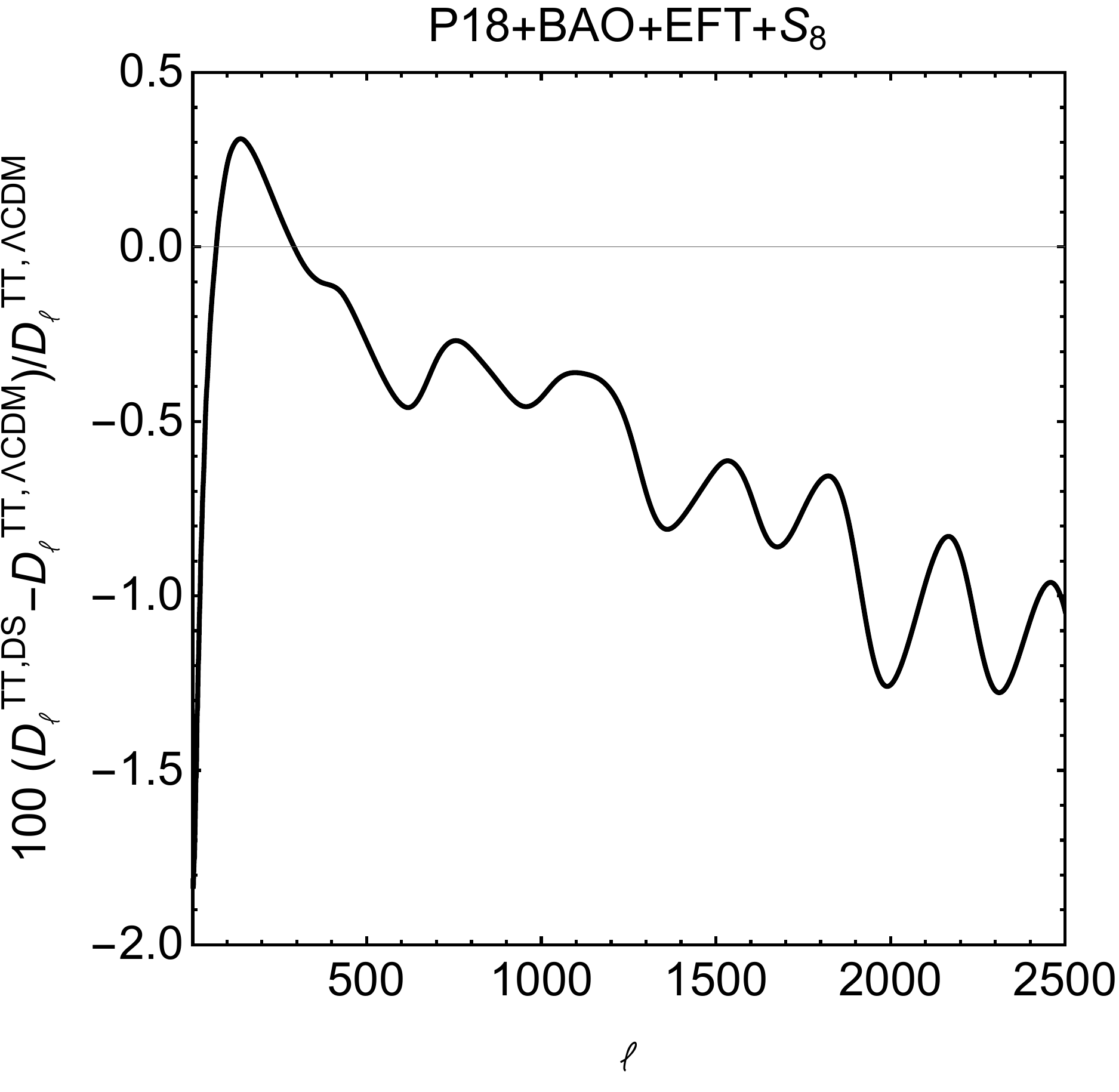}
	\caption{\small {\it Left}: CMB temperature anisotropy power spectra for $\Lambda$CDM (solid, black) and the DS model. {\it Right}: Residuals, multiplied by $10^2$. The curves have been obtained using the best fit values of cosmological parameters obtained with the P18+BAO+EFT+S$_8$ dataset, reported in Table~\ref{table:resultsEPB}. \label{fig:DlPBES}}
\end{figure*}

\begin{table*}[b]
\begin{tabular} {| l | c| c| c|}
\hline\hline
 \multicolumn{1}{|c|}{ Parameter} &  \multicolumn{1}{|c|}{~~~$\Lambda$CDM~~~} &  \multicolumn{1}{|c|}{~~~~~DS~~~~~} &  \multicolumn{1}{|c|}{~~~~NEDE~~~~}\\
\hline\hline
$100 \omega_b              $ & $2.248~(2.25)^{+0.013}_{-0.013}   $ & $2.274~(2.28)^{+0.020}_{-0.026}   $ & $2.258~(2.257)^{+0.017}_{-0.019}   $\\
$\omega_{cdm }             $ & $0.11828~(0.11864)^{+0.00082}_{-0.00084}$ & $0.1191~(0.12)^{+0.0025}_{-0.0035}$ & $0.1203~(0.1209)^{+0.0011}_{-0.0024}$\\
$\ln 10^{10}A_s            $ & $3.039~(3.041)^{+0.014}_{-0.014}   $ & $3.050~(3.047)^{+0.015}_{-0.015}   $ & $3.043~(3.039)^{+0.012}_{-0.015}   $\\
$n_{s }                    $ & $0.9679~(0.9679)^{+0.0036}_{-0.0036}$ & $0.9738~(0.9748)^{+0.0065}_{-0.0083}$ & $0.9722~(0.9691)^{+0.0048}_{-0.0062}$\\
$\tau_{reio }              $ & $0.0534~(0.054)^{+0.0068}_{-0.0068}$ & $0.0557~(0.0545)^{+0.0071}_{-0.0071}$ & $0.0540~(0.0492)^{+0.0060}_{-0.0072}$\\
$H_0\,[\text{km/s/Mpc}]                       $ & $68.12~(67.95)^{+0.38}_{-0.38}     $ & $69.37~(70.02)^{+0.85}_{-1.4}      $ & $68.85~(68.73)^{+0.48}_{-0.90}     $\\
\hline
$F_{dde}                   $ &- & $< 0.127~[95\%]~(0.073)$ & $< 0.078~[95\%]~(0.028)$\\
$z_{dde}                   $ &- & $5055~(4440)^{+1300}_{-1600}  $ & $5214~(5022)^{+1700}_{-3200}  $\\
$r_a\equiv\Omega_a/\Omega_{\text{dm}}                       $ &- & $< 0.069~[95\%]~(0.037)$ &-\\
\hline
$\log_{10}z_a$ &- & fixed to: 4.2 &-\\
$S_8                       $ & $0.8097~(0.8147)^{+0.0090}_{-0.0091}$ & $0.788~(0.783)^{+0.016}_{-0.015}   $ & $0.8124~(0.8154)^{+0.0096}_{-0.0095}$\\
\hline
Tension with SH$_0$ES & $4.0\, \sigma $ & $2.8\, \sigma $ & $3.5\, \sigma $\\
Tension with S$_8$ & $2.6\, \sigma $ & $1.4\, \sigma $ & $2.7\, \sigma $\\
\hline
$\chi^2_{DS}-\chi^2_{\Lambda CDM}$ & - & $-7.7$ & $+0.5$\\
\hline
\end{tabular}
  \caption{The mean (best-fit in parenthesis) $\pm1\sigma$ error of the cosmological parameters obtained by fitting $\Lambda$CDM, our DS model, and the NEDE model to the cosmological datasets: Planck18+BAO+EFT+S$_8$. Upper bounds are presented at 95\% CL.}
  \label{table:resultsEPBS8}
\end{table*}

\begin{table}[h!]
\centering
{\renewcommand{\arraystretch}{1.25} 
{\begin{tabular}{|c | c | c | c |}
\hline
Dataset & $\Lambda$CDM & DS & NEDE \\
\hline
 Planck highl TTTEEE  & 2354.64 & 2355.63 & 2354.72\\
\hline
  Planck lowl EE   & 395.95 & 396.00 & 395.72\\
\hline
  Planck lowl TT  & 22.84 & 22.32 & 22.79\\
\hline
 Planck lensing  & 9.48 & 10.25 &  9.99\\
\hline
   eft with bao highz NGC & 66.94 & 66.27 &  66.86\\
\hline
  eft with bao highz SGC & 61.50 & 61.15 & 62.04  \\
\hline
   eft with bao lowz NGC & 70.44 & 69.96 & 69.85   \\
   \hline
  bao smallz 2014 & 1.43 & 1.77 & 1.59  \\
  \hline
  S$_8$ prior & 9.88 & 2.10 & 10.11\\
\hline
Total  & 2993.12 & 2985.46 & 2993.66\\
\hline
\end{tabular}}
}
\caption{\small Contributions to the total $\chi^2_{\rm eff}$ for individual data sets, for the best-fits of the $\Lambda$CDM, DS, and NEDE models. These were obtained when performing a fit to the data sets P18+BAO+EFT+S$_8$.\label{chi2EPBS8}}
\end{table}

\begin{figure*}[]
\includegraphics[width=\textwidth]{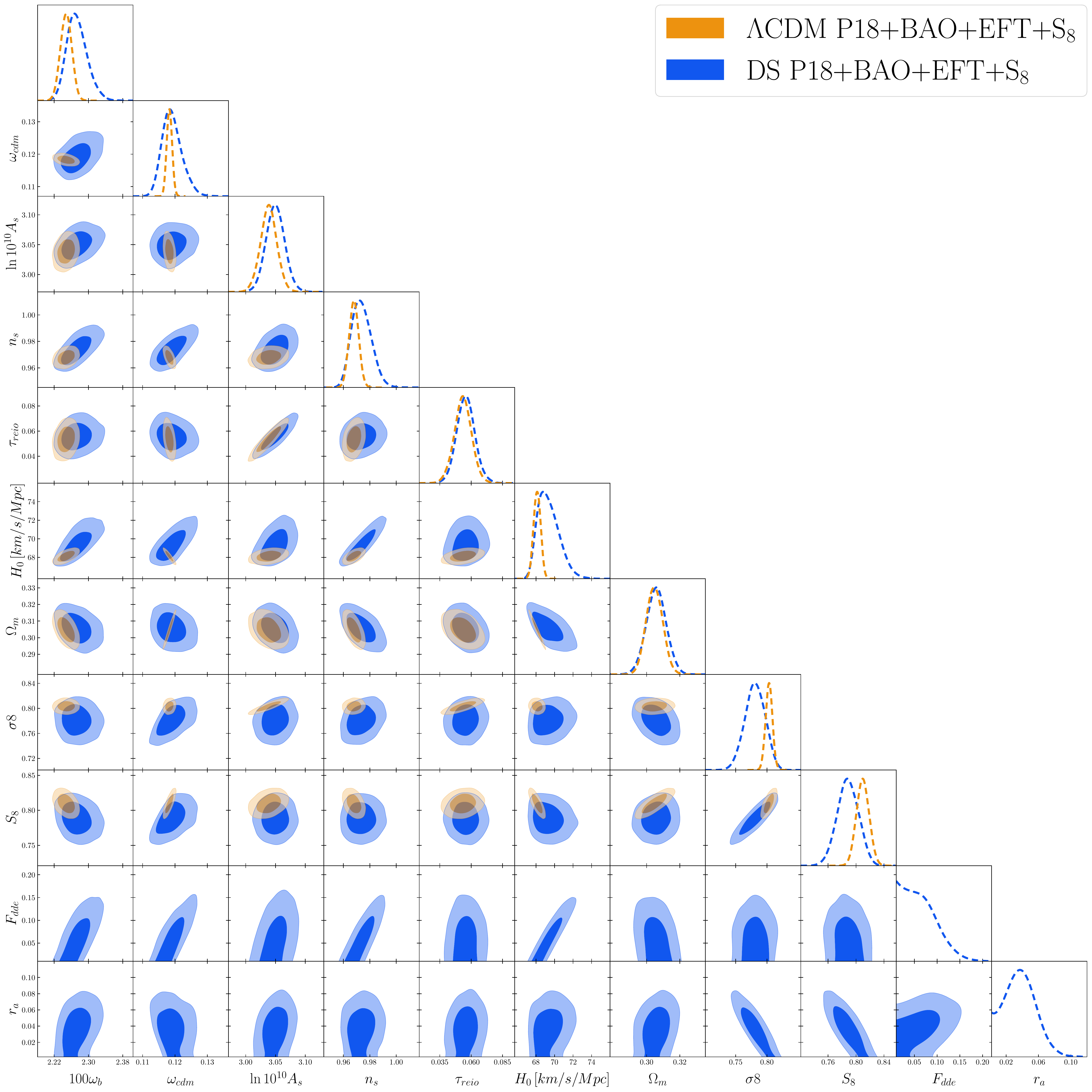}
\caption{Comparison of posterior distributions of $\Lambda$CDM parameters for $\Lambda$CDM and the Dark Sector model, fit to the Planck18+BAO+EFT+S$_8$ data set.}
\label{fig:ELvPBS8}
\end{figure*}
\FloatBarrier


\pagebreak

\subsection*{P18+BAO+EFT+S$_8$+SN+H$_0$}

\begin{figure*}[t]
	\centering
	\includegraphics[width=0.45\linewidth]{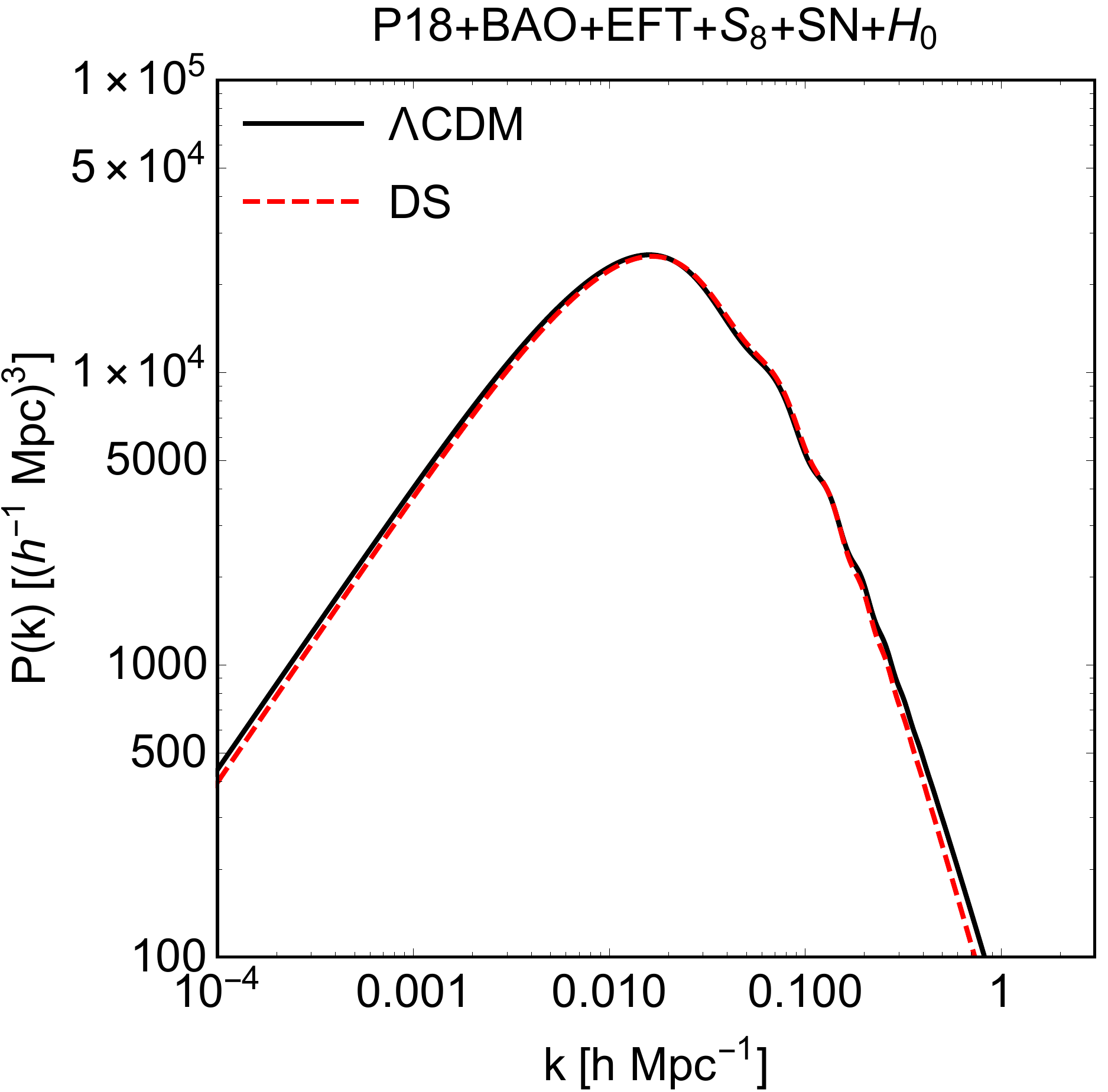}
	\caption{\small Linear matter power spectra at $z=0$ for $\Lambda$CDM (solid, black) and the DS model (dashed, red) for the best fit values of cosmological parameters obtained with the P18+BAO+EFT+S$_8$+SN+H$_0$ dataset.  \label{fig:PkPBESH}}
\end{figure*}

The inclusion of the SH$_0$ES prior and Pantheon data leads to a further significant improvement of the fit for the DS model compared to $\Lambda$CDM. This is mostly due to a dramatically better fit to both the $S_8$ and $H_0$ priors, see Table~\ref{chi2EPBS8H0}. The residual tensions in the values of these parameters are both at the $\sim 1.2-1.3\sigma$ level for the DS model, see Table~\ref{table:resultsEPBS8H0} . Significant preference for the DDE component making up $\sim 13\%$ of the energy density at $z\sim 4,300$ now arises at $4\sigma$. Moderate preference for $\sim 5\%$ of ULA DM arises at $2\sigma$.

The significance of the ULA component further increases with this dataset, and as a consequence the matter power spectrum is suppressed at small scales as compared to $\Lambda$CDM, see Fig.~\ref{fig:PkPBESH}. Nonetheless, the temperature anisotropy power spectrum for the DS model is still almost indistinguishable from that obtained in the $\Lambda$ model, with the maximal size of the residuals being $\Delta D^{\text{TT}}\simeq 1\%$ for the modes of interest, see Fig.~\ref{fig:DlPBESH}.

The NEDE model also features an improved $\chi^2$ compared to $\Lambda$CDM due to a better fit to the SH$_0$ES prior, albeit much less significant than for the DS model. The residual $H_0$ and $S_8$ tensions are at the $\sim 2\sigma$ and $2.7\sigma$ level respectively in the NEDE model.

In order to take into account the number of additional parameters introduced by the DS and NEDE models, we compute the value of $\Delta$AIC for both models with respect to the $\Lambda$CDM model, using~Table~\ref{chi2EPBS8H0}. The NEDE model adds two additional parameters, whereas the DS model adds three (four if also $z_a$ is counted, although it is fixed in our MCMC analysis). We obtain $\Delta$AIC$_{\text{DS}}= -11.9~(9.9)$. This translates to strong evidence for the DS model according to the revised Jeffreys' scale of~\cite{Trotta:2017wnx}. For comparison, we obtain $\Delta$AIC$_{\text{NEDE}}= -4.1$ for the NEDE model, which corresponds to weak evidence instead.

\begin{figure*}[t]
	\centering
	\includegraphics[width=0.45\linewidth]{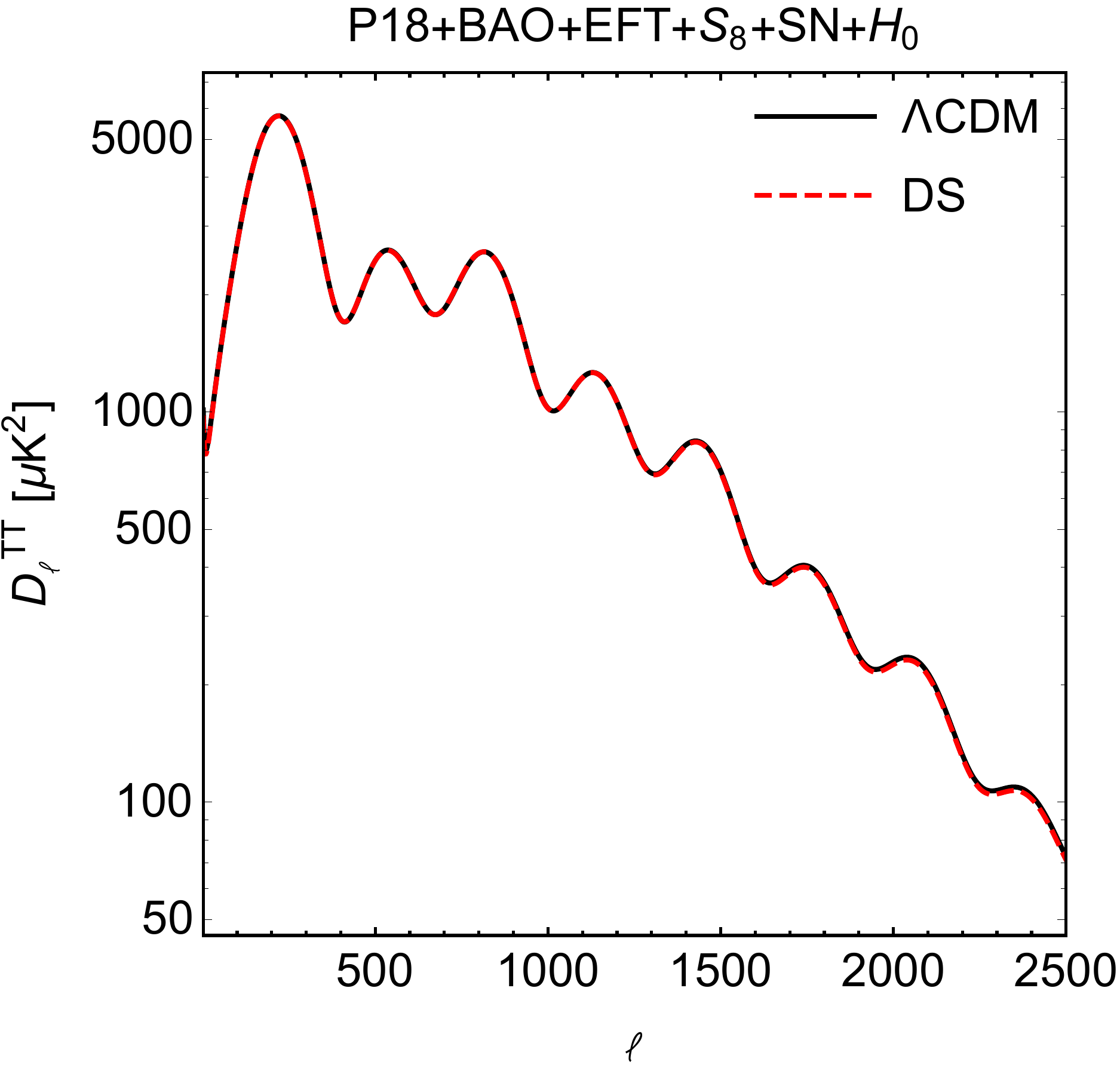}
	\, \,
	\includegraphics[width=0.45\linewidth]{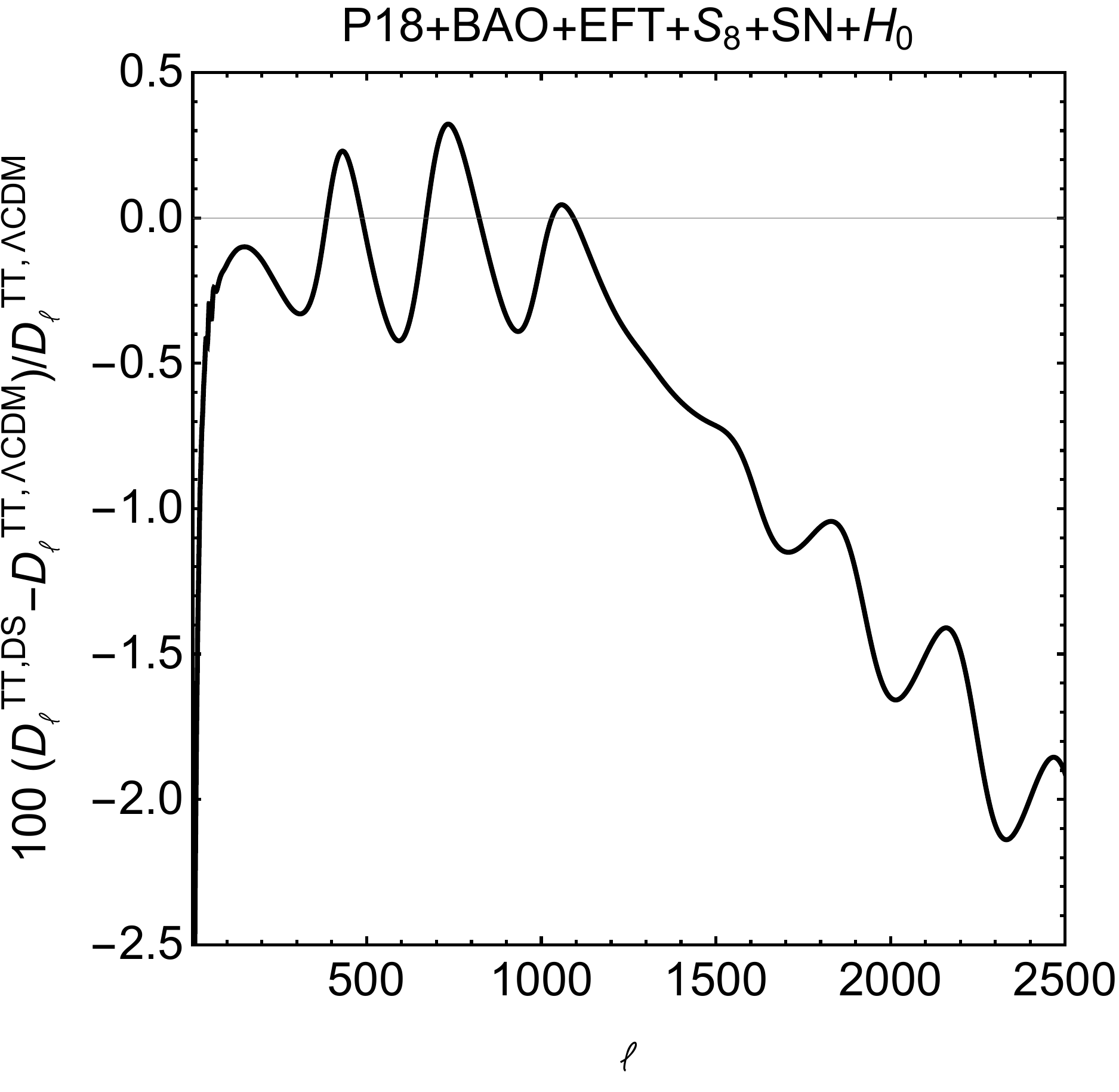}
	\caption{\small {\it Left}: CMB temperature anisotropy power spectra for $\Lambda$CDM (solid, black) and the DS model. {\it Right}: Residuals, multiplied by $10^2$. The curves have been obtained using the best fit values of cosmological parameters obtained with the P18+BAO+EFT+S$_8$+SN+H$_0$ dataset, reported in Table~\ref{table:resultsEPB}. \label{fig:DlPBESH}}
\end{figure*}

\begin{table*}[hbt!]
\begin{tabular} {| l | c| c| c|}
\hline\hline
 \multicolumn{1}{|c|}{ Parameter} &  \multicolumn{1}{|c|}{~~~$\Lambda$CDM~~~} &  \multicolumn{1}{|c|}{~~~~~DS~~~~~} &  \multicolumn{1}{|c|}{~~~~NEDE~~~~}\\
\hline\hline
$100 \omega_b              $ & $2.257~(2.266)^{+0.013}_{-0.013}   $ & $2.303~(2.295)^{+0.023}_{-0.025}   $ & $2.281~(2.283)^{+0.019}_{-0.022}   $\\
$\omega_{cdm }             $ & $0.11745~(0.1169)^{+0.00080}_{-0.00080}$ & $0.1235~(0.1238)^{+0.0030}_{-0.0029}$ & $0.1236~(0.1239)^{+0.0030}_{-0.0038}$\\
$\ln 10^{10}A_s            $ & $3.044~(3.037)^{+0.014}_{-0.015}   $ & $3.062~(3.057)^{+0.015}_{-0.015}   $ & $3.051~(3.058)^{+0.014}_{-0.014}   $\\
$n_{s }                    $ & $0.9703~(0.9722)^{+0.0036}_{-0.0036}$ & $0.9860~(0.9828)^{+0.0065}_{-0.0066}$ & $0.9821~(0.9797)^{+0.0072}_{-0.0074}$\\
$\tau_{reio }              $ & $0.0568~(0.0559)^{+0.0068}_{-0.0074}$ & $0.0574~(0.0562)^{+0.0069}_{-0.0077}$ & $0.0554~(0.0553)^{+0.0064}_{-0.0074}$\\
$H_0\,[\text{km/s/Mpc}]                       $ & $68.54~(68.77)^{+0.36}_{-0.37}     $ & $71.56~(70.99)^{+0.98}_{-0.98}     $ & $70.5~(70.2)^{+1.0}_{-1.2}        $\\
\hline
$F_{dde}                   $ &- & $0.124~(0.123)^{+0.034}_{-0.029}   $ & $0.073~(0.07)^{+0.036}_{-0.038}   $\\
$z_{dde}                   $ &- & $4749~(4894)^{+640}_{-820}  $ & $5058~(4648)^{+840}_{-1600}  $\\
$r_a\equiv\Omega_a/\Omega_{\text{dm}}                       $ &- & $0.048~(0.052)^{+0.017}_{-0.017}   $ &-\\
\hline
$\log_{10}z_a$ &- & fixed to: 4.2 &-\\
$S_8                       $ & $0.8016~(0.7932)^{+0.0085}_{-0.0087}$ & $0.784~(0.789)^{+0.014}_{-0.014}   $ & $0.813~(0.821)^{+0.011}_{-0.011}   $\\
\hline
Tension with SH$_0$ES & $3.7\, \sigma $ & $1.4\, \sigma $ & $2.0\, \sigma $\\
Tension with S$_8$ & $2.2\, \sigma $ & $1.2\, \sigma $ & $2.7\, \sigma $\\
\hline
$\chi^2_{DS}-\chi^2_{\Lambda CDM}$ & - & $-17.9$ & $-6.6$\\
\hline
\end{tabular}
  \caption{The mean (best-fit in parenthesis) $\pm1\sigma$ error of the cosmological parameters obtained by fitting $\Lambda$CDM, our DS model, and the NEDE model to the cosmological datasets: Planck18+BAO+EFT+S$_8$+SN+H$_0$. Upper bounds are presented at 95\% CL. } 
  \label{table:resultsEPBS8H0}
\end{table*}

\begin{table}[h!]
\centering
{\renewcommand{\arraystretch}{1.25} 
{\begin{tabular}{|c | c | c | c |}
\hline
Dataset & $\Lambda$CDM & DS & NEDE \\
\hline
 Planck highl TTTEEE  & 2360.31 & 2355.40 & 2353.76\\
\hline
  Planck lowl EE   & 396.09 & 396.30 & 396.04\\
\hline
  Planck lowl TT  &  22.07 & 21.48 &  21.07\\
\hline
 Planck lensing  & 12.18 & 9.57 &  9.36\\
\hline
   eft with bao highz NGC & 64.80 & 68.50 &  66.16\\
\hline
  eft with bao highz SGC & 62.97 & 60.46 & 63.05  \\
\hline
   eft with bao lowz NGC & 70.34 & 70.51 & 70.18   \\
   \hline
     Pantheon & 1027.57 & 1027.10 & 1026.91   \\
   \hline
  bao smallz 2014 & 2.37 & 1.44 & 2.09  \\
  \hline
  S$_8$ prior & 4.05 & 3.19 & 11.92\\
    \hline
  SH$_0$ES 2019 & 13.70 & 4.57 & 7.13\\
\hline
Total  & 4036.44 & 4018.54 & 4028.32\\
\hline
\end{tabular}}
}
\caption{\small Contributions to the total $\chi^2_{\rm eff}$ for individual data sets, for the best-fits of the $\Lambda$CDM, DS, and NEDE models. These were obtained when performing a fit to the data sets P18+BAO+EFT+S$_8$+SN+H$_0$.\label{chi2EPBS8H0}}
\end{table}

\begin{figure*}[b]
\includegraphics[width=0.8\textwidth]{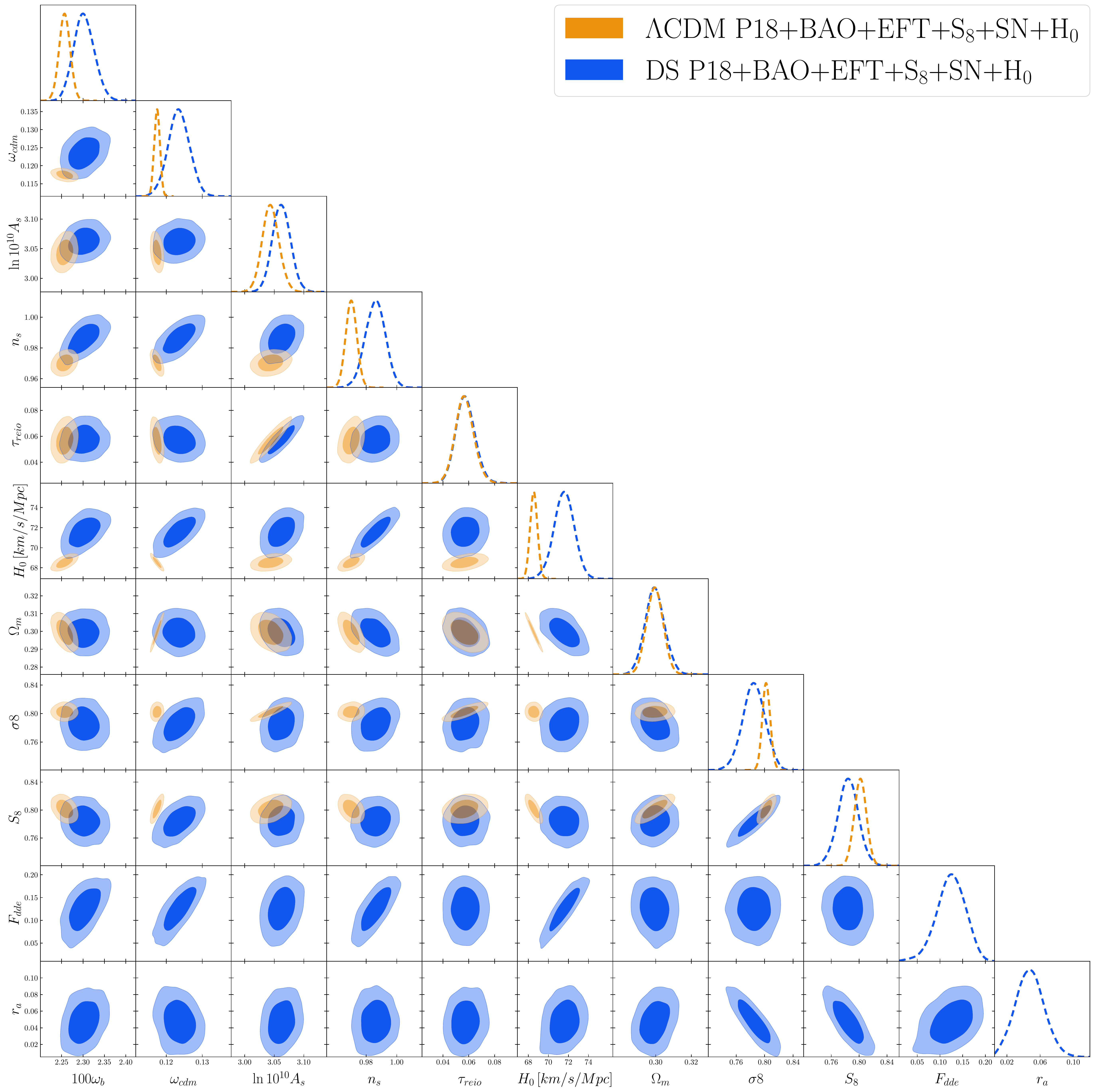}
\caption{Comparison of posterior distributions of $\Lambda$CDM parameters for $\Lambda$CDM and the Dark Sector model, fit to the full Planck18+BAO+EFT+S$_8$+SN+H$_0$ data set.}
\label{fig:ELvPBS8H0}
\end{figure*}

\FloatBarrier

\subsection*{P18+BAO}

\begin{table*}[hbt!]
  
\begin{tabular} {| l | c| c| c|}
\hline\hline
 \multicolumn{1}{|c|}{ Parameter} &  \multicolumn{1}{|c|}{~~~$\Lambda$CDM~~~} &  \multicolumn{1}{|c|}{~~~~~DS~~~~~} &  \multicolumn{1}{|c|}{~~~~NEDE~~~~}\\
\hline\hline
$100 \omega_b              $ & $2.239~(2.241)^{+0.013}_{-0.013}   $ & $2.267~(2.277)^{+0.022}_{-0.026}   $ & $2.262~(2.278)^{+0.020}_{-0.024}   $\\
$\omega_{cdm }             $ & $0.11956~(0.11941)^{+0.00093}_{-0.00093}$ & $0.1241~(0.1261)^{+0.0031}_{-0.0044}$ & $0.1245~(0.127)^{+0.0026}_{-0.0043}$\\
$\ln 10^{10}A_s            $ & $3.048~(3.051)^{+0.014}_{-0.014}   $ & $3.057~(3.051)^{+0.015}_{-0.015}   $ & $3.056~(3.056)^{+0.014}_{-0.016}   $\\
$n_{s }                    $ & $0.9657~(0.9667)^{+0.0037}_{-0.0037}$ & $0.9761~(0.9784)^{+0.0074}_{-0.0089}$ & $0.9754~(0.9816)^{+0.0068}_{-0.0085}$\\
$\tau_{reio }              $ & $0.0565~(0.059)^{+0.0067}_{-0.0074}$ & $0.0565~(0.0518)^{+0.0068}_{-0.0075}$ & $0.0563~(0.054)^{+0.0065}_{-0.0075}$\\
$H_0\,[\text{km/s/Mpc}]                       $ & $67.56~(67.62)^{+0.41}_{-0.42}     $ & $69.3~(69.3)^{+1.0}_{-1.4}        $ & $69.08~(69.91)^{+0.88}_{-1.3}      $\\
\hline
$F_{dde}                   $ &- & $< 0.137~[95\%]~(0.077)$ & $< 0.13~[95\%]~(0.088)$\\
$z_{dde}                   $ &- & $5168~(5452)^{+1100}_{-1300}  $ & $5271~(5295)^{+1200}_{-1600}  $\\
$r_a\equiv\Omega_a/\Omega_{\text{dm}}                       $ &- & $< 0.032~[95\%]~(0.005)$ &-\\
\hline
$\log_{10}z_a$ &- & fixed to: 4.2 &-\\
$S_8                       $ & $0.827~(0.827)^{+0.011}_{-0.011}   $ & $0.827~(0.838)^{+0.016}_{-0.013}   $ & $0.837~(0.84)^{+0.012}_{-0.012}   $\\
\hline
Tension with SH$_0$ES & $4.4\, \sigma $ & $2.7\, \sigma $ & $3.0\, \sigma $\\
Tension with S$_8$ & $3.3\, \sigma $ & $3.1\, \sigma $ & $3.6\, \sigma $\\
\hline
$\chi^2_{DS}-\chi^2_{\Lambda CDM}$ & - & $-4.0$ & $-1.6$\\
\hline
\end{tabular}
  \caption{The mean (best-fit in parenthesis) $\pm1\sigma$ error of the cosmological parameters obtained by fitting $\Lambda$CDM, our DS model, and the NEDE model to the cosmological datasets: Planck18+BAO. Upper bounds are presented at 95\% CL. }
  \label{table:resultsPB}
\end{table*}

With the inclusion of only Planck18 and BAO data, the goodness-of-fit of the DS sector improves very mildly over $\Lambda$CDM, with a decrease in $\chi^2$ of $-4$. However, the DS model reduces the tensions with SH$_0$ES data and the $S_8$ prior down to around 3$\sigma$.

The NEDE model exhibits an even milder improvement in $\chi^2$ compared to $\Lambda$CDM. The tension with SH$_0$ES data is similarly reduced to close to 3$\sigma$, but the tension with the $S_8$ prior is worsened to 3.6$\sigma$ when compared to $\Lambda$CDM.

\begin{table}[h!]
\centering
{\renewcommand{\arraystretch}{1.25} 
{\begin{tabular}{|c | c | c | c |}
\hline
Dataset & $\Lambda$CDM & DS & NEDE \\
\hline
 Planck highl TTTEEE  & 2350.44 & 2348.37 & 2351.21\\
\hline
  Planck lowl EE   & 397.17 & 395.76 & 395.95\\
\hline
  Planck lowl TT  & 23.17 & 21.71 & 21.44\\
\hline
 Planck lensing  & 8.77 & 9.19 & 9.21\\
\hline
 bao fs boss dr12 & 6.72 & 7.39 & 6.63\\ 
\hline
  bao smallz 2014 & 1.17 & 1.05  & 1.36\\
\hline
Total  & 2787.44 & 2783.47  & 2785.80\\
\hline
\end{tabular}}
}
\caption{\small Contributions to the total $\chi^2_{\rm eff}$ for individual data sets, for the best-fits of the $\Lambda$CDM, DS, and NEDE models. These were obtained when performing a fit to the data sets P18+BAO.\label{chi2PB}}
\end{table}

\begin{figure*}[hbt!]
\includegraphics[width=\textwidth]{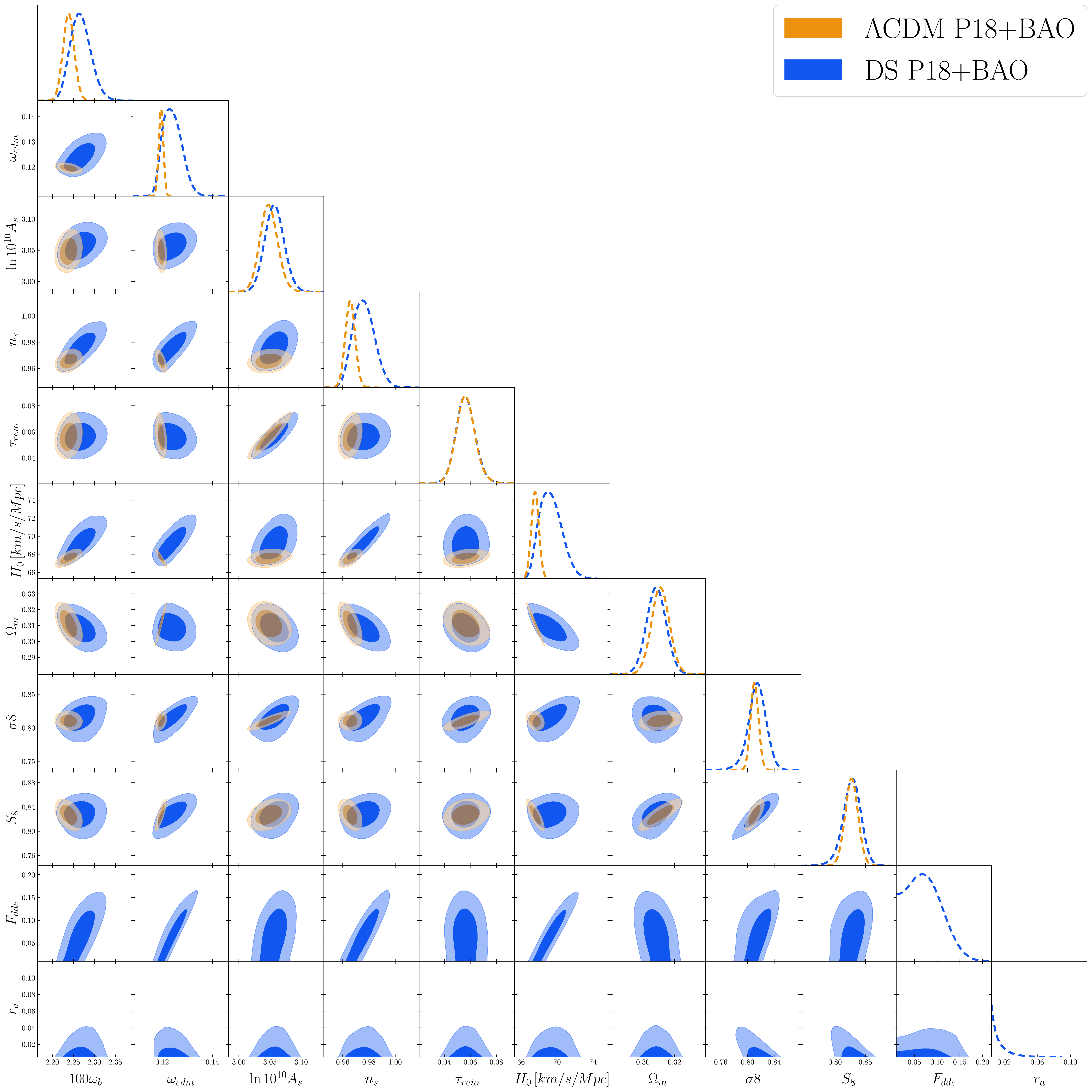}
\caption{Comparison of posterior distributions of $\Lambda$CDM parameters for $\Lambda$CDM and the Dark Sector model, fit to the Planck18+BAO data set.}
\label{fig:LvPB}
\end{figure*}
\FloatBarrier
\pagebreak
\subsection*{P18+BAO+S$_8$}

\begin{table*}[hbt!]
\begin{tabular} {| l | c| c| c|}
\hline\hline
 \multicolumn{1}{|c|}{ Parameter} &  \multicolumn{1}{|c|}{~~~$\Lambda$CDM~~~} &  \multicolumn{1}{|c|}{~~~~~DS~~~~~} &  \multicolumn{1}{|c|}{~~~~NEDE~~~~}\\
\hline\hline
$100 \omega_b              $ & $2.248~(2.254)^{+0.013}_{-0.013}   $ & $2.275~(2.295)^{+0.021}_{-0.028}   $ & $2.260~(2.252)^{+0.016}_{-0.020}   $\\
$\omega_{cdm }             $ & $0.11831~(0.11861)^{+0.00084}_{-0.00084}$ & $0.1201~(0.1214)^{+0.0026}_{-0.0036}$ & $0.1206~(0.1183)^{+0.0013}_{-0.0026}$\\
$\ln 10^{10}A_s            $ & $3.041~(3.042)^{+0.014}_{-0.014}   $ & $3.052~(3.061)^{+0.015}_{-0.016}   $ & $3.044~(3.057)^{+0.013}_{-0.014}   $\\
$n_{s }                    $ & $0.9680~(0.9697)^{+0.0036}_{-0.0036}$ & $0.9751~(0.9783)^{+0.0067}_{-0.0087}$ & $0.9726~(0.9686)^{+0.0048}_{-0.0065}$\\
$\tau_{reio }              $ & $0.0543~(0.0523)^{+0.0070}_{-0.0069}$ & $0.0564~(0.0583)^{+0.0073}_{-0.0073}$ & $0.0544~(0.0616)^{+0.0061}_{-0.0073}$\\
$H_0\,[\text{km/s/Mpc}]                       $ & $68.11~(68.08)^{+0.39}_{-0.39}     $ & $69.51~(70.32)^{+0.90}_{-1.4}      $ & $68.90~(68.28)^{+0.52}_{-0.92}     $\\
\hline
$F_{dde}                   $ &- & $< 0.135~[95\%]~(0.099)$ & $< 0.082~[95\%]~(0.003)$\\
$z_{dde}                   $ &- & $5024~(5145)^{+1300}_{-1500}  $ & $5216~(4377)^{+1600}_{-2900}  $\\
$r_a\equiv\Omega_a/\Omega_{\text{dm}}                       $ &- & $< 0.066~[95\%]~(0.052)$ &-\\
\hline
$\log_{10}z_a$ &- & fixed to: 4.2 &-\\
$S_8                       $ & $0.8101~(0.8134)^{+0.0091}_{-0.0091}$ & $0.793~(0.786)^{+0.016}_{-0.014}   $ & $0.8141~(0.8152)^{+0.0097}_{-0.0096}$\\
\hline
Tension with SH$_0$ES & $4.0\, \sigma $ & $2.7\, \sigma $ & $3.4\, \sigma $\\
Tension with S$_8$ & $2.6\, \sigma $ & $1.6\, \sigma $ & $2.8\, \sigma $\\
\hline
$\chi^2_{DS}-\chi^2_{\Lambda CDM}$ & - & $-6.2$ & $+0.5$\\
\hline
\end{tabular}
  \caption{The mean (best-fit in parenthesis) $\pm1\sigma$ error of the cosmological parameters obtained by fitting $\Lambda$CDM, our DS model, and the NEDE model to the cosmological datasets: Planck18+BAO+S$_8$. Upper bounds are presented at 95\% CL. }
  \label{table:resultsPBS8}
\end{table*}

The addition of the $S_8$ prior leads the DS model to improve over $\Lambda$CDM in goodness-of-fit in a more substantial way, $\Delta\chi^2=-6$.  In addition, the fit of the DS model reduces the $S_8$ tension down to a 1.6$\sigma$ disagreement. In contrast, the same dataset results in $\Lambda$CDM exhibiting a residual 2.6$\sigma$ disagreement with the $S_8$ prior.

The NEDE model, on the other hand, has an increased tension with SH$_0$ES data due to the inclusion of the $S_8$ prior, and a $\chi^2$ which is now greater than that of $\Lambda$CDM. In addition, the tension with the $S_8$ prior is barely reduced, still close to the 3$\sigma$ level.

\begin{table}[h!]
\centering
{\renewcommand{\arraystretch}{1.25} 
{\begin{tabular}{|c | c | c | c |}
\hline
Dataset & $\Lambda$CDM & DS & NEDE \\
\hline
 Planck highl TTTEEE  & 2354.17 & 2353.33 & 2351.77\\
\hline
  Planck lowl EE   & 395.75 & 396.83 & 397.95\\
\hline
  Planck lowl TT  & 22.49 & 21.90 & 23.01\\
\hline
 Planck lensing  & 9.56 & 9.42 & 9.00\\
\hline
 bao fs boss dr12 & 5.94 & 7.09 & 5.91 \\ 
\hline
  bao smallz 2014 & 1.53 & 1.40  & 1.72\\
\hline
 $S_8$ Prior & 9.45 & 2.72 & 10.03\\
\hline
Total  & 2798.90 & 2792.69  & 2799.40\\
\hline
\end{tabular}}
}
\caption{\small Contributions to the total $\chi^2_{\rm eff}$ for individual data sets, for the best-fits of the $\Lambda$CDM, DS, and NEDE models. These were obtained when performing a fit to the data sets P18+BAO+S$_8$.\label{chi2PBS8}}
\end{table}

\begin{figure*}[hbt!]
\includegraphics[width=\textwidth]{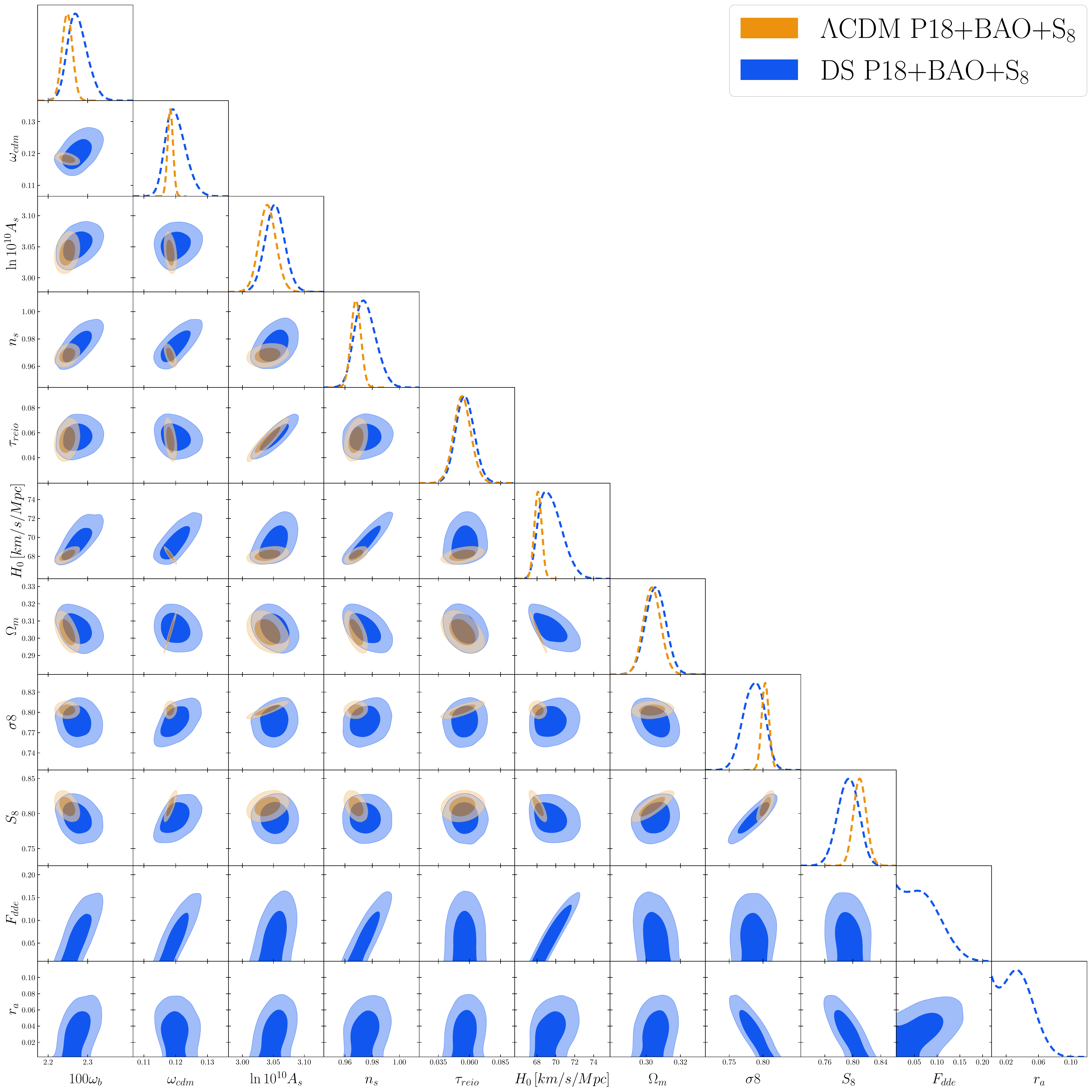}
\caption{Comparison of posterior distributions of $\Lambda$CDM parameters for $\Lambda$CDM and the Dark Sector model, fit to the Planck18+BAO+S$_8$ data set.}
\label{fig:LvPBS8}
\end{figure*}
\FloatBarrier
\pagebreak
\subsection*{P18+BAO+S$_8$+SN+H$_0$}

\begin{table*}[hbt!]
\begin{tabular} {| l | c| c| c|}
\hline\hline
 \multicolumn{1}{|c|}{ Parameter} &  \multicolumn{1}{|c|}{~~~$\Lambda$CDM~~~} &  \multicolumn{1}{|c|}{~~~~~DS~~~~~} &  \multicolumn{1}{|c|}{~~~~NEDE~~~~}\\
\hline\hline
$100 \omega_b              $ & $2.257~(2.254)^{+0.013}_{-0.013}   $ & $2.305~(2.304)^{+0.023}_{-0.026}   $ & $2.284~(2.281)^{+0.020}_{-0.022}   $\\
$\omega_{cdm }             $ & $0.11749~(0.11794)^{+0.00081}_{-0.00082}$ & $0.1244~(0.1233)^{+0.0031}_{-0.0030}$ & $0.1249~(0.1265)^{+0.0032}_{-0.0032}$\\
$\ln 10^{10}A_s            $ & $3.045~(3.039)^{+0.014}_{-0.014}   $ & $3.065~(3.075)^{+0.015}_{-0.015}   $ & $3.054~(3.063)^{+0.013}_{-0.014}   $\\
$n_{s }                    $ & $0.9703~(0.9706)^{+0.0036}_{-0.0036}$ & $0.9872~(0.9894)^{+0.0064}_{-0.0065}$ & $0.9841~(0.9869)^{+0.0068}_{-0.0068}$\\
$\tau_{reio }              $ & $0.0574~(0.0531)^{+0.0068}_{-0.0074}$ & $0.0582~(0.0622)^{+0.0070}_{-0.0077}$ & $0.0552~(0.0585)^{+0.0062}_{-0.0074}$\\
$H_0\,[\text{km/s/Mpc}]                       $ & $68.53~(68.3)^{+0.37}_{-0.37}     $ & $71.7~(72.0)^{+1.0}_{-1.0}        $ & $70.8~(71.2)^{+1.0}_{-1.0}        $\\
\hline
$F_{dde}                   $ &- & $0.130~(0.139)^{+0.035}_{-0.030}   $ & $0.086~(0.101)^{+0.035}_{-0.030}   $\\
$z_{dde}                   $ &- & $4802~(4712)^{+640}_{-800}  $ & $4822~(4509)^{+650}_{-1100}  $\\
$r_a\equiv\Omega_a/\Omega_{\text{dm}}                       $ &- & $0.045~(0.061)^{+0.020}_{-0.018}   $ &-\\
\hline
$\log_{10}z_a$ &- & fixed to: 4.2 &-\\
$S_8                       $ & $0.8026~(0.8058)^{+0.0089}_{-0.0089}$ & $0.789~(0.778)^{+0.016}_{-0.015}   $ & $0.816~(0.825)^{+0.010}_{-0.010}   $\\
\hline
Tension with SH$_0$ES & $3.7\, \sigma $ & $1.3\, \sigma $ & $1.8\, \sigma $\\
Tension with S$_8$ & $2.3\, \sigma $ & $1.4\, \sigma $ & $2.8\, \sigma $\\
\hline
$\chi^2_{DS}-\chi^2_{\Lambda CDM}$ & - & $-18.3$ & $-8.6$\\
\hline
\end{tabular}
  \caption{The mean (best-fit in parenthesis) $\pm1\sigma$ error of the cosmological parameters obtained by fitting $\Lambda$CDM, our DS model, and the NEDE model to the cosmological datasets: Planck18+BAO+S$_8$+SN+H$_0$. Upper bounds are presented at 95\% CL. }
  \label{table:resultsPBS8H0}
\end{table*}

The addition of supernovae data from Pantheon and the SH$_0$ES prior allows for a further reduction in the $S_8$ tension in the DS model, as well as an analogous reduction of the $H_0$ tension down to the 1.3$\sigma$ level. Notice that $\Lambda$CDM retains an almost 4$\sigma$ $H_0$ tension, and still more than 2$\sigma$ disagreement with the $S_8$ prior. Equally notable is the dramatic improvement in the fit of the DS model to the data, with a $\Delta\chi^2 = -17$ as compared to $\Lambda$CDM.

The NEDE model also exhibits an improvement in $\chi^2$ over $\Lambda$CDM, but less dramatically. This can be seen to be due to the fact that, with this dataset, the NEDE model relieves the $H_0$ tension less efficiently than the DS model, and more significantly it still retains a relatively large tension with the $S_8$ prior near 3$\sigma$.

\begin{table}[h!]
\centering
{\renewcommand{\arraystretch}{1.25} 
{\begin{tabular}{|c | c | c | c |}
\hline
Dataset & $\Lambda$CDM & DS & NEDE\\
\hline
 Planck highl TTTEEE  & 2356.17 & 2357.17 & 2354.71\\
\hline
  Planck lowl EE   & 395.80 & 397.83 & 396.25\\
\hline
  Planck lowl TT  & 22.25 & 20.88 & 21.02\\
\hline
 Planck lensing  & 10.60 & 9.57 & 9.67\\
\hline
 Pantheon & 1027.16 & 1026.89 & 1027.08\\
\hline
 bao fs boss dr12 & 5.84 & 6.77 & 7.15\\ 
\hline
  bao smallz 2014 & 1.80 & 2.00 & 2.53 \\
\hline
 $S_8$ Prior & 7.15 & 1.48 & 11.40\\
\hline
  SH$_0$ES 2019 & 16.27 & 2.12 & 4.60\\
\hline
Total  & 3843.03 & 3824.72 & 3834.41 \\
\hline
\end{tabular}}
}
\caption{\small Contributions to the total $\chi^2_{\rm eff}$ for individual data sets, for the best-fits of the $\Lambda$CDM, DS, and NEDE models. These were obtained when performing a fit to the data sets P18+BAO+S$_8$+SN+H$_0$.\label{chi2PBS8H0}}
\end{table}

\begin{figure*}[hbt!]
\includegraphics[width=\textwidth]{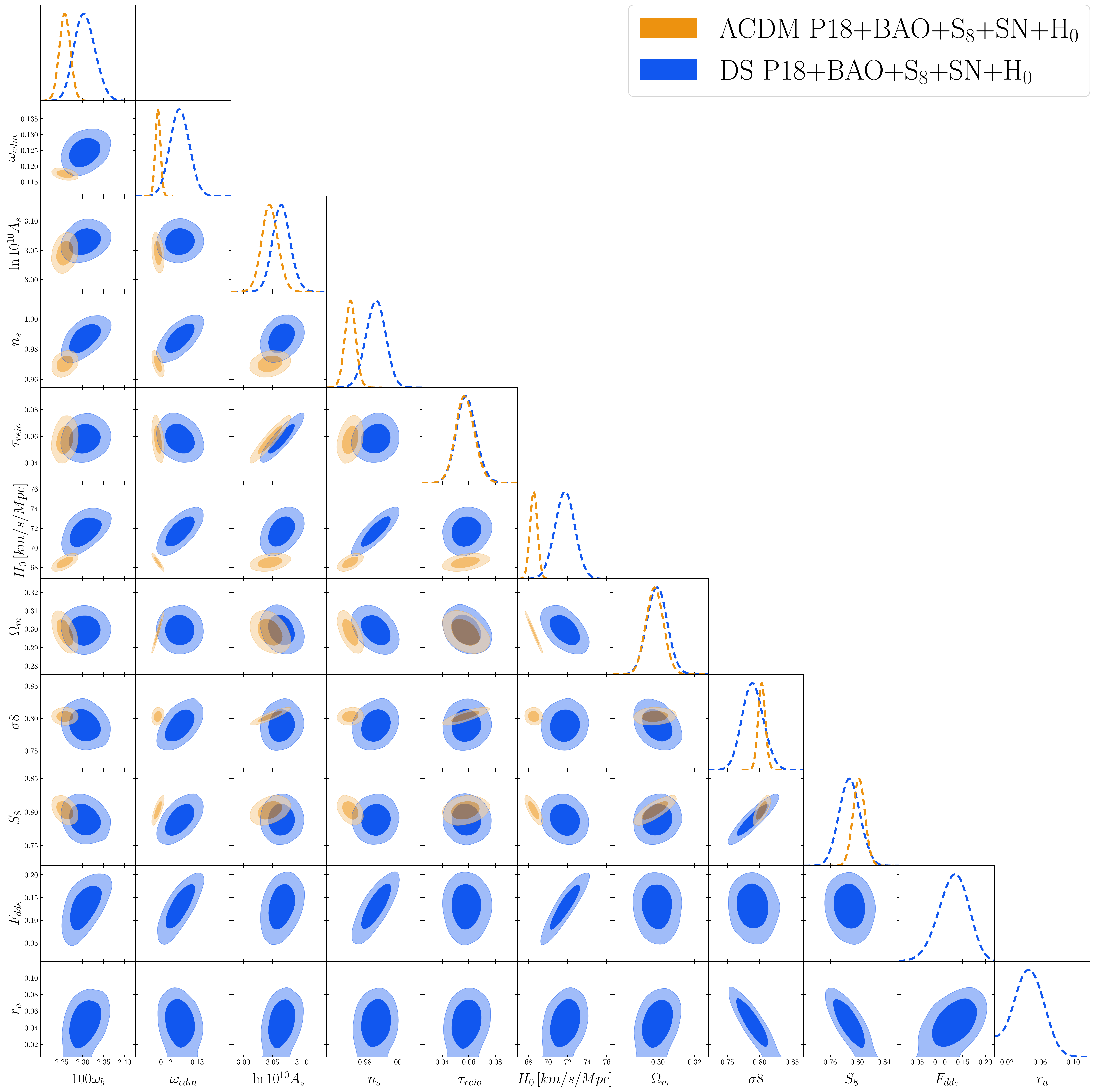}
\caption{Comparison of posterior distributions of $\Lambda$CDM parameters for $\Lambda$CDM and the Dark Sector model, fit to the full Planck18+BAO+S$_8$+SN+H$_0$ data set.}
\label{fig:LvPBS8H0}
\end{figure*}
\FloatBarrier

\end{widetext}

\end{document}